\documentclass[debug]{rmaa}
\usepackage{subfig}
\usepackage{graphicx}
\usepackage{longtable}
\usepackage[colorlinks]{hyperref}   % page numbers and '\ref's become clickable
\usepackage{color}

\usepackage{paralist}

\usepackage{psfrag,color}

\usepackage[utf8]{inputenc}

\title{Study of Blazar activity in 10 year Fermi-LAT data and implications for TeV neutrino expectations} 

\author{
 J. R. Sacahui,\altaffilmark{1} 
  A. V. Penacchioni,\altaffilmark{2,8}
   A. Marinelli,\altaffilmark{3}
   A. Sharma \altaffilmark{4}
   M. Castro,\altaffilmark{5,9}
  J. M. Osorio,\altaffilmark{6}
 and M. A. Morales\altaffilmark{7}}

\altaffiltext{1}{Instituto de Investigaci\'on en Ciencias F\'isicas y Matem\'aticas, USAC, Guatemala.}
\altaffiltext{2}{Instituto de F\'isica de La Plata (IFLP)- CONICET, La Plata, Argentina.}
\altaffiltext{3}{Instituto Nazionale di Fisica Nucleare, Sezione di Napoli, Italy.}
\altaffiltext{4}{Dept. of Physics and Astronomy, Uppsala University, Sweden.}
\altaffiltext{5}{Instituto Nacional de Pesquisas Espaciais (INPE)- Divis\~ao de Astrof\'isica, Brazil.}
\altaffiltext{6}{Instituto de Astronom\'ia, UNAM, M\'exico.}
\altaffiltext{7}{Instituto Balseiro, Universidad Nacional de Cuyo, Argentina.}
\altaffiltext{8}{Dept. of Physics, University of La Plata, Argentina.}
\altaffiltext{9}{Universidade Estadual de Campinas (UNICAMP)-Instituto de Computação, Campinas, Brazil.}

\shortauthor{Sacahui, J. R. et al.}
\shorttitle{Blazar activity}

\fulladdresses{
\item M. A. Morales: Instituto Balseiro, Universidad Nacional de Cuyo - CNEA, 8400 S. C. Bariloche, Río Negro, Argentina. (marco.morales@ib.edu.ar).

\item J. M. Osorio: Instituto de Astronomía, Universidad National Autónoma de México, Mexico 04510, Mexico. (jmosorio@astro.unam.mx).

\item  M. Castro: Instituto Nacional de Pesquisas Espaciais (INPE)- Divis\~ao de Astrof\'isica, 12227-010, S\~ao Jos\'e dos Campos, S\~ao Paulo, Brazil. Universidade Estadual de Campinas (UNICAMP)-Instituto de Computação, 13083-852, Campinas, São Paulo, Brazil.  (manuel.castro@inpe.br).

\item  A. Sharma: Dept. of Physics and Astronomy, Uppsala University, Box 516, S-75120 Uppsala, Sweden. (ankur.sharma@physics.uu.se).

\item A. Marinelli: Instituto Nazionale di Fisica Nucleare, Sezione di Napoli, Complesso Universitario di Monte S. Angelo, Via Cintia ed.G, Napoli, 80126 Italy. (antonio.marinelli@na.infn.it).

\item A. V. Penacchioni: Instituto de F\'isica de La Plata (IFLP)- CONICET, La Plata, Argentina. Dept. of Physics, University of La Plata, c.c.~67, (1900) La Plata, Argentina. (la,
  ana.penacchioni@fisica.unlp.edu.ar).
    
\item J. R. Sacahui: Instituto de Investigaci\'on en Ciencias F\'isicas y Matem\'aticas, ed. T-1, Ciudad Universtiaria, zona 12, Guatemala.
  (jrsacahui@profesor.usac.edu.gt).}

\listofauthors{J. R. Sacahui, A. V. Penacchioni, A. Marinelli, A. Sharma, M. Castro, J. M. Osorio \& M. A. Morales}
\indexauthor{Sacahui, J. R.}
\indexauthor{Penacchioni, A. V.}
\indexauthor{Marinelli, A.}
\indexauthor{Sharma, A.}
\indexauthor{Castro, M.}
\indexauthor{Osorio, J. M.}
\indexauthor{Morales, M. A.}

\abstract{Blazars are the most active extragalactic gamma-ray sources. They show sporadic bursts of activity, lasting from hours to months. In this work we present a 10-year analysis of a sample of bright sources detected by Fermi-LAT (100 MeV - 300 GeV). Using 2-week binned lightcurves (LC) we estimated the Duty Cycle (DC): fraction of time that the source spends in an active state.
The objects present different DC values, with an average of  $22.74\%$ and $23.08 \%$ when considering (and not) the Extragalactic Background Light ( EBL). Additionally we study the so called ``blazar sequence'' trend for the sample of selected blazars in the ten years of data. This analysis constrains a possible counterpart of sub-PeV neutrino emission during the quiescent states, leaving the possibility to explain the observed IceCube signal during the flaring states.}

\resumen{Los blazares son las fuentes de rayos gamma m\'as activas. Presentan per\'iodos esporádicos de actividad, con duraciones de horas a meses. En este trabajo presentamos un an\'alisis de una muestra de fuentes brillantes detectadas por Fermi-LAT (100 MeV- 300 GeV) durante una d\'ecada. Usando curvas de luz con episodios de 2 semanas estimamos el ciclo de actividad (DC): la fracci\'on de tiempo que la fuente pasa en el estado activo. Los objetos presentan diferentes valores del DC, con promedio de $22.74\%$ y $23.08 \%$ cuando se considera y no se considera el fondo de luz difusa. Adicionalmente estudiamos la tendencia conocida como "secuencia de blazares". Este an\'alisis restringe una posible contraparte de neutrinos con energías de sub-PeV durante los estados estacionarios, dejando la posibilidad de explicar las observaciones de IceCube durante per\'iodos de fase activa.}

\addkeyword{Astrophysical Neutrinos}
\addkeyword{Blazars}
\addkeyword{Duty Cycle}
\addkeyword{Extragalactic gamma-rays}
\addkeyword{Fermi-LAT}

\graphicspath{{./Fig4/}, {./Fig5/}}
\begin{document}
% Typeset article header
\maketitle

\section{Introduction}
\label{sec:intro}
Blazars are the most variable and luminous type of Active Galactic Nuclei (AGN). They are among the most powerful emitters in the Universe, and they are thought to be powered by material falling onto a supermassive ($10^6$ - $10^{10}$ $M_{\odot}$)  black hole (BH) at the center of the host galaxy through an accretion disc. The strong, non-thermal electromagnetic emission is generally detected in all observable bands, from radio to gamma-rays, and is dominated by a relativistic jet pointing in the direction of the observer (when the jet is not pointing in this direction, they are catalogued as radio galaxies). Blazars thus reveal the energetic processes occurring in the center of active galaxies. The most compelling mechanism for the emission in the range from radio to optical is synchrotron emission, while gamma-rays are produced via Inverse Compton (IC) by the same relativistic electrons producing the synchrotron emission \citep{Sikora94,2009ApJ...704...38S}. 

Many blazars were first identified as irregular variable stars in our own galaxy. Their luminosity changes in periods that go from days to years, but without a pattern. With the development of radio astronomy many bright radio sources were discovered, most of them with optical counterparts. This led to the discovery of quasars and blazars in the late 50's. The first redshift measurement was done for blazar 3C 273 \citep{1963Natur.197.1040S}. The Fermi satellite, launched in 2008, paved the way for the beginning of a new blazar era thanks to its Large Area Telescope (LAT), which detects sources in the $0.1$ - $300$ GeV energy range \citep{atwood09} with much higher sensitivity than its predecessors. Blazars present high variability in this energy range, with periods in which the flux increases considerably \citep{2009ApJ...700..597A}. Detailed studies of blazar spectra, and in particular their spectral variability \citep{1996rxte.prop10356U,1996ASPC..110..391U}, are fundamental tools to determine the physical processes responsible for particle acceleration and emission in the jet.

Blazars are often divided into two subclasses: BL Lacs \citep{stickel91} and flat spectrum radio quasars (FSRQs) \citep{angel80}, although up to date there is no complete agreement in the selection criteria. BL Lacs are named after their prototype, BL Lacertae. They are characterized by rapid, large-amplitude flux variability and significant optical polarization. One of their defining features is the weakness or absence of emission lines; therefore their redshifts can only be determined from features in their host galaxies' spectra. Besides, their spectra are dominated by a relatively featureless non-thermal emission continuum over the entire electromagnetic range. The nearby BL Lacs, which are associated to elliptical galaxies with typical absorption spectra, sometimes have narrow emission lines, while the more distant BL Lacs, whose host galaxies have never been detected, present broad emission lines. All known BL Lacs are associated with core-dominated radio sources.    

Another type of source are Flat Spectrum Radio Quasars (FSRQ). They are usually more distant, more luminous, and have stronger emission lines than BL Lacs but, due to their similar continuum properties, they are collectivelly called blazars, together with BL Lacs. There are three suggestions regarding their classification: i) FSRQ evolve into BL Lacs, ii) they are different manifestations of the same physical object, and iii) BL Lacs are gravitationally micro-lensed FSRQ \citep{1995PASP..107..803U}. An example of FSRQ is the source 3C 279 (also known as PKS 1253-05). It presents very high variability in the visible, radio and X-ray bands. It is also one of the brightest sources in the gamma-ray sky monitored by the Fermi Space Telescope. 

Blazars have been strong candidate sources of astrophysical neutrinos since 2017, when IceCube reported the first neutrino EHE event (IC-170922A) \citep{Kopper17}  with energy $\sim$ 290 TeV associated with a flare from blazar TXS 0506+056 \citep{eaat1378}. This opened the possibility to establish a link between the neutrino signal and the electromagnetic emission at different energy ranges, which is of key importance in the search for the privileged target of hadronic interaction processes as well as for a pion decay component at VHE.

However, the blazar contribution to the diffuse astrophysical flux measured by IceCube is still a matter of debate~\citep{Aartsen:2016lir}. The diffuse Galactic contribution~\citep{2015ApJ...815L..25G} is constrained by recent analyses from IceCube and ANTARES \citep{2017ApJ...849...67A,2017PhRvD..96f2001A} to just $8.5\%$ of the full sky measured neutrino flux, and large room remains for other AGNs~\citep{2001PhRvL..87v1102A,2016MNRAS.457.3582P,2016PhRvD..94j3006M} and Starbursts galaxies~\citep{2014JCAP...09..043T,2020MNRAS.493.5880P}. In this work we investigate whether the long term observation, which is mainly characterized by the quiescent states, is well described by a leptonic scenario. On the other hand, by concentrating on a possible hadronic component during the flaring states, we remark the importance of the estimation of a DC factor to obtain long term neutrino expectations.

In this paper we analyze a sample of 38 bright blazars: \textbf 19 FSRQ and  \textbf 19 BL Lacs from the Fermi 3FGL catalog \citep{2015ApJS..218...23A}, through 10 years of data (from 2008 to 2018). We made a first selection of the blazars with the highest flux values and, from them, we selected the ones with 'good' LC (not dominated by upper limits). The paper is organized as follows: in Section \ref{sec:data analysis} we analyse the LC for each source and identify the active states. Following two different procedures, we distinguish the steady-state phase from the active phase for each of the sources of the sample. In Section \ref{sec:duty cycle} we calculate the duty cycle (DC) according to the baseline flux obtained with each procedure. In Section \ref{sec:blazar_sequence} we discuss about a possible negative correlation between the blazar's synchrotron peak frequency and the gamma-ray luminosity.
In Section \ref{sec:multimessenger} we consider the DC value and the blazar sequence trend from Section \ref{sec:blazar_sequence} as useful tools to better understand the physical processes at play in the potential neutrino sources. In Section \ref{sec:results} we show our results. Finally, in Section \ref{sec:conclusions} we present our discussion and conclusions.

\section{Data analysis}\label{sec:data analysis}

\begin{figure}
\centering
%\hspace*{-4cm}
\includegraphics[width=0.7\linewidth]{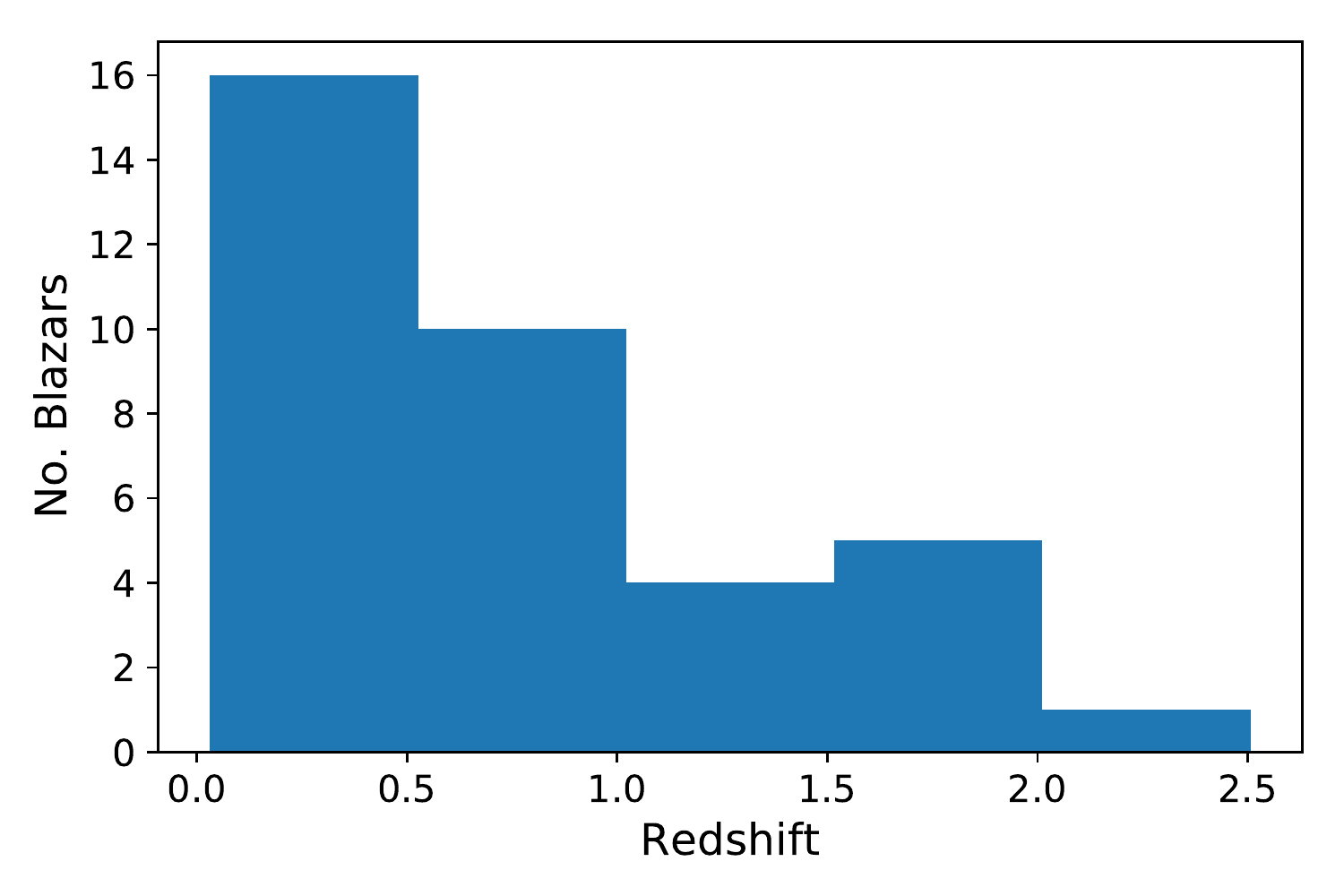}
%\vspace*{-2cm}
\caption{Redshift distribution of the sample of blazars considered in this work.}
\label{fig:HistoRedshift}
\end{figure}

We selected 38 of the blazars with highest flux values from the Fermi 3FGL catalog \citep{2015ApJS..218...23A} as listed in Table \ref{tab:sample+redshift}. Among these, there are 2 with no reported redshift. Figure \ref{fig:HistoRedshift} shows the redshift distribution of the sample. The redshift and spectral type of each source were taken from \citep{2015ApJ...810...14A}. In order to perform the data analysis we made use of the tool \textit{Enrico} \citep{2013Enrico-soft_Sanchez} from the \textit{Fermi Science Tools} (FSSC) software, version v11r5p3\footnote{http://fermi.gsfc.nasa.gov/ssc/data/analysis/software} \citep{2018AAS...23115009A}. The analysis was performed in the energy range of 0.1 - 300 GeV, in a RoI (region of interest) with radius of 20$^\circ$, considering the isotropic and Galactic diffuse emission components (iso\_P8R2\_SOURCE\_v6\_v06.txt and gll\_iem\_v06.fits\footnote{http://fermi.gsfc.nasa.gov/ssc/data/access/lat/BackgroundModels.html}) falling within the RoI. For the cases in which we considered the Extragalactic Background Light (EBL) absorption, we used the model by \citet{Franceschini08} as  reference.

\begin{figure}[h]
\centering
\includegraphics[width=\linewidth]{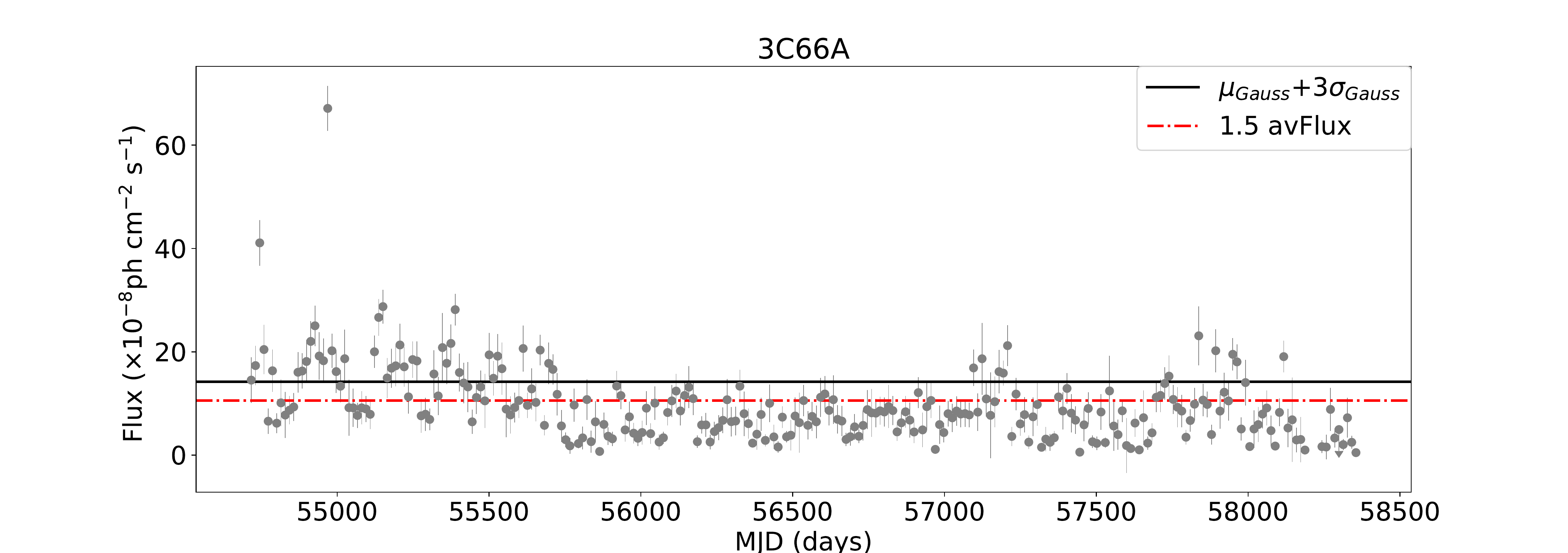}
\includegraphics[width=\linewidth]{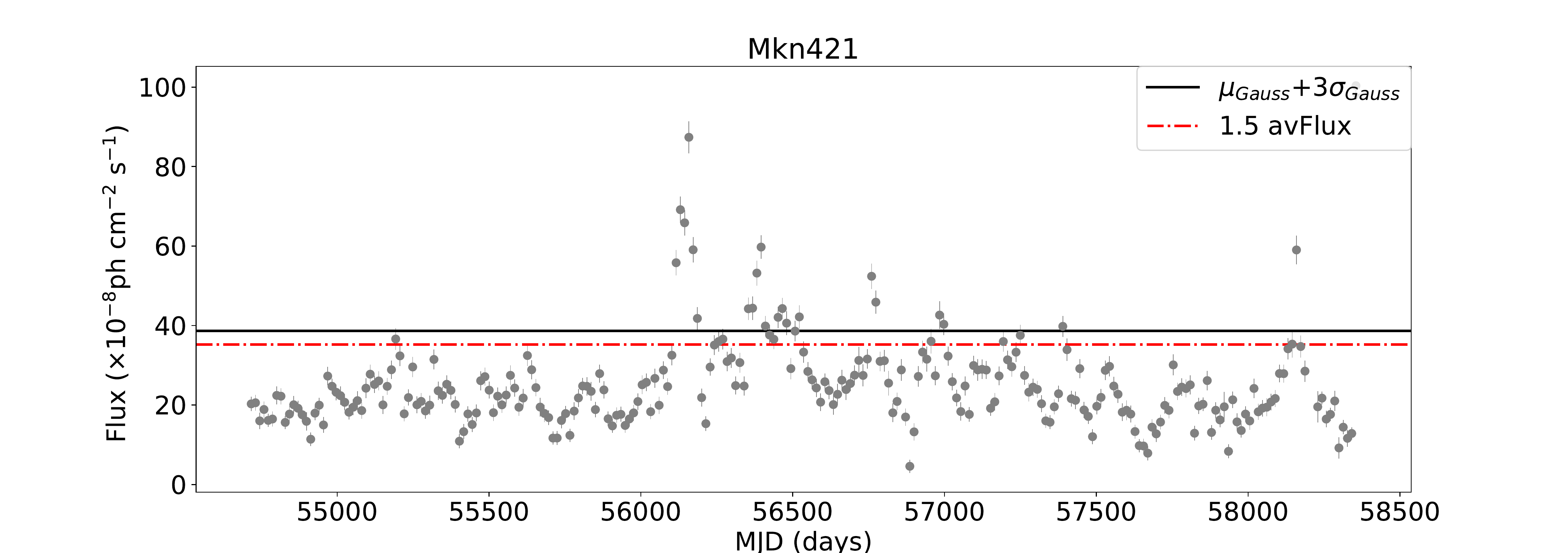}
\caption{Flux light curves (considering EBL absorption) for the sources 3C 66A and Mkn 421 using a two-week binning. The red dashed-dotted line and the black solid line the threshold flux $F_{thr}$ calculated according to Vercellone's criterion and the $\mu_G+3 \sigma$ criterion (Tluczykont).}
\label{fig:2weekLC-3C}
\end{figure}

We generated the LC for the 10-year period with two-month, one-month, two-week and one-week time bins. From these, we selected the two-week binning, since it allowed us to see periods of activity with good resolution while minimizing the number of upper limits (UL). Figure \ref{fig:2weekLC-3C} shows the LC for the sources 3C 66A, a BL Lac in the  Andromeda constellation, and Mkn 421, a BL Lac located in the constellation Ursa Major. We consider them as benchmark sources since they have been intensively studied in the last decade, they are relatively nearby sources ($z=0.444$ \citep{1978bllo.conf..176M} and $z=0.03$ \citep{deVaucouleurs91}, respectively) and both flux and spectral information are available over the entire electromagnetic spectrum.

\begin{table}[!]\centering
\small
\centering
\begin{tabular}{|l|l|l|l|l|}
\hline
3FGL name & Other name & Classification& Redshift & Spectral Type \\
\hline
J1015.0+4925&	1H 1013+498	   & BL Lac &    0.212 & PowerLaw \\
J1229.1+0202&   3C 273         & FSRQ   &    0.16 & LogParabola \\
J1256.1-0547&   3C 279         & FSRQ   &    0.5362 & LogParabola\\
J2254.0+1608&   3C 454.3     & FSRQ   &    0.859 & PLExpCutoff \\
J0222.6+4301&	3C 66A	       & BL Lac &    0.444 & LogParabola\\
J1058.5+0133&   4C +01.28	   & BL Lac &    0.89 & LogParabola \\
J1224.9+2122&   4C +21.35      & FSRQ   &    0.435 & LogParabola \\
J0237.9+2848&   4C +28.07      & FSRQ   &    1.206 & LogParabola \\
J1635.2+3809&   4C +38.41      & FSRQ   &    1.8139 & LogParabola \\
J0238.6+1636&   AO 0235+164    & BL Lac &    0.94 &LogParabola \\
J1522.1+3144&   B2 1520+31     & FSRQ   &    1.4886 & LogParabola\\
J2202.7+4217&   BL Lacertae    & BL Lac &    0.686 & LogParabola\\
J2001.1+4352&   MG4 J2001.1+4352& BL Lac&    -  & PowerLaw   \\
J1104.4+3812&	Mkn421         & BL Lac &    0.031 & PowerLaw \\
J1653.9+3945&	Mkn 501        & BL Lac &	 0.0337 & PowerLaw \\
J1555.7+1111&	PG 1553+113	   & BL Lac &    0.5  & LogParabola \\
J0428.6-3756&   PKS 0426-380   & BL Lac &    1.111 & LogParabola \\
J0449.4-4350&	PKS 0447-439   & BL Lac &	 0.205 & PowerLaw \\
J0457.0-2324&   PKS 0454-234   & FSRQ   &    1.003 & LogParabola\\
J0538.8-4405&	PKS 0537-441   & BL Lac &	 0.892 & LogParabola \\
J0730.2-1141&   PKS 0727-11    & FSRQ   &    1.591 & LogParabola\\
J0808.2-0751&	PKS 0805-07    & FSRQ   &    1.8369 & LogParabola \\
J1246.7-2547&   PKS 1244-255   & FSRQ   &    0.638 & LogParabola \\
J1427.0+2347&	PKS 1424+240   & BL Lac & 	 - &LogParabola    \\
J1504.4+1029&	PKS 1502+106   & FSRQ   &    1.8379 & LogParabola \\
J1512.8-0906&	PKS 1510-08    & FSRQ   &    0.3599 & LogParabola \\
J1625.7-2527&	PKS 1622-253   & FSRQ   &    0.7860 & LogParabola \\
J1833.6-2103&	PKS 1830-211   & FSRQ   &    2.507 & LogParabola\\
J2158.8-3013&	PKS 2155-304   & BL Lac &    0.116 & LogParabola\\
J2236.5-1432&	PKS 2233-148   & BL Lac &    0.3250 & LogParabola\\
J2329.3-4955&	PKS 2326-502   & FSRQ   &    0.518 & LogParabola\\
J1427.9-4206&	PKS B1424-418  & FSRQ   &    1.5220 & LogParabola \\
J1802.6-3940&	PMN J1802-3940 & FSRQ   &    1.319 & LogParabola\\
J2345.2-1554 &	PMN J2345-1555 & FRSQ   &  0.621 & LogParabola \\
J0112.1+2245&	S2 0109+22     & BL Lac &    0.265 & LogParabola \\
J0721.9+7120&	S5 0716+71     & BL Lac &    0.300 & LogParabola\\
J0509.4+0541&	TXS 0506+056   & BL Lac &	 0.3365 & PowerLaw \\
J0521.7+2113&	TXS 0518+211   & BL Lac &  	 0.108 & PowerLaw \\
\hline
\end{tabular}
\caption{List of our sample of blazars. The redshift value ($z$) is indicated when available, and marked with '-' when unknown.}
\label{tab:sample+redshift}
\end{table}

\section{Duty Cycle}\label{sec:duty cycle}

The duty cycle is the level of activity of a source. It is obtained by calculating the fraction of time that the source is in the active phase (or flaring state, with the flux above a given threshold) with respect to the total observation period $T_{obs}$ as a function of the flux:
\begin{equation}
\rm{Duty \, cycle}=DC=\frac{\sum_i t_i}{T_{obs}}.
\end{equation}
Here, $t_i$ is each of the time intervals the source spends in the  flaring state \citep{2004MNRAS.353..890V}. The baseline phase (or steady state) is defined as the fraction of time the flux remains below a given threshold.

In order to separate the active phase from the steady-state phase, we need to define a threshold value for the flux. There are different criteria to do this.

\subsection{Vercellone's criterion} 

Following the work of \citet{2004MNRAS.353..890V}, we define the  threshold flux $F_{\rm{thr}}$ as $1.5$ times the average flux of the source:

\begin{equation}
    F_{\rm{thr}}=1.5 \, \, \overline{F}
\end{equation}
(see red-dotted lines in Figure \ref{fig:2weekLC-3C}). A source is in the active state when the flux of the \textit{i}th time bin is $> F_{\rm{thr}}$ and the  $1 \sigma$ uncertainty of the \textit{i}th time bin is smaller than the difference between the datapoint and $ F_{\rm{thr}}$ \citep{2001ApJ...556..738J}.

The average flux $\overline{F}$ is defined as the weighted mean of all the detections and UL, with weight $w(\sigma) = 1/\sigma$, where $\sigma$ is the uncertainty of the measurement. In the case of UL, $\sigma$ is equal to the UL value. $\overline{F}$ can be decomposed into the fluence from the baseline states (flux below $F_{\rm{thr}}$) and the fluence from the flaring states (flux above $F_{\rm{thr}}$) , respectively:
    \begin{equation}\label{DC}
	    \overline{F} \times T_{\rm{obs}}= F_{\rm{baseline}} \times T_{\rm{baseline}} +\sum_i F_{\rm{flare},i} \, t_i .
    \end{equation}
    Here, $T_{{\rm{baseline}}}$ is the total time the source spends in the baseline state, $F_{\rm{baseline}}$ is the flux of the baseline state, i.e., the weighted average flux of the non-flaring state (below the threshold $F_{\rm{thr}}$), $F_{\rm{flare},i}$ is the average flux of the \textit{i}th flare and $t_i$ is the duration of each flare. Note that, for Vercelone's criterion, the value of $F_{\rm{baseline}}$ is fixed once we fix the value of $F_{\rm{thr}}$.

\subsection{$3 \sigma$ or Tluczykont criterion}

\begin{figure}[h!]
\centering
\includegraphics[width=\linewidth]{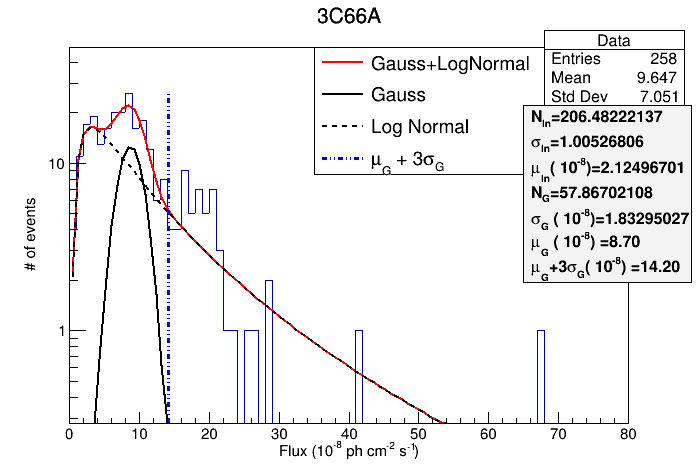}
\includegraphics[width=\linewidth]{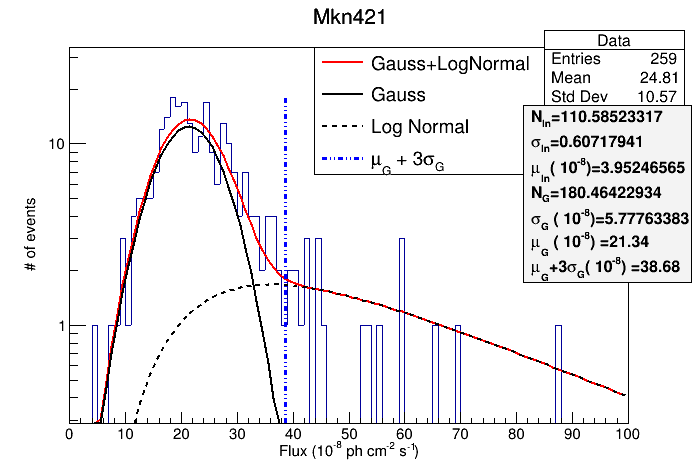}
\caption{Flux distribution for the source 3C 66A and Mkn 421 considering EBL absorption. The black dotted line is the fit with the Log-Normal distribution, the black solid line is the fit with the Gaussian distribution, the red solid line is the fit with the sum of both distributions and the blue dotted line represents the threshold value.}
\label{fig:hist_3C66A}
\end{figure}

According to \citet{2010A&A...524A..48T}, we organized the data into a flux histogram, as shown in Figure \ref{fig:hist_3C66A}.
%, a binned representation of the flux distribution: with the number of flux states in each bin on the y-axis, and the range of fluxes on the x-axis.
A flux state is defined as the integrated flux value in each time bin, obtained from the source LC. We then organized the data into a flux histogram (flux vs. number of events with that value of the flux) and fit them with the sum of a Gaussian and Log-Normal distributions\footnote{Here, $\rm{x}$ represents the values of the flux for a given blazar, while $f_G(\rm{x})$ and $f_{LN}(\rm{x})$ represent the number of occurrences (number of time intervals that have a flux $\rm{x}$).}, $f_G+f_{LN}$. The Gaussian
    \begin{equation}
	    f_G(\rm{x})= \frac{A_G}{\sigma_G \sqrt{2 \pi}} \rm{exp}\left(\frac{-(x-\mu_G)^2}{2 \, \sigma_G^2}\right)
    \end{equation} 
    represents the baseline state and the Log-Normal 
    \begin{equation}
	    f_{LN}(\rm{x})=\frac{A_{LN}}{x \sigma_{LN} \sqrt{2 \pi}} \rm{exp}\left(\frac{-(log(x)-\mu_{LN})^2}{2 \, \sigma_{LN}^2}\right)
    \end{equation}
    describes the active-state phases. The parameters $\mu_G$, $\sigma_G$ and $A_G$ are the mean, standard deviation and normalization for the Gaussian distribution, while $\mu_{LN}$, $\sigma_{LN}$ and $A_{LN}$ are the mean, standard deviation and normalization for the Log-Normal distribution. The parameters of the fits of the entire sample can be found in Appendix \ref{sec:fit-tbl}.  
    
     According to Tluczykont's criterion, the source threshold flux $F_{thr}$ is a variable in the range from $3 \, \sigma_G$ to $\mu_G +3 \, \sigma_G$. We define the flux $F_{baseline}$ as the mean value of all the non-flaring states:
    \begin{equation}
       F_{baseline}= \frac{\int_0 ^{F_{thr}} x \, f_{G}(x) \, dx}{\int_0 ^{F_{thr}} f_{G}(x) \, dx}. 
    \end{equation}
    %Its value lies in the range between $0$ and $\mu_G$ \citep{2014ApJ...782..110A,2014arXiv1401.3348P}. We define the source threshold flux as $F_{\rm{thr}}= F_{baseline} + 3 \, \sigma_G$ so that  $F_{\rm{thr}}$ has a value in the range from $3 \, \sigma_G$ to $\mu_G +3 \, \sigma_G$.

    If the flux is $> F_{\rm{thr}}$ the source is considered to be in the active phase. 
    We recall that \begin{equation}
	 T_{\rm{baseline}}=T_{\rm{obs}} - T_{\rm{flare}},
\end{equation}
with $T_{\rm{flare}}=\sum_i t_i$. 

Following the work of \citet{2014ApJ...782..110A}, we calculated the average flare flux

\begin{equation}
\label{eq:Fflare}
\langle F_{flare} \rangle=\frac{\int_{F_{\rm{thr}}}^{F_{\rm{max}}} x \, f_{LN}(x) \, dx}{\int_{F_{\rm{thr}}}^{F_{\rm{max}}} f_{LN}(x) \, dx}  
\end{equation}
as a function of $F_{\rm{thr}}$ and $F_{\rm{max}}$, the maximum flux detected. 

In order to obtain the average flux of the flaring states we need to replace $F_{flare}$ by $\langle F_{flare} \rangle$. 
Substituting these quantities in Eq. (\ref{DC}) we obtain

\begin{equation}
\label{eq:DC}
DC=\frac{T_{\rm{flare}}}{T_{\rm[obs]}}=\frac{\overline{F}-F_{\rm{baseline}}}{\langle F_{\rm{flare}} \rangle -F_{\rm{baseline}}}.
\end{equation}

\begin{figure}
\centering
\includegraphics[width=0.49\linewidth]{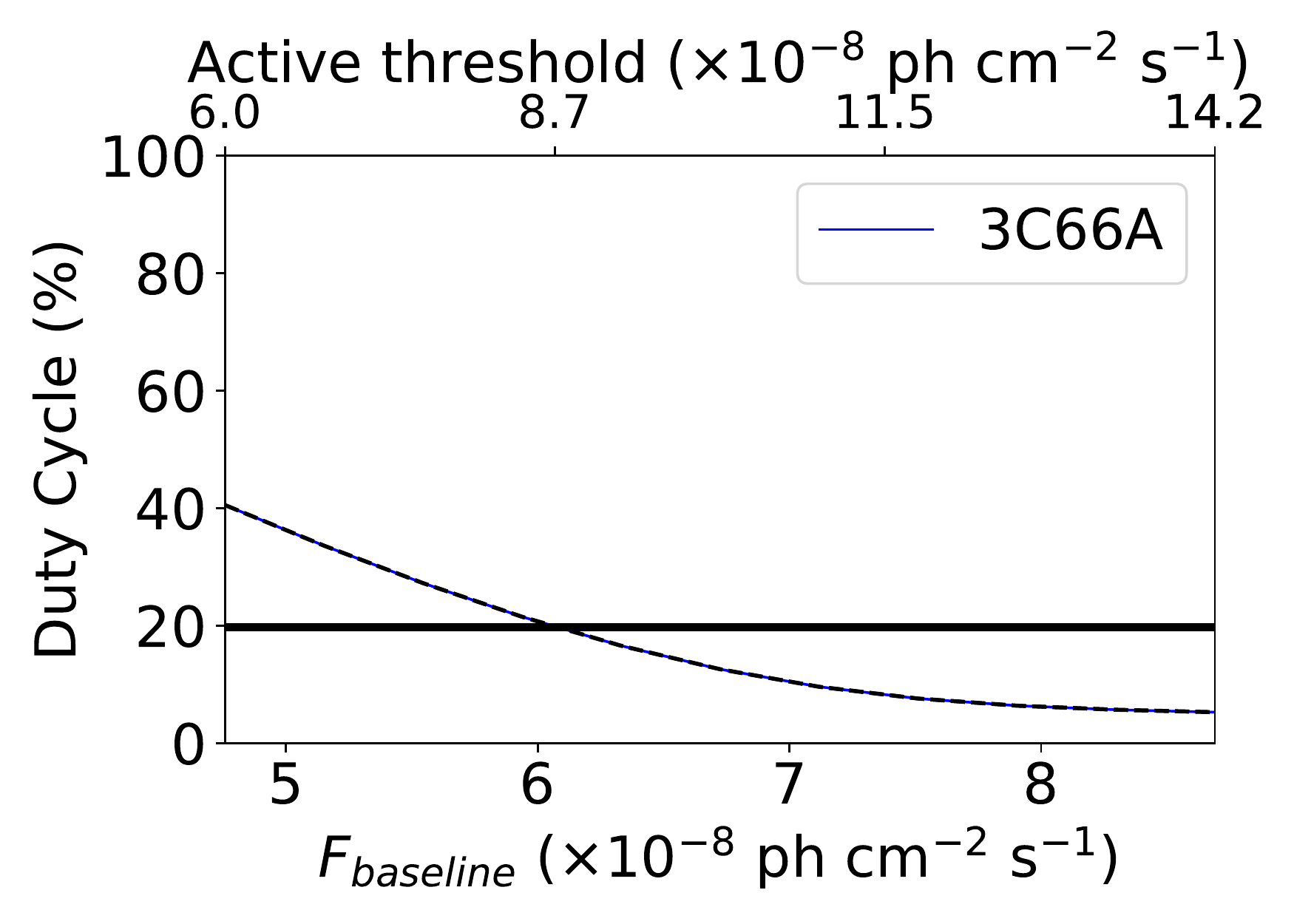}
\includegraphics[width=0.49\linewidth]{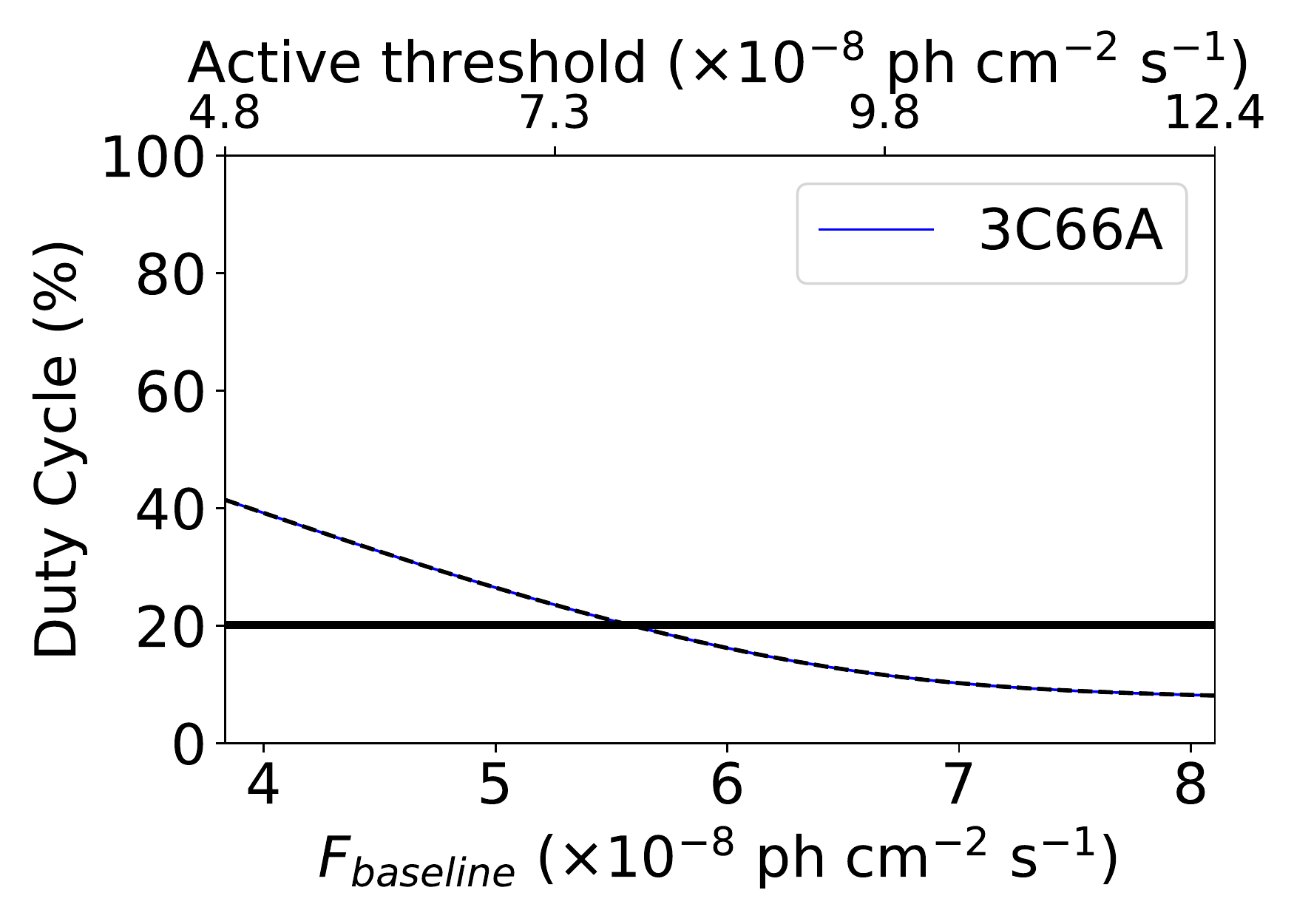}\\
\includegraphics[width=0.49\linewidth]{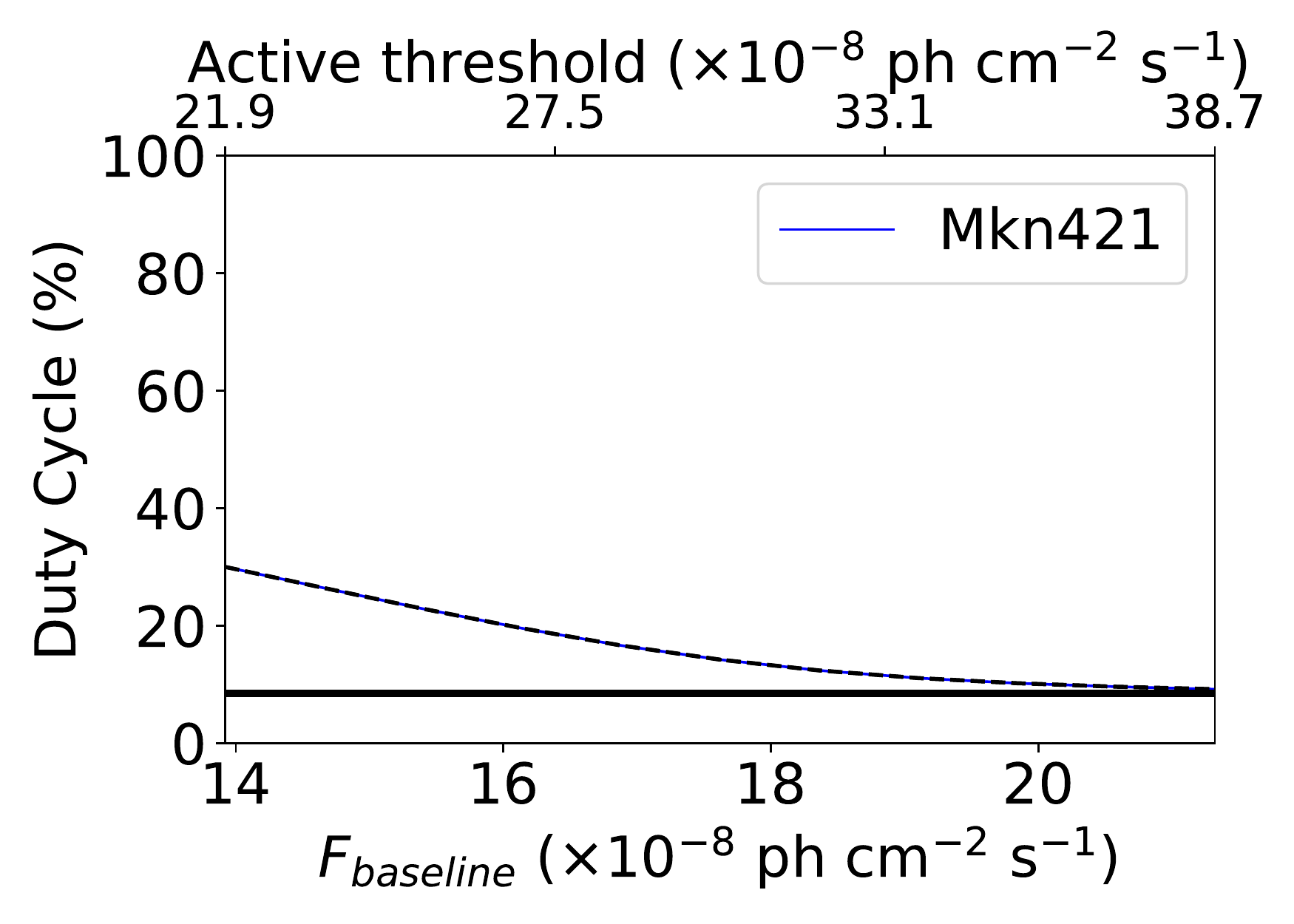}
\includegraphics[width=0.49\linewidth]{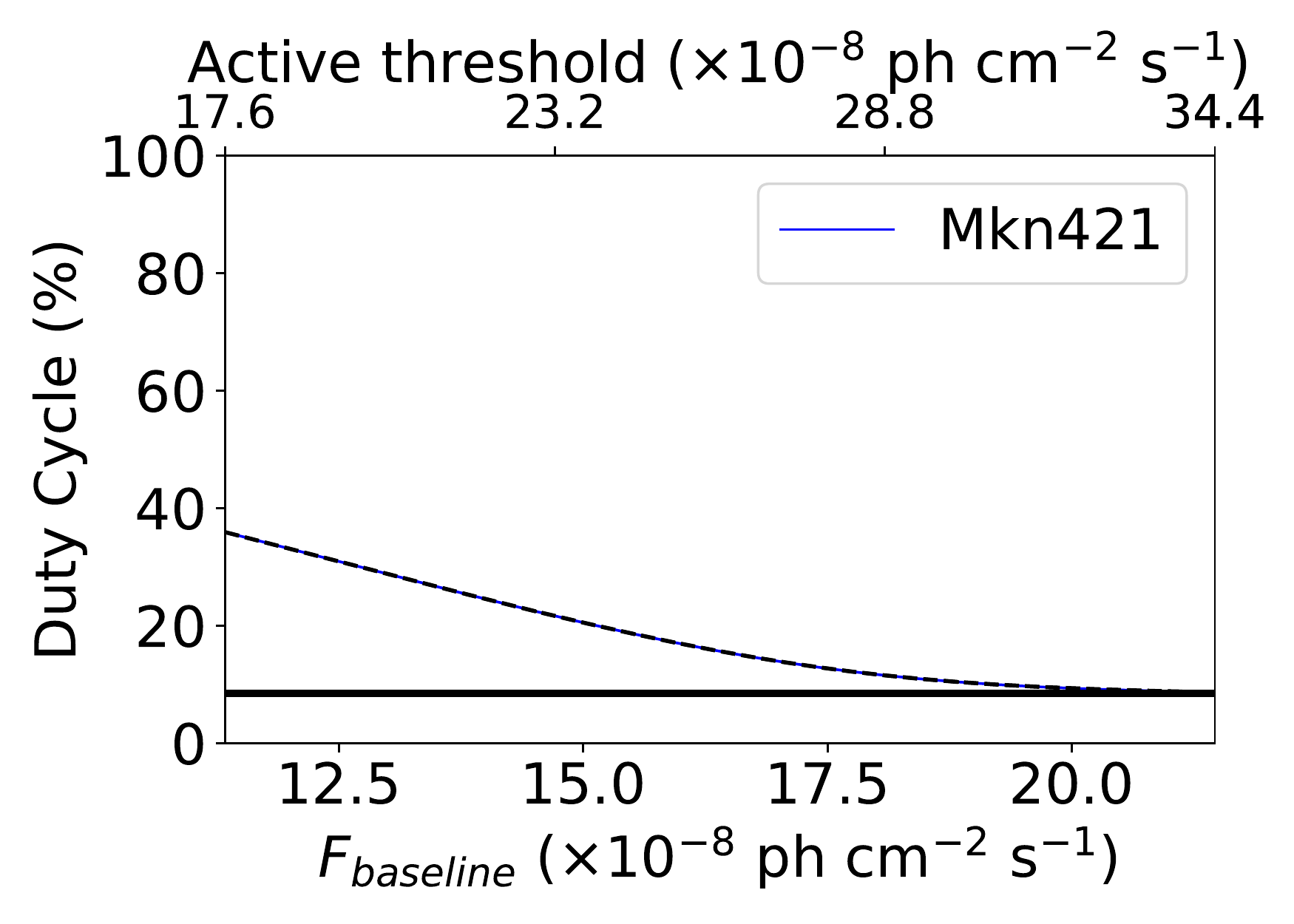}

\caption{DC vs. flux with (left panels) and without (right panels) EBL absorption, for 3C 66A (upper panels) and Mkn 421 (lower panels). The blue (black) lines represent the DC range inferred from Tluczykont (Vercellone's) criterion.}
\label{fig:DC_3C66A}
\end{figure}

Figure \ref{fig:DC_3C66A} shows the plots of the DC obtained for 3C 66A and Mkn 421. In particular, the value of the DC obtained for Mrk 421 (av DC= 24.7\%) is consistent with the results reported by \citep{Pat14,2014ApJ...782..110A} using 3 year Milagro data in the TeV energy range. The DC plots for the whole sample are shown in Appendix \ref{sec:dc-ap}.

The threshold flux calculated for 3C 66A is $F_{\rm{thr}}=1.4 \times 10^{-7}$ ph cm$^{-2}$ s$^{-1}$ (Tluczykont criterion, black solid line in Figure \ref{fig:2weekLC-3C}) and $F_{\rm{thr}}= 1.06 \times 10^{-7}$ ph cm$^{-2}$ s$^{-1}$ (Vercellone's criterion, red dashed-dotted line in Figure \ref{fig:2weekLC-3C}). In the case of Mkn 421, $F_{\rm{thr}}= 3.87 \times 10^{-7}$ ph cm$^{-2}$ s$^{-1}$ (Tluczykont criterion) and $F_{\rm{thr}}= 3.52 \times 10^{-7}$ ph cm$^{-2}$ s$^{-1}$ (Vercellone's criterion). We recall that these values of $F_{\rm{thr}}$ have been calculated considering EBL absorption.

\section{Blazar sequence}\label{sec:blazar_sequence}
\citet{Fossati:1998zn}, \citet{Kubo:1998vm} and \citet{Donato:2001ge} observed a negative correlation between the peak frequency of the synchrotron emission from blazars and their bolometric luminosities (traced through their radio luminosities in their observations). It was named the \textit{observational blazar sequence}, and was thought to be an evidence in favor of leptonic emission from the source. This was later revealed to be a result of selection bias, although the existence of a physical blazar sequence~\citep{ghisellini98} was explained due to the different cooling experienced by the electrons under different environments in BL Lacs and FSRQs.

\begin{figure}[h]
\centering
\includegraphics[width=0.8\linewidth]{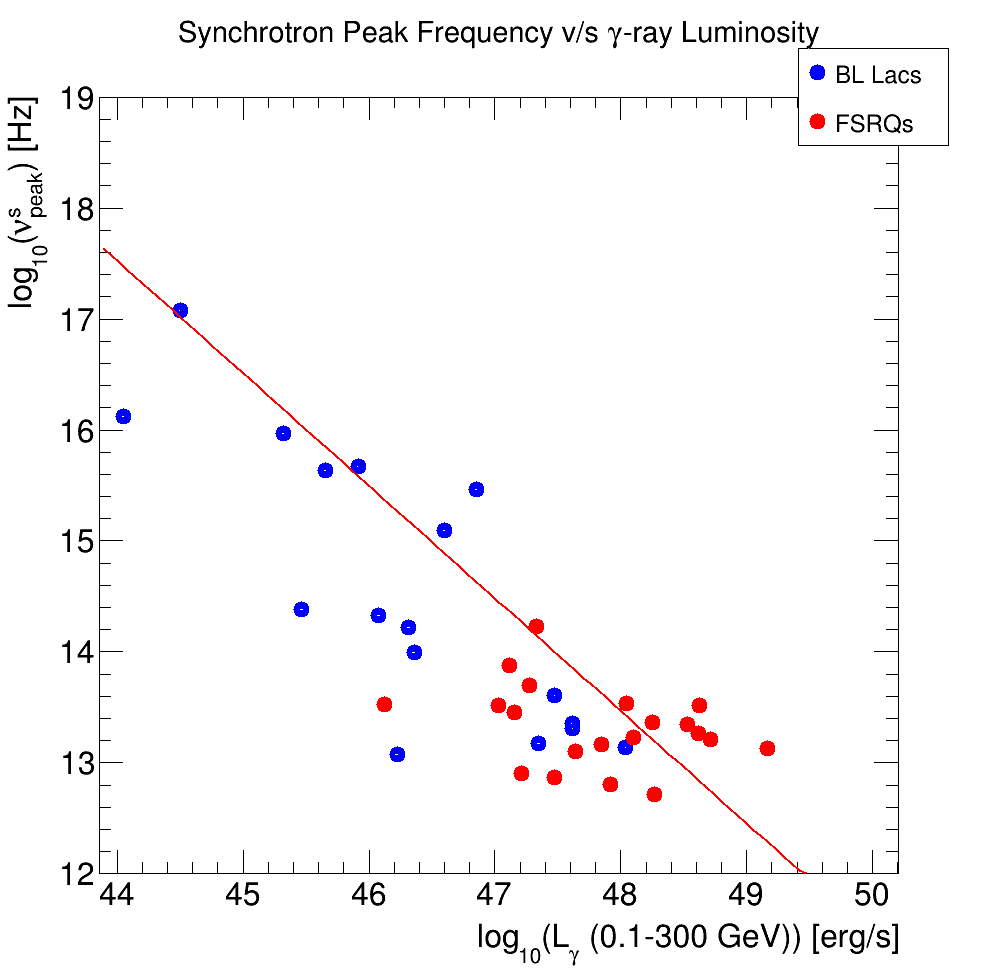}
\caption{Correlation between the synchrotron peak frequency and gamma-ray luminosity (0.1 - 300 GeV) for our sample. BL Lacs are shown in blue and FSRQs in red. The red line represents the best fit for the combined sub-sets.}
\label{fig:sync_peak_lum}
\end{figure}

In the case of blazars, the gamma-ray luminosity can be a better indicator of the integrated bolometric luminosity than the radio luminosity \citep{Fossati:1998zn, 2LAC_bolo_lum}. It is thus possible to explore this relationship between the two variables for our sample, using Fermi-LAT data accumulated over 10 years. Figure \ref{fig:sync_peak_lum} shows the integrated gamma-ray luminosity ($L_{\gamma}$) in the 0.1 - 300 GeV range, averaged over 10 years of observation, vs. the synchrotron peak frequency (${\nu^s}_{\rm{peak}}$) for each source. The value of ${\nu^s}_{\rm{peak}}$ has been obtained from the 3FHL catalog~\citep{2017ApJS..232...18A} when available. Luminosities have been calculated assuming a $\Lambda$CDM Universe with standard cosmological parameters ($H_0$ = $67.8$, $\Omega_m$ = $0.308$, $\Omega_\lambda$ = $0.692$).

The two sub-samples of BL Lacs and FSRQs show a negative correlation trend, the FSRQs having higher luminosity and lower peak frequency on average as compared to the set of BL Lacs. The combined sample of BL Lacs and FSRQs has a Pearson coefficient of -0.3, with a best-fit slope of $-1.0025 \pm 0.0003$. This moderate correlation trend can be interpreted in the context of leptonic emission favored in the inner part of the jet, providing information on the sub-PeV neutrino emission from this region~ \citep{2014PhRvD..90b3007M}. Other authors \citep
{2013HEAD...1311401F} argue that there is a correlation between Compton dominance and the peak frequency of the synchrotron component, which is also consistent with the blazar sequence. 

\section{The Multimessenger case and very high energy neutrino emission}\label{sec:multimessenger}

\begin{figure}[h]
\centering
\includegraphics[width=\linewidth]{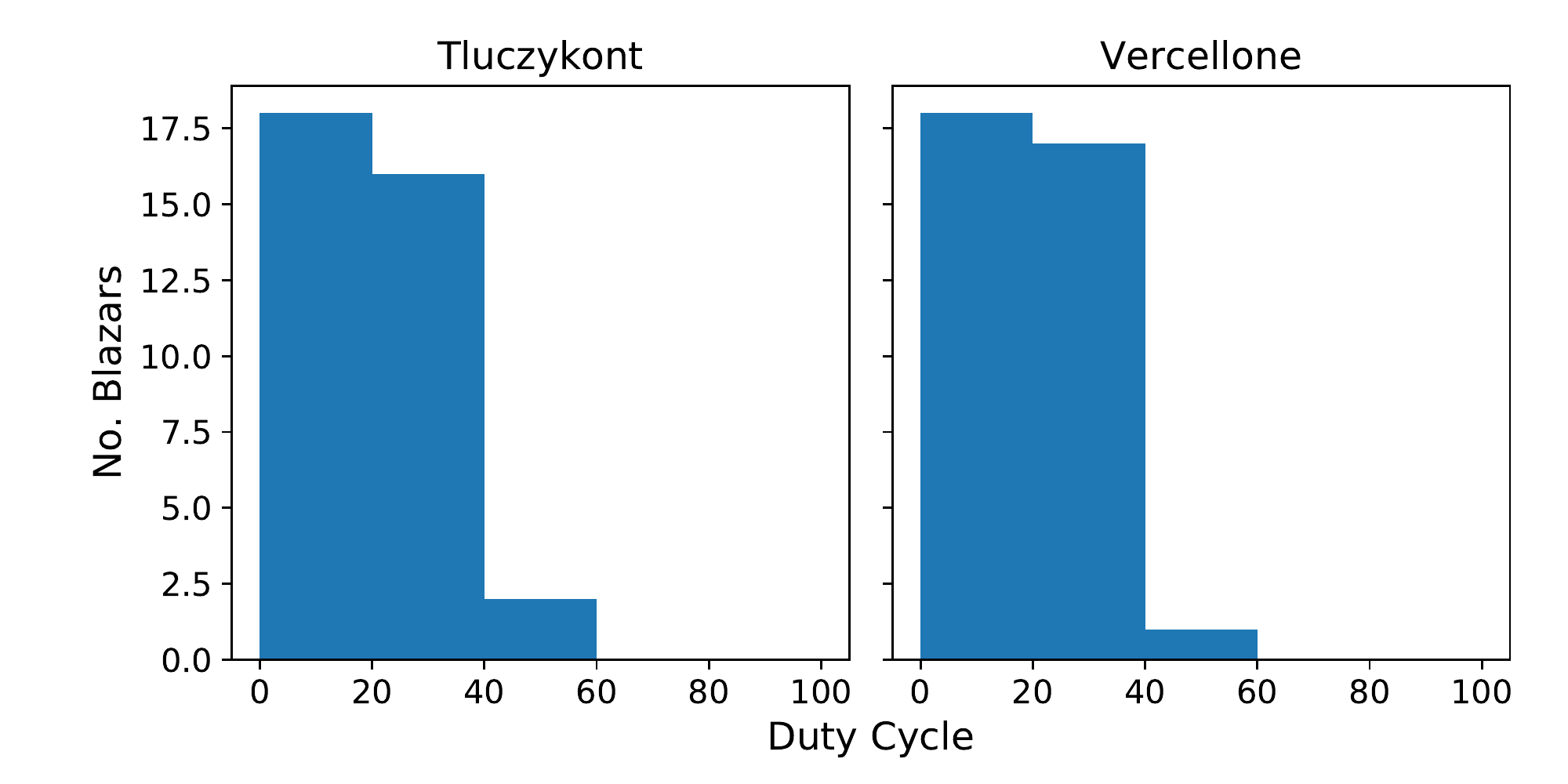}
\includegraphics[width=\linewidth]{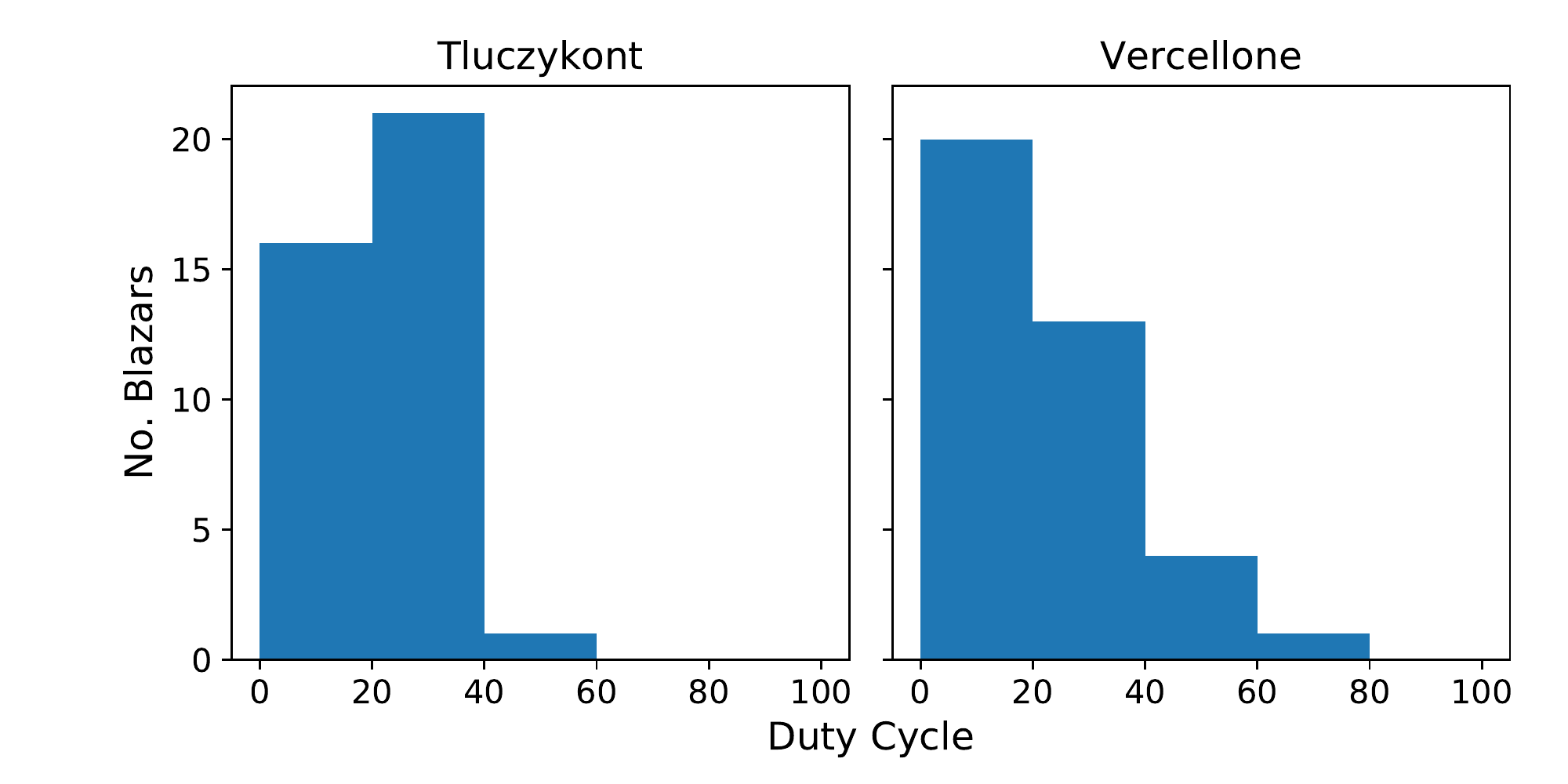}
\caption{DC distribution of our samples according to Tluczykont's (left panels) and Vercellone's (right panels) criteria, considering EBL absortion (upper panels) and not considering it (lower panels).}
\label{fig:Histo_DC}
\end{figure}

In this work, two important multimessenger aspects of gamma-ray emission from blazars were analyzed: the DC value and the blazar sequence trend. 
In particular, they can be used to better understand the physical processes responsible for the gamma-ray emission and to constrain a possible VHE neutrino counterpart.
The first parameter represents a duty factor of the blazar activity, and it is considered a crucial factor whenever a long term unblinded analysis is done with a neutrino telescope, taking into account one or more VHE emitting blazars. In particular, under the assumption that the hadronic emission can be amplified during the VHE flaring states, it is worth using this value to set the expected neutrino observations. Figure \ref{fig:Histo_DC} shows the value of average DCs of the samples obtained with 10 years of Fermi-LAT data. Taking them into account, the known ULs of full-sky neutrino emission expected from blazars can change by a factor $> 2$. In fact, the full sky upper limit reported by IceCube collaboration for blazars considering the unblinded point-like searches~\citep{Aartsen:2016lir} does not consider the DC for the weighting factor $(w_{\gamma/\nu})$ when deducing possible neutrino flux from the gamma-ray emission. 

More in detail, the contribution to the diffuse astrophysical neutrino flux observed by IceCube~\citep{2017ApJ...835..151A,Aartsen:2016xlq,2020arXiv200109520I} attributed to blazars will decrease from 27\% to 7\% in~\citet{Aartsen:2016lir}, for energies $> 10$ TeV, and from 40-80\% to 10-20\% in~\citet{IceCube:2018cha} for energies $> 200$ TeV, when an average DC of 26\% is considered. This decreasing factor on the expected neutrino flux will be quite independent of the blazar gamma-ray luminosity reached during the flaring periods even though the possibility to observe it during such activity period is strictly dependent of that, under the assumption of an amplified hadronic emission on active states.\\
While the DC can represent a temporal constraint on the hadronic activity from BL-lacs and FSRQs, the blazar sequence trend can constrain the emitting process as well as the energy range of VHE neutrinos produced, limiting the production of sub-PeV neutrinos~\citep{Murase:2014foa} by the interaction of jet accelerated protons with the ambient photon fields. The trend shown in Figure \ref{fig:sync_peak_lum} can be explained assuming that a self-synchrotron Compton (SSC) emission scenario takes place in the inner part of the jet. 
%This result cannot be generalized since the photopion interaction can happen also with the photons present in broad line region (BLR) or in the dust torus regions of the blazar. 
Assuming the results of~\citep{Murase:2014foa}, Figure \ref{fig:sync_peak_lum} favors FSRQs with respect to BL Lacs as the main candidates for neutrino production in the range 1-100 PeV.
Moreover, it is worth highlighting the fact that the blazar sequence reported here takes into account the full Fermi-LAT dataset, which is dominated by the quiescent periods and does not exclude a different emission scenario during the flaring periods. If that were the case, Figure \ref{fig:Corn_2d_dc_lum_tflare} shows that we cannot discriminate which class of blazars, BL Lacs or FSRQs, can produce a higher neutrino fluence over a longer observational period.

%\begin{figure}
%\centering
%\includegraphics[width=\linewidth]{tflare_dc_flum_2D.png}
%\caption{Left panel: DC vs. flare duration. Middle panel: Flare luminosity vs. flare duration. Right panel: Flare luminosity vs. DC. The whole sample of BL Lacs and FSRQs is included in all the plots. There is no clear correlation among these quantities. The DC in these plots was calculated using Tluczykont criterion.}
%\label{fig:2d_dc_lum_tflare}
%\end{figure}

\section{Results}\label{sec:results}

Tables \ref{tab:DC_fit_all_withEBL} and \ref{tab:DC_fit_all_withoutEBL} show the DC values for all the sources with and without EBL absortion, according to Tluczykont and Vercellone's criteria. We calculated the DC in the 0.1-300 GeV energy range for $\sim$ 10 years of Fermi-LAT data (2008 - 2018). The table includes: for Tluczykont criterion, the DC obtained considering the maximum baseline flux (DC$_{bl}$\%), the DC obtained assuming 0 as the baseline flux (DC$_{0}$\%), and the average DC for each source (DC$_{av}$\%). For Vercellone's criterion the DC value and its error. The distribution of the values presented in the tables is shown in Figure \ref{fig:Histo_DC}.

\begin{table}
\centering

\small
\caption{DC values calculated for each source considering EBL absorption. For objects without a known redshift, DC has not been calculated}
\label{tab:DC_fit_all_withEBL}
%\begin{tabular}{@{}l@{\:}l@{\:}l@{\:}l@{}} % manual @ spacing to prevent this being too wide for a page
\begin{tabular}{|l|c|c|c||c|c|}
\hline
Source Name & \multicolumn{3}{c ||}{Tluczykont} &   \multicolumn{2}{c |}{Vercellone}  \\
           & DC$_{bl}$\% & DC$_{0}$\% & DC$_{av}$\% &DC \% & error DC \% \\
\hline
1H 1013+498  & 8.59 & 18.18 & 13.39 &	5.81 &	0.01 \\
3C 273    & 15.73 & 19.47 & 17.6 & 17.05 &	0.03 \\
3C 279	  & 14.32 & 15.94 & 15.13 & 22.69 & 0.04  \\ 
3C 454.3	& 16.20 & 16.82 & 16.51 & 11.63 & 0.02  \\ 
3C 66A	  & 5.29 & 40.53 & 22.91 & 19.77 & 0.04  \\
4C +01.28 & 37.12 & 38.73  & 37.93  & 32.56 & 0.06 \\
4C +21.35 & 12.06 & 12.66 & 12.36 & 27.31 & 0.06\\
4C +28.07 & 14.47 & 15.96 & 15.21 & 23.05 & 0.05 \\
4C +38.41 & 20.47 & 22.49 & 21.48 & 28.96 & 0.06 \\
AO 0235+164 & 18.66 & 20.68 & 19.67 & 22.48 & 0.04 \\
B2 1520+31 & 29.00 & 35.86 & 32.43 & 21.58 & 0.05 \\
BL Lacertae & 38.86  & 46.33 & 42.60 & 26.36 & 0.5  \\
MG4 J2001.1+4352 & - & - & - & - & - \\
Mkn 421  & 9.21 & 30.01 & 19.61 & 8.49 & 0.02 \\
Mkn 501   & 15.03  & 37.15  & 26.09  & 7.39 & 0.01 \\
PG 1553+113	 & 4.11 & 27.40 & 15.76	& 3.46 & 0.01\\
PKS 0426-380   & 40.62 & 43.57  & 42.10 & 25.1 & 0.05  \\
PKS 0447-439   & 16.7 & 29.31 & 23.01 & 12.69 & 0.03  \\
PKS 0454-234   & 35.2 & 37.03 & 36.12 & 25.49  & 0.05  \\
PKS 0537-441   & 29.43 & 34.52  & 31.97 & 26.54  & 0.05  \\
PKS 0727-11	   & 16.95 & 23.43 & 20.19 & 19.52 & 0.04 \\
PKS 0805-07	   & 12.75 & 13.44 & 13.09 & 20.08 & 0.04\\
PKS 1244-255 & 20.64 & 27.46 & 24.05 & 18.99 &0.04 \\
PKS 1424+240   & - & - &	- & - & - \\
PKS 1502+106   & 35.32 & 40.84 & 38.08	& 23.17 & 0.05 \\
PKS 1510+089   & 16.64	& 20.14 & 18.39	& 19.62& 0.04 \\
PKS 1622-253   & 31.86	& 33.38 & 32.62	& 28.4& 0.06 \\
PKS 1830-211   & 16.11	& 23.56 &	19.83	& 12.79 & 0.03 \\
PKS 2155-304  & 11.53	& 20.31 & 15.92	& 14.62& 0.03 \\
PKS 2233-148   & 12.25	& 13.13 & 12.69	& 22.83& 0.05 \\
PKS 2326-502   & 14.87	& 23.12 & 18.99	& 41.09 & 0.08 \\
PKS B1424-418   & 14.57	& 21.51 & 18.04	& 19.18& 0.04  \\
PMN J1802-3940   & 21.75 & 22.22 &	21.98	& 30.24& 0.06 \\
PMN J2345-1555   & 19.39 & 28.6 &	24.00	& 16.33& 0.04 \\
S2 0109+22	   & 2.75 & 58.24  & 30.49  & 25  & 0.05   \\ 
S5 0716+71	   & 0.7 & 15.89 & 8.3 & 16.02 & 0.03   \\ 
TXS 0506+056   & 14.16 & 26.3 & 20.23 & 13.51 & 0.03  \\
TXS 0518+211   & 13.54 & 26.41 & 19.98 & 13.33 & 0.03  \\
\hline
Mean values & 18.24   & 27.24   &  22.74  & 20.09  &   \\ \hline
\end{tabular}
\end{table}

\begin{table}
\centering

\small
\caption{DC values calculated for each source without EBL absorption.}
\label{tab:DC_fit_all_withoutEBL}
%\begin{tabular}{@{}l@{\:}l@{\:}l@{\:}l@{}} % manual @ spacing to prevent this being too wide for a page
\begin{tabular}{|l|c|c|c||c|c|}
\hline
Source Name & \multicolumn{3}{c ||}{Tluczykont} &   \multicolumn{2}{c |}{Vercellone}  \\
           & DC$_{bl}$\% & DC$_{0}$\% & DC$_{av}$\% &DC \% & error DC \% \\
\hline
1H 1013+498  & 11.46 & 40.14 & 25.80	 & 3.49	 & 0.01	 \\
3C 273    & 16.63 & 21.03 & 18.83 & 16.99  &	0.03 \\
3C 279	  & 14.04 & 14.99 & 14.52 & 7.72  & 0.02  \\ 
3C 454.3  & 16.2 & 16.82 & 16.51 & 11.63 & 0.02  \\ 
3C 66A	  & 8.12  & 41.42  & 24.77 & 20.16 & 0.04   \\
4C +01.28 & 31.52  & 32.12 & 31.82 & 32.56 & 0.07  \\
4C +21.35 & 15.54 & 16.70 & 16.12 & 4.64 & 0.01 \\
4C +28.07 & 19.39 & 23.24 & 21.32 & 23.83 & 0.05 \\
4C +38.41 & 12.86 & 14.93 & 13.89 & 4.26 & 0.01  \\
AO 0235+164 & 18.37 & 20.87 & 19.62 & 26.25 & 0.05 \\
B2 1520+31 & 35.27 & 43.21 & 39.24 & 22.78 & 0.04 \\
BL Lacertae & 41.20 & 49.60 & 45.4 & 26.36 & 0.05   \\
MG4 J2001.1+4352 & 7.40 & 13.24 & 10.32 & 13.18 & 0.03  \\
Mkn 421  & 8.67 & 35.93 & 22.3 & 8.49 & 0.02 \\
Mkn 501        & 12.97 & 33.92 & 23.45 & 6.98 & 0.01 \\
PG 1553+113	   & 13.52 & 22.35 & 17.93	& 11.58  & 0.02 \\
PKS 0426-380   & 37.2 & 41.14 & 39.17 & 53.46 &0.1   \\
PKS 0447-439   & 13.59 & 27.59 & 20.59 & 12.69 & 0.03  \\
PKS 0454-234   & 33.71 & 40.53 & 37.12 & 21.24 & 0.05  \\
PKS 0537-441   & 28.36 & 31.41 & 29.88 & 55.21 & 0.1  \\
PKS 0727-11	   & 27.75  & 37.59 & 32.67 & 18.38 & 0.04 \\
PKS 0805-07	   & 2.32 & 3.49 & 2.9 & 39.15 & 0.08 \\
PKS 1244-255 & 11.33 & 16.95 & 14.14 & 62.16 & 0.1  \\
PKS 1424+240   & 16.24	& 25.36 & 20.80	& 40.15 & 0.08\\
PKS 1502+106   & 16.34	& 17.68 & 17.53	& 18.08& 0.04\\
PKS 1510+089   & 13.80	& 16.10 & 14.95	& 15.77 & 0.03 \\
PKS 1622-253   & 32.17	& 33.78 & 32.98	& 28 & 0.06 \\
PKS 1830-211  & 19.38	& 25.96 & 22.67	& 14.01& 0.03 \\
PKS 2155-304   & 13.44	& 23.43 &	18.44 & 13.46 & 0.03 \\
PKS 2233-148   & 17.81	& 20.03 & 18.92	& 21.2& 0.04 \\
PKS 2326-502   & 22.75	& 24.98 &	23.87	& 41.08 & 0.08\\
PKS B1424-418   & 14.31	& 19.41 & 16.86	& 17.83 & 0.04 \\
PMN J1802-3940   &	23.82 & 24.32 & 24.07	& 32.13& 0.07 \\
PMN J2345-1555 & 26.37 & 33.71 & 29.67 & 36.54 & 0.07 \\
S2 0109+22	  & 0.84 & 53.99 & 27.41  & 25 & 0.05   \\ 
S5 0716+71	   &  0.70 & 15.89  & 8.30  & 16.02 & 0.03   \\ 
TXS 0506+056   & 19.96 & 35.59  & 27.78 & 10.81 &0.02   \\
TXS 0518+211   & 26.78 & 42.29 & 34.54 & 12.74 & 0.03  \\
\hline
Mean values & 18.47  & 27.67  & 23.08  & 22.27  &   \\ \hline
\end{tabular}
\end{table}

\begin{figure}
\centering

\includegraphics[width=0.48\linewidth]{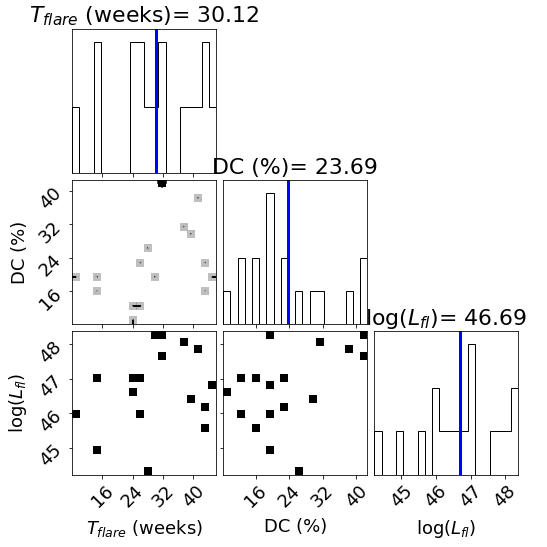}
\includegraphics[width=0.48\linewidth]{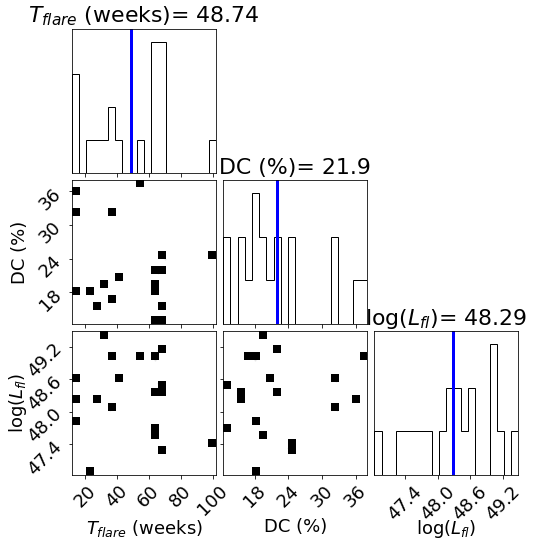}

\caption{Relation between flare duration ($T_{flare}$), Duty Cycle (DC) and flare luminosity ($L_{ff}$) for BL Lac type (left panel) and FSRQ type blazars (right panel).}
\label{fig:Corn_2d_dc_lum_tflare}
\end{figure}

Following Tluczykont criterion, the average DC for the sample when considering EBL absorption (and not) is 22.74\% (23.08\%). According to Vercellone's criterion we obtained values of 20.09\% (22.27\%). The blazars PKS 0426-380 and BL Lacertae have the highest DC values (considering EBL absorption), with an average DC of $\sim$ $42\%$ and $43\%$, respectively (Tluczykont's criterion). On the ther hand, the lowest value of DC obtained corresponds to blazar S5 0716+71 with a DC$_{av}$ value of $\sim 8\%$ considering EBL absorption (Tluczykont's criterion).

At this point, we must consider the following: choosing a flare flux threshold in terms of the time average flux (Vercellone's criterion) does not allow a direct comparison of DC between sources to be made, because the time average flux is influenced by the level of activity. In the case of a highly active source, the time average flux, and consequently its flare flux threshold, would be much higher than the baseline flux, so the DC only refers to the highest flux states. On the contrary, for a less active source the time average flux  is  close  to  the  baseline  flux,  so  the  DC  refers  to almost all the flaring states. Since the DC of an active source includes only the highest flux-flaring states, it is possible to obtain a DC value smaller than the one for a less active source  and  erroneously  conclude  that  the  latter is more active than the former. In general, Tluczykont criterion leads to a higher DC average value than Vercellone's criterion.  

%\begin{figure}
%\centering
%\hspace*{-4cm}
%\includegraphics[width=\linewidth]{bllac_fsrq_3d_stat.png}
%\vspace*{-2cm}
%\caption{In this 3D plot, for each source category (BL Lac or FSRQ) we present the flare luminosity during the most luminous flare vs. the average DC obtained with both Tluczykont (Tlz) and Vercellone (Vc) criteria, vs. the average duration of the gamma-ray flare.}
%\label{fig:3d_stats}
%\end{figure}

Figure \ref{fig:Corn_2d_dc_lum_tflare} shows the average DC and flare luminosity (Tluczykont criterion) as a function of the major flare duration, together with flare luminosity vs DC. As seen from the figure,  there is no significant difference between the values of DC for BL Lacs and FSRQ. We selected the major flares as the ones presenting the greatest number of consecutive temporal bins with the highest fluence in the 10-year dataset. We searched for possible correlations between these variables. However, no correlation was found. Furthermore, we found no correlation between DC and the source redshift, independently of the criterion used.

\section{Discussion and conclusion}\label{sec:conclusions}
We have presented the results of 10 years of Fermi-LAT data in the energy range of 0.1-300 GeV for a sample of 38 bright blazars following the two criteria presented in Section \ref{sec:duty cycle}. 
In particular, we have estimated the gamma-ray variability of the BL Lacs and FSRQs present in the sample as well as the gamma-ray fluence during the activity periods. We have searched for the possible origin of this emission when looking at the entire observation time, dominated by quiescent periods, advocating a more enhanced hadronic component during the flaring periods.
The average DC for the whole sample following Tluczykont's criterion varies in the range [$\sim$8-43]\% considering EBL absorption. This implies that blazars present a wide range of activity levels. Ongoing work aims to make a complete spectral analysis of these sources. 

When correlating the gamma-ray activity to the hadronic emission from the sources, the obtained DC becomes relevant also for the calculation of the expected neutrino signal over a long observational period, as discussed in~\citep{2019arXiv190913198M}. In particular, the average DC values reported in Figure \ref{fig:Corn_2d_dc_lum_tflare} can be used as a scaling factor for the expected neutrino fluence.
It is noteworthy however that the variance of the DC inferred considering EBL and not considering it is negligible, which would imply that the DC is not affected by the EBL absorption at these energies and distances.
Even though the sample is small, a quick analysis shows that there is no correlation between DC and redshift (nor between DC and luminosity) for this sample of blazars. However, further analysis is required in order to make a stronger statement.
Finally, we find that, for this sample, there is no notorious difference between the DC values inferred for FSRQ and BL Lacs (see Figure \ref{fig:Corn_2d_dc_lum_tflare}). 

{\bf Acknowledgments}

\textbf{JRS} acknowledges DIGI-USAC for financial support by grant 4.8.63.4.44 and DGAPA-UNAM IG101320. \textbf{AVP} acknowledges the National Research Council of Argentina (CONICET) by the grant PIP 616, and by the Agencia Nacional de Promoci\'on Cient\'ifica y Tecnol\'ogica (ANPCYT) PICT 140492. \textbf{AM} acknowledges Istituto Nazionale di Fisica Nucleare (INFN), Ministero dell' Universit\'a e della Ricerca (MIUR), PRIN 2017 program (Grant NAT-NET 2017W4HA7S) Italy. \textbf{MC} acknowledges  Funda\c c\~ao de Ampara \`a Pesquisa do Estado de S\~ao Paulo (FAPESP) for   financial   support   under   grants\\ \#2015/25972-0 and \#2017/00968-6.

\bibliography{references}

\begin{appendices}

\section{Fit parameters}
\label{sec:fit-tbl}
{\small
\begin{longtable}{|l|cccc|cccc|}
%\label{tab:DC_fit_all_withEBL}
%\small
\caption{LN+G distribution fit parameters. Flux is in units $\times 10^{-8}$. }\\

%\begin{tabular}{|l|cccc|cccc|}
\hline
Source name &  \multicolumn{4}{c |}{EBL Correction} &\multicolumn{4}{c |}{No EBL Correction}   \\
    &   $\mu_{LN}$ & $\sigma_{LN}$ & $\mu_G$ & $\sigma_G$ &   $\mu_{LN}$ & $\sigma_{LN}$ & $\mu_G$ & $\sigma_G$ \\
\hline
1H 1013+498 &  3.43	& 0.93 & 	4.87	& 1.83 & 2.97 & 1.02 & 6.31 & 1.54\\   
3C 273   &	4.55&	1.23	&		12.99	&	5.07 & 4.47 & 1.18 & 13.05 & 4.53 \\ 
3C 279  &	6.22&	1.49	&		21.91	&	11.58 & 6.33 & 1.55 & 15.14 & 9.96 \\ 
3C 66A  &	2.12&	1.01	&		8.70	&	1.83 & 2.35 & 1.1 & 8.11 & 1.42\\ 
4C +01.28 	&3.27&	1.65	&		0.96&	2.36 & 3.65 & 1.12 & 0.69 & 3.00\\ 
4C +21.35  &	17.48	&4.42	&		8.25&	10.91 & 10.56 & 3.77 & 4.78 & 1.83 \\ 
4C +28.07	&	5.09	&	1.66	&		8.83	&13.90 & 5.18 & 1.81 & 11.59 & 7.69\\ 
4C +38.41 &	5.21	&	1.01&		13.70	&	13.74 & 4.94 & 1.72 & 6.66 & 2.34\\ 
AO 0235+164 &	5.72	&	1.93		&4.67&	2.33 & 5.91 &2.13 & 5.08 & 2.36\\ 
B2 1520+31	&	3.58	&	0.75&	 9.26 &	8.03 & 4.18 & 0.65 & 15.03 & 8.17\\ 
BL Lacertae &	3.95	& 1.04		& 8.58 &	5.19 & 3.89 & 1.05 & 8.47 &4.52\\ 
MG4 J2001.1+4352 &-&	-&-		&	- & 4.52 & 2.31 & 4.35 & 1.92\\ 
Mkn421	&	3.95&	0.60&		21.34&	5.77 & 4.34 &1.13 & 21.48 & 4.32\\
Mkn 501  1&	2.58	&	0.51&	6.06&	2.16 & 2.74 & 0.42 & 8.13 & 3.37\\
PG 1553+113 &	3.17	&	0.11&		12.53	&	4.13 & 2.87 & 0.73 & 4.99 & 2.65 \\ 
PKS 0426-380 &3.88&	0.78	&		4.69&	7.74 & 3.84 & 0.72 & 6.39 & 8.16\\ 
PKS 0447-439  &	3.70	&	0.96& 	8.60	&3.78 & 3.35 & 0.91 & 9.15 & 4.00\\ 
PKS 0454-234   &	3.81&	1.05&		2.91	&8.15 & 3.63   & 0.62 & 8.65 & 5.28\\ 
PKS 0537-441  &	4.67&	1.45	&	 6.41	&	2.88 & 4.60 & 1.29 & 5.07 & 3.14\\ 
PKS 0727-11  &	4.38	&	1.09&		12.44	&7.73 & 4.51 & 0.69 & 23.08 & 12.21\\ 
PKS 0805-07 &	4.46	&	1.83		&	1.87	&3.78 & 0.75 & 1.45 & 5.01 & 4.23\\
PKS 1244-255 &	4.22	&	1.12		&	7.80	&3.12 & 0.60 & 1.05 & 7.05 & 3.26\\
PKS 1424+240 &	-&	-&	-	&	- & 2.69 & 0.76 & 5.43 & 3.53\\ 
PKS 1502+106 &	5.17	&	1.67&		8.73&3.82 & 5.53 & 1.83 & 4.93 & 1.72\\ 
PKS 1510-089 	&	7.14	&1.69 & 	37.06&	21.16 & 6.29 & 1.48 & 25.86 & 11.14\\ 
PKS 1622-253 	&	0.89	&	0.22		&2.20&1.06 & 8.59 & 2.18 & 23.6 & 11.85\\ 
PKS 1830-211   &	5.17	&	0.89&	31.88&	11.93 & 4.04 & 1.26 & 12.52 & 3.56\\ 
PKS 2155-304   &	3.37	&	0.55 &  	8.10&3.70 & 3.22 & 0.60 & 7.82 & 3.56\\ 
PKS 2233-148  	&	4.81	&	1.98&  	2.15	&	1.61 & 2.74 & 1.56 & 2.07  & 1.73\\ 
PKS 2326-502   &	5.77	&	1.70	&	13.35&	2.17 & 19.19 & 4.81 & 4.52 & 1.24\\ 
PKS B1424-418  &4.32&0.69 &	 31.46	&22.91 & 5.34 & 0.77 & 35.14 & 21.78\\ 
PMN J1802-3940  &	12.08&	2.68 	&	13.64	&7.35 & 22.19 & 4.28 & 12.65 &6.66\\ 
PMN J2345-1555 	& 3.32 & 1.17	& 5.9	& 1.8 & 3.32 & 1.08 & 5.32 & 1.90\\ 
S2 0109+22 	&	2.21&	1.23	&		9.38&1.46 & 2.38 & 1.25 & 9.56 &1.50\\ 
S5 0716+71 	&	3.30	&	1.56 &		13.60&	6.16 & 3.3 & 1.56 & 13.60 & 6.17\\ 
TXS 0506+056 &	2.29	&0.91 & 	6.60	&	3.13 & 2.59 & 0.78 & 5.13 & 1.63\\ 
TXS 0518+211 &	2.63	&	0.83	&	11.70	&5.97 & 3.42 & 0.97 & 7.71 & 1.78\\ 
\hline
%\end{tabular}
\end{longtable}
}

\section{Duty Cycle Plots}
\label{sec:dc-ap}

%%%%%%%%%%%%%%%%%FIG CON EBL
\begin{figure*}
\centering
\subfloat{
\includegraphics[width=0.29\textwidth]{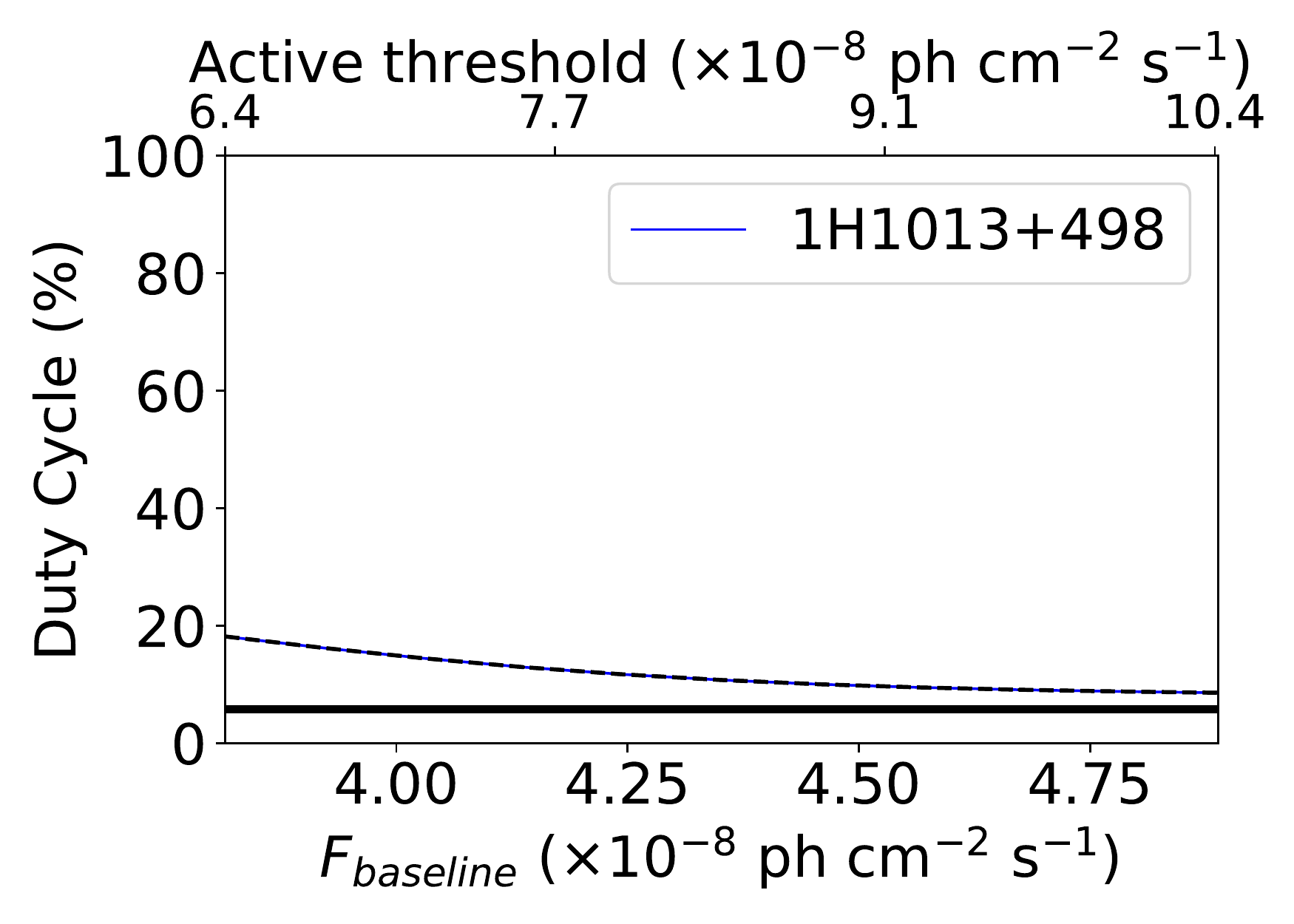}}
\subfloat{
\includegraphics[width=0.29\textwidth]{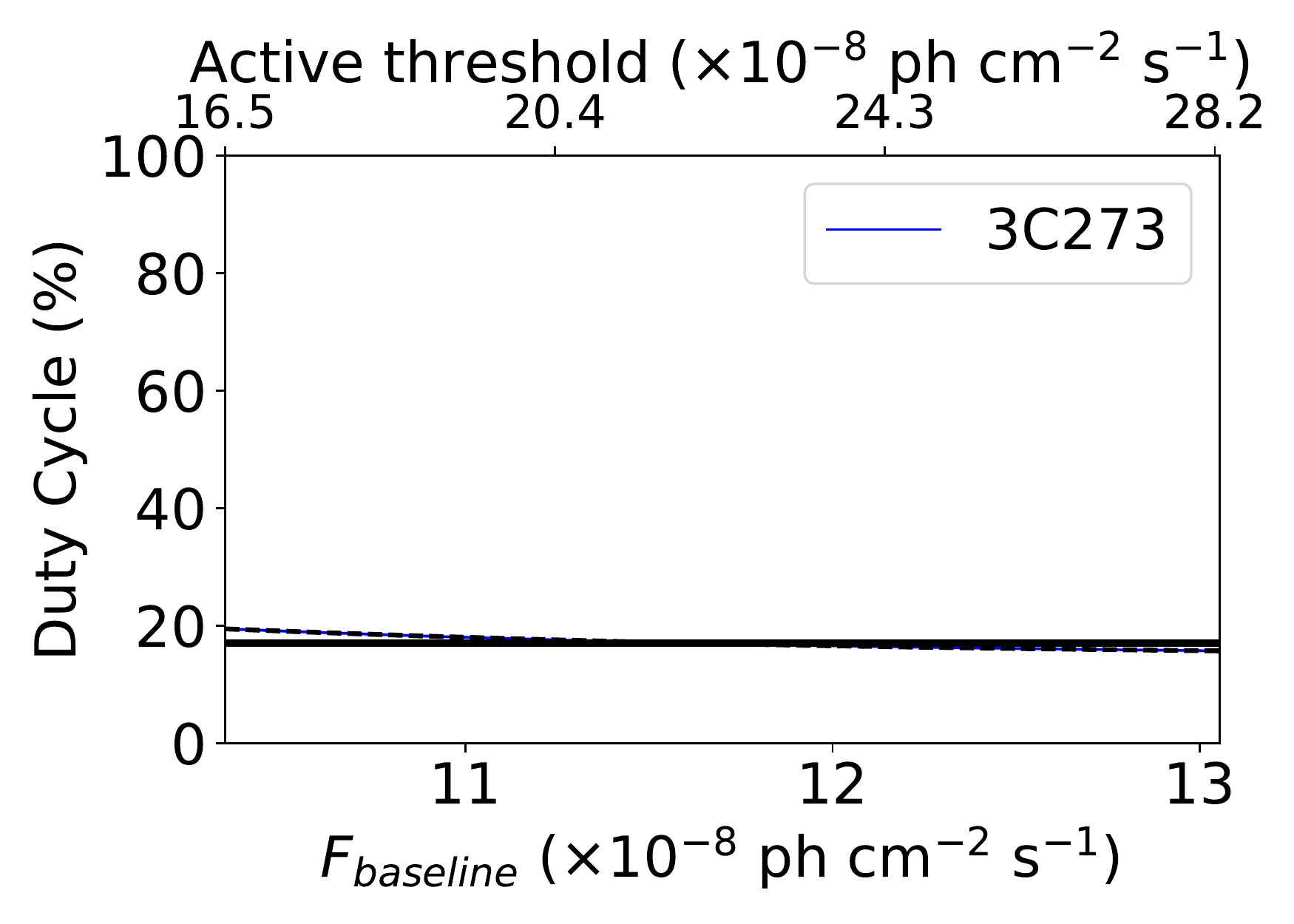}}
\subfloat{
\includegraphics[width=0.29\textwidth]{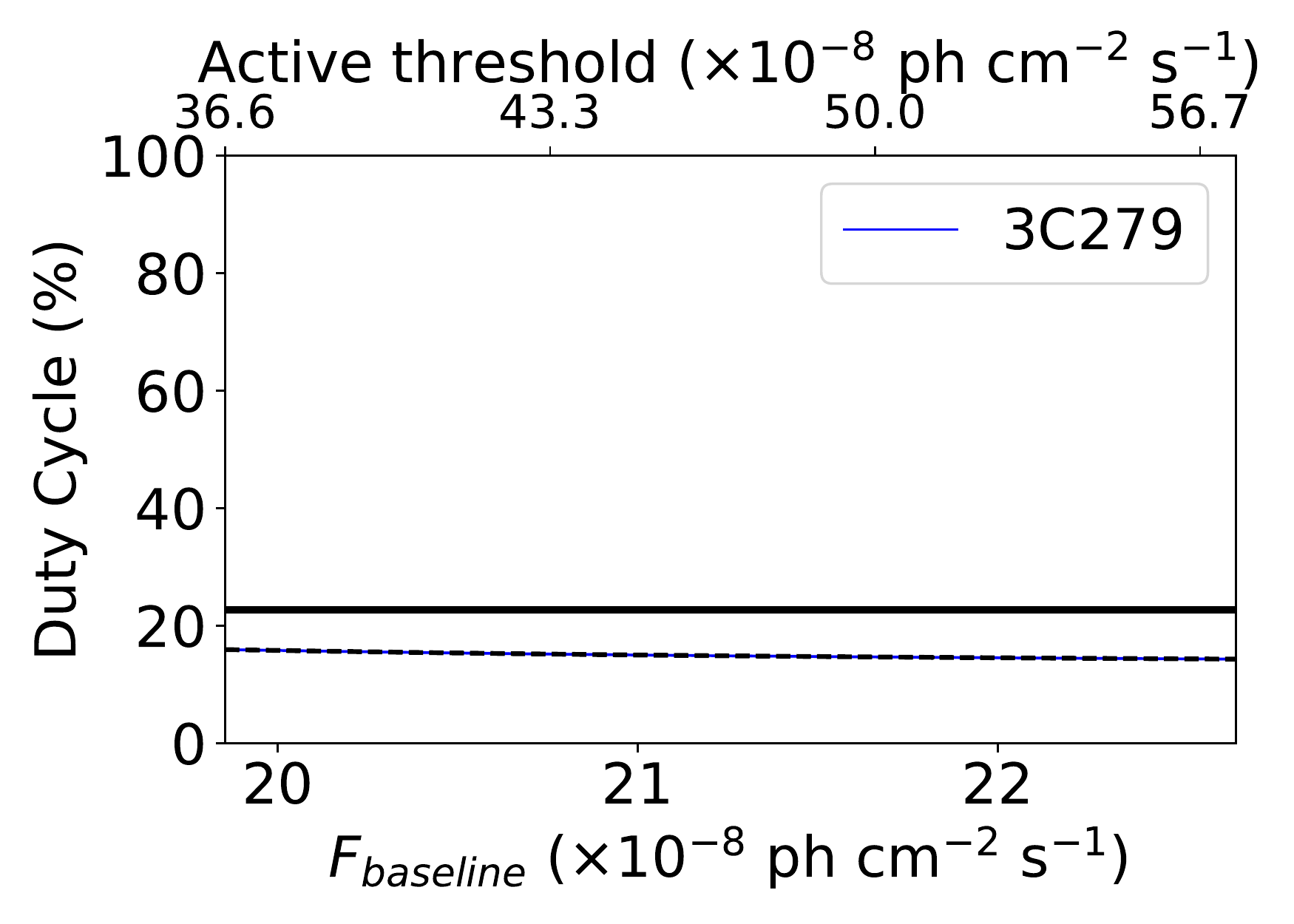}}
\qquad  %%%%%%%

\subfloat{
\includegraphics[width=0.29\textwidth]{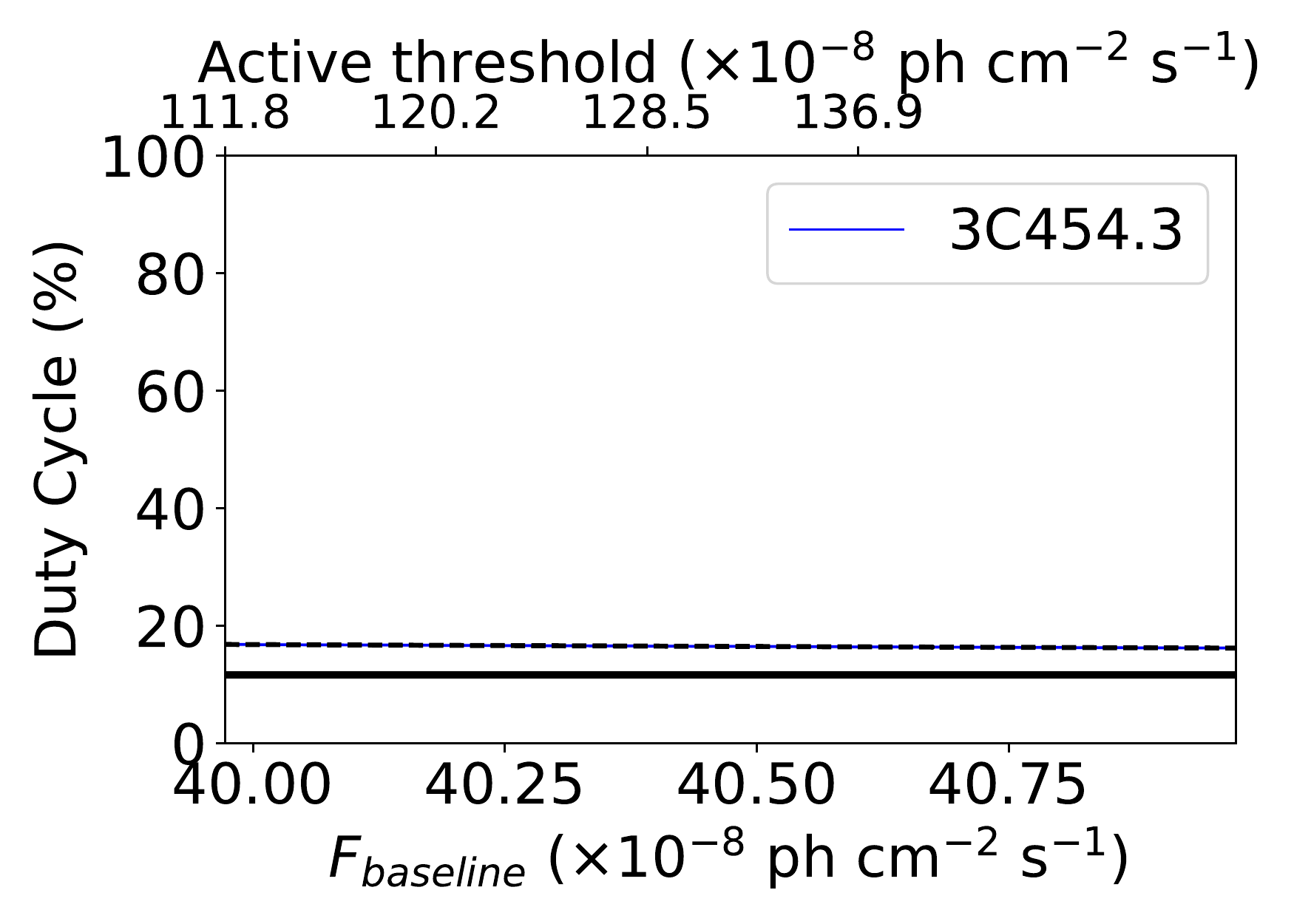}}
\subfloat{
\includegraphics[width=0.29\textwidth]{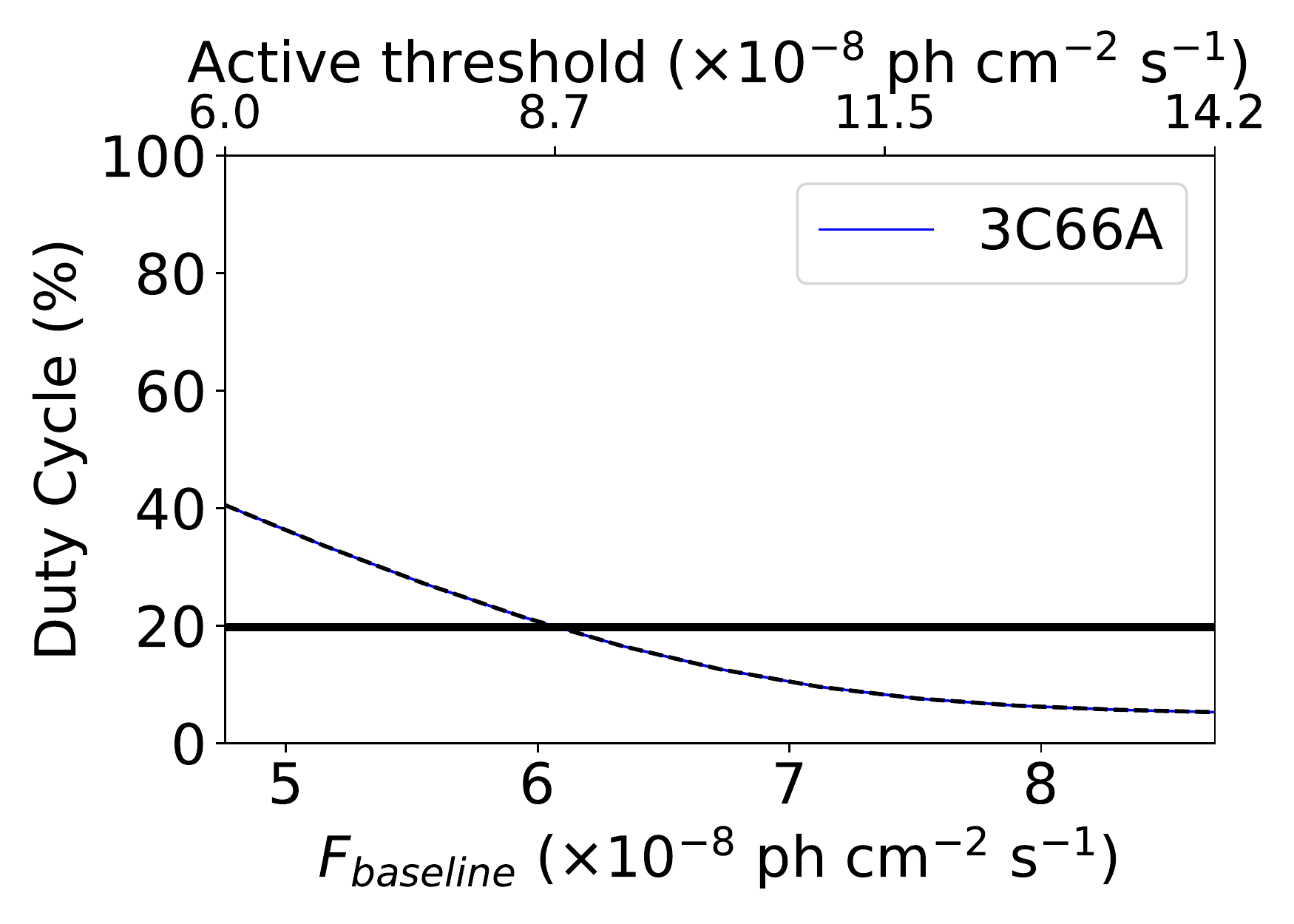}}
\subfloat{
\includegraphics[width=0.29\textwidth]{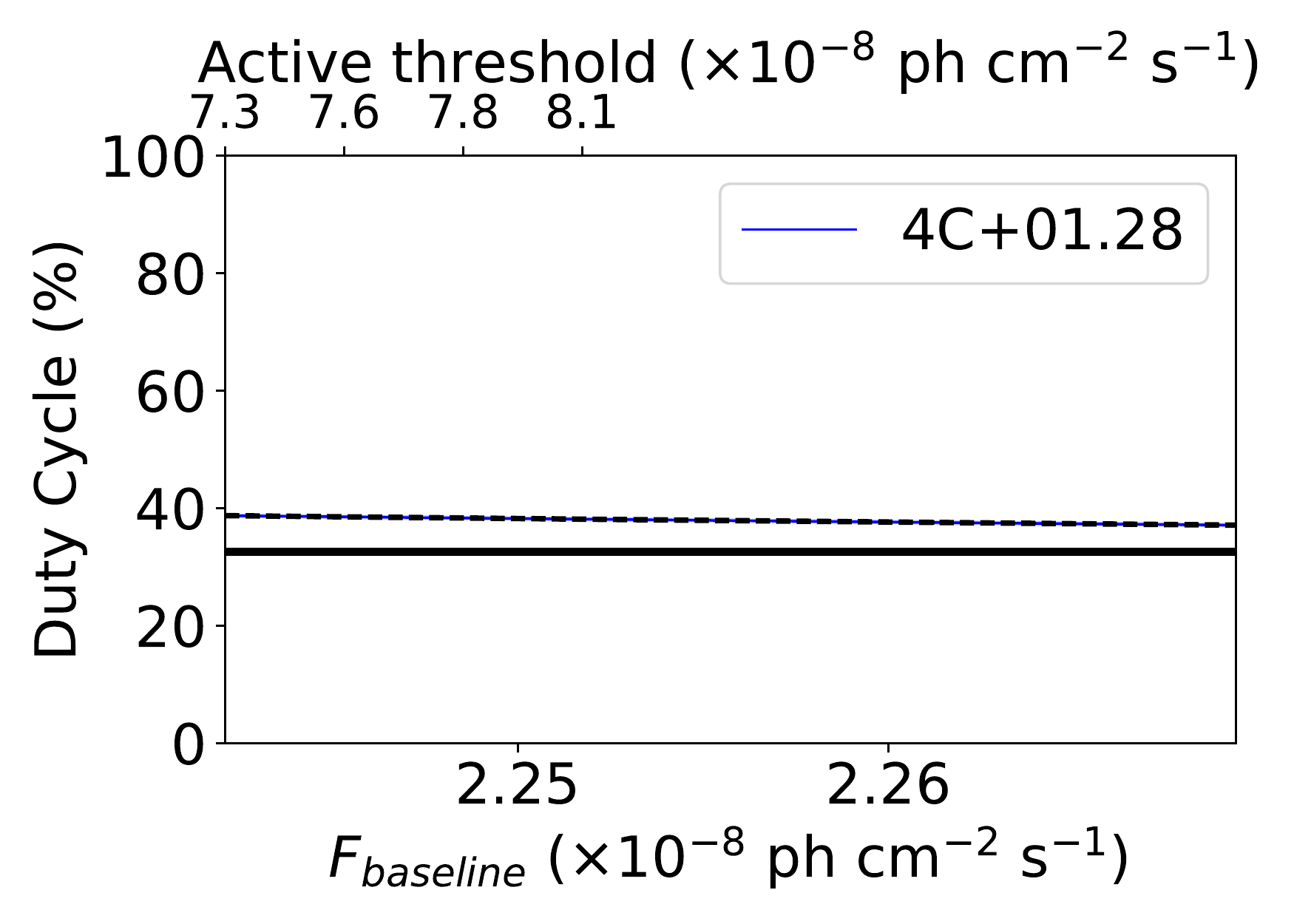}}
\qquad  %%%%%%%
\subfloat{
\includegraphics[width=0.29\textwidth]{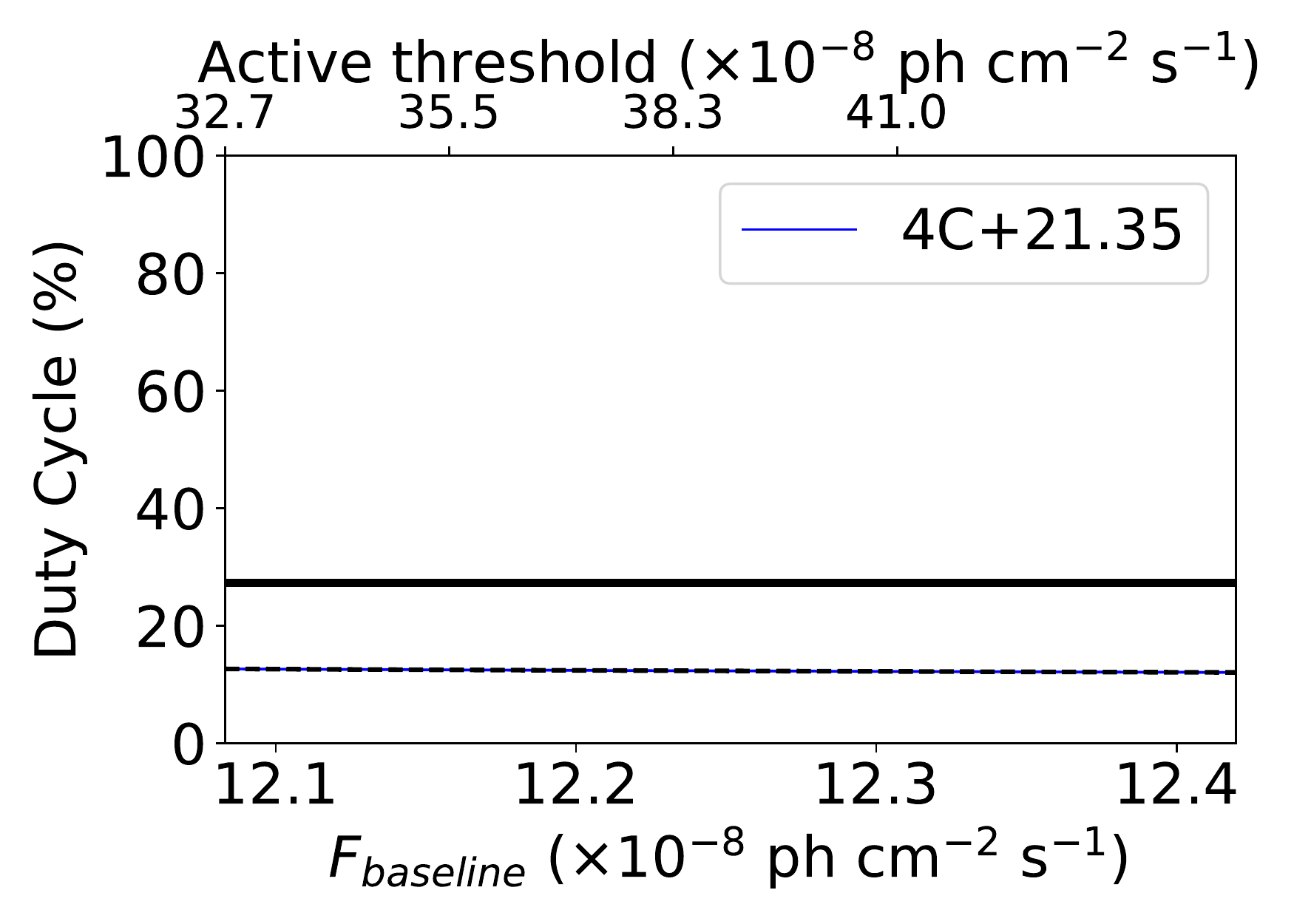}}
\subfloat{
\includegraphics[width=0.29\textwidth]{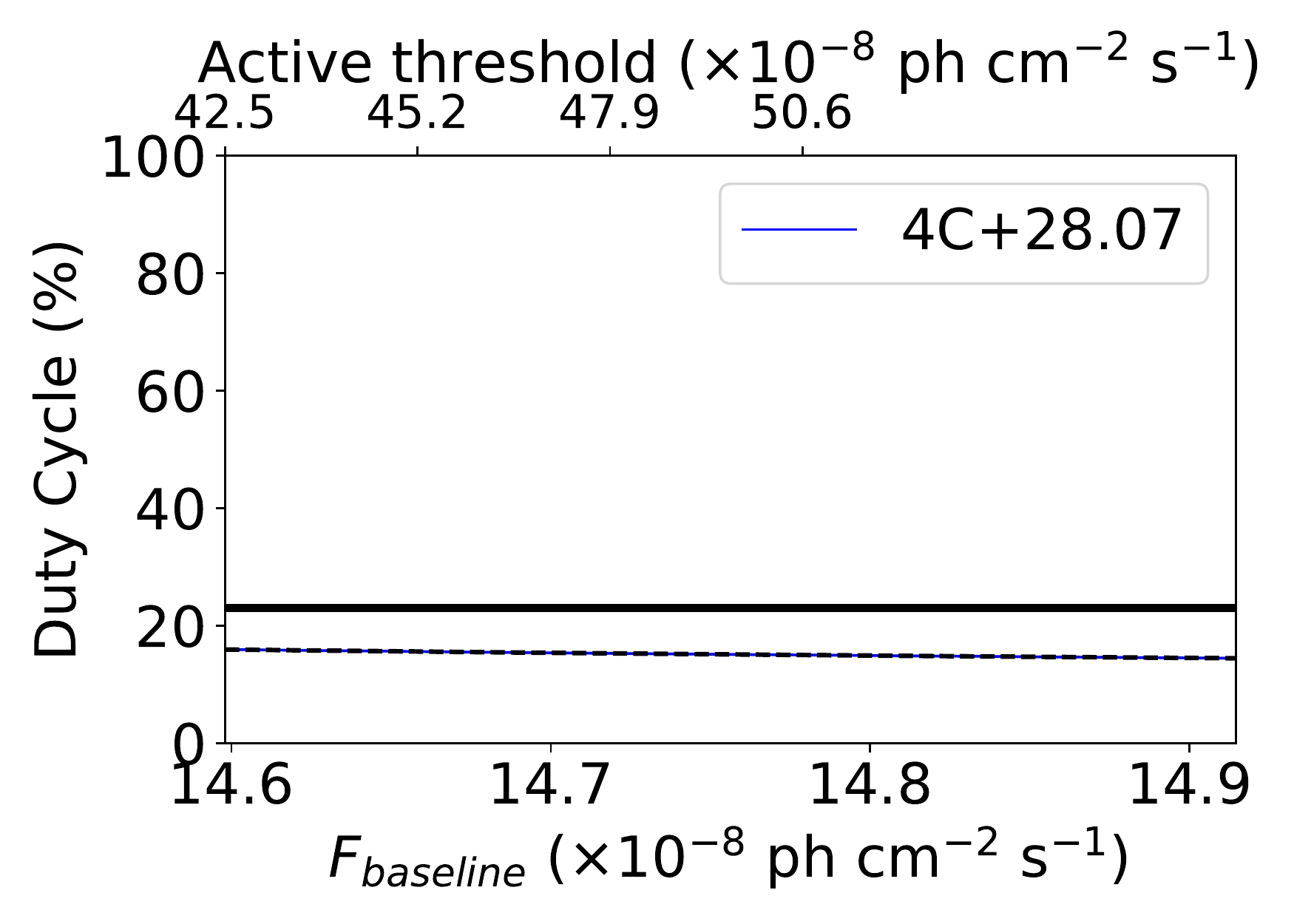}}
\subfloat{
\includegraphics[width=0.29\textwidth]{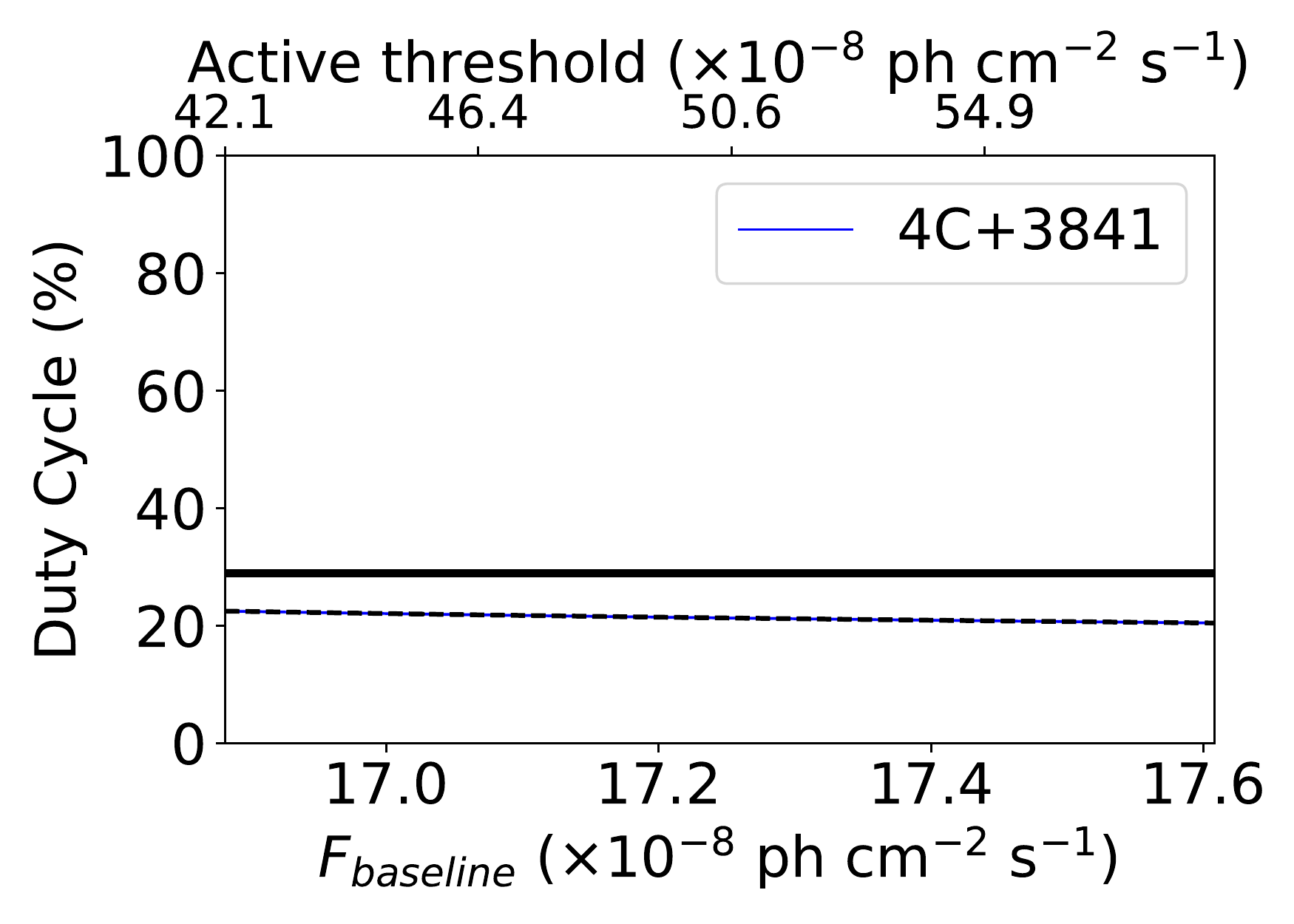}}
\qquad  %%%%%%%
\subfloat{
\includegraphics[width=0.29\textwidth]{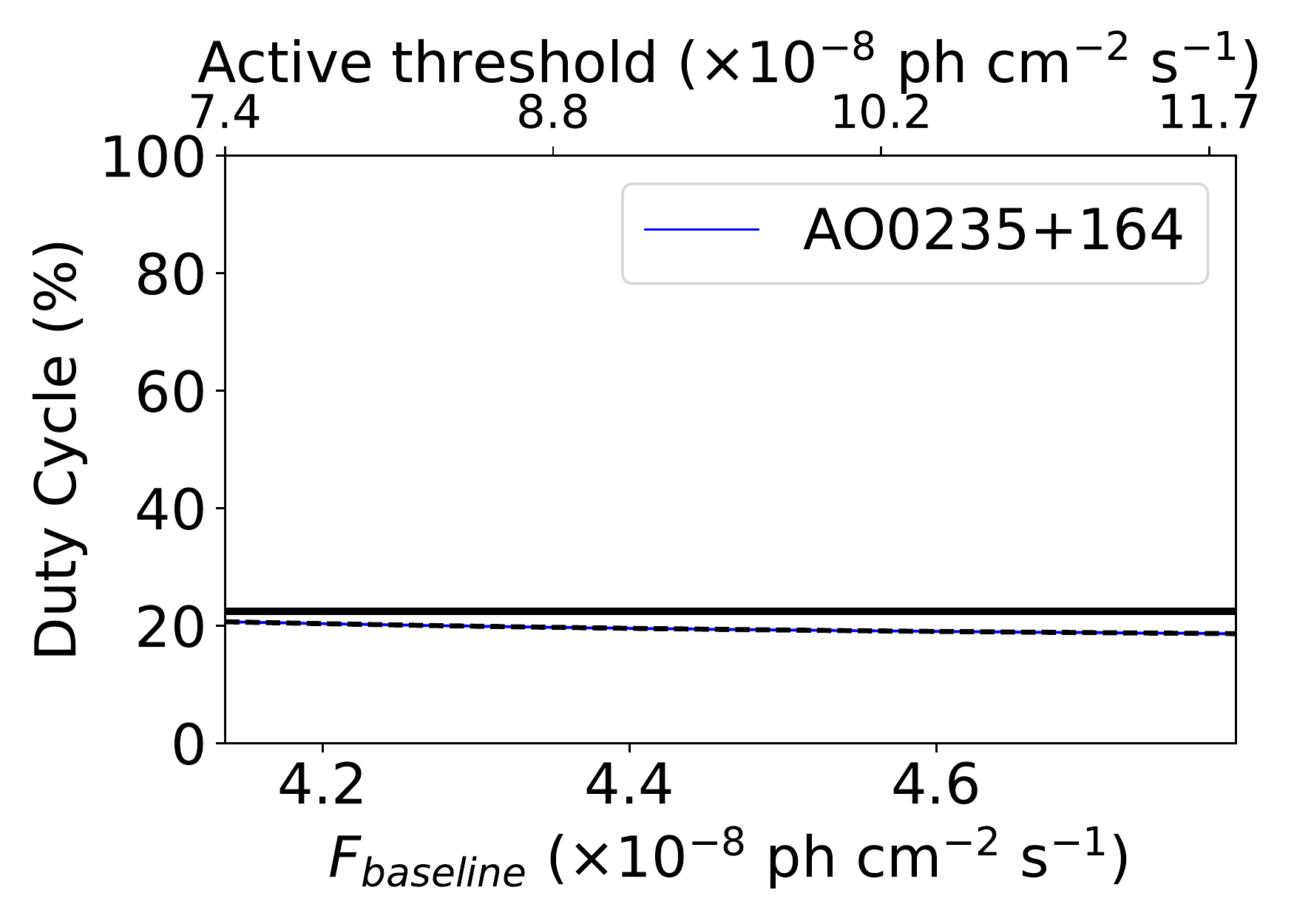}}
\subfloat{
\includegraphics[width=0.29\textwidth]{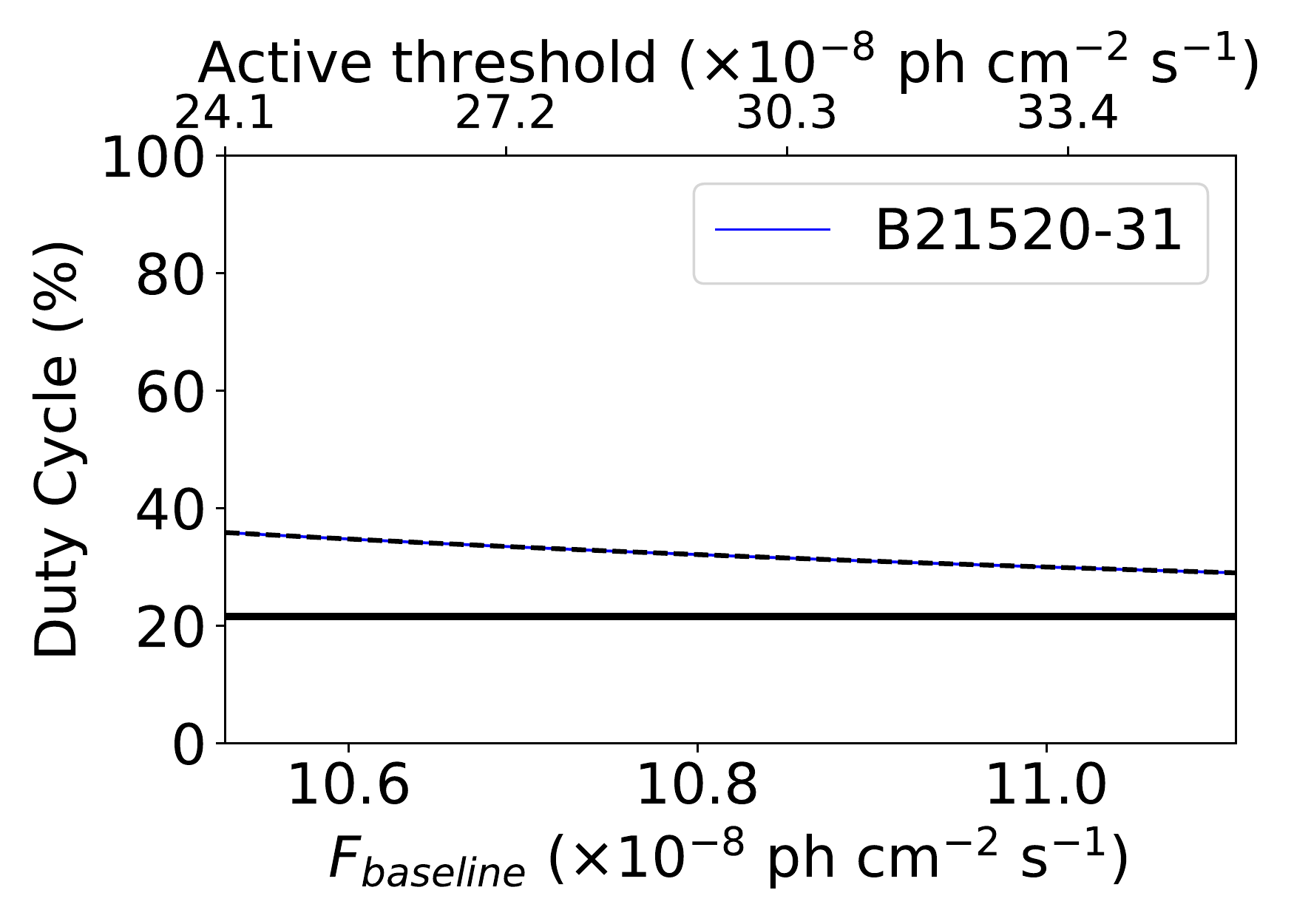}}
\subfloat{
\includegraphics[width=0.29\textwidth]{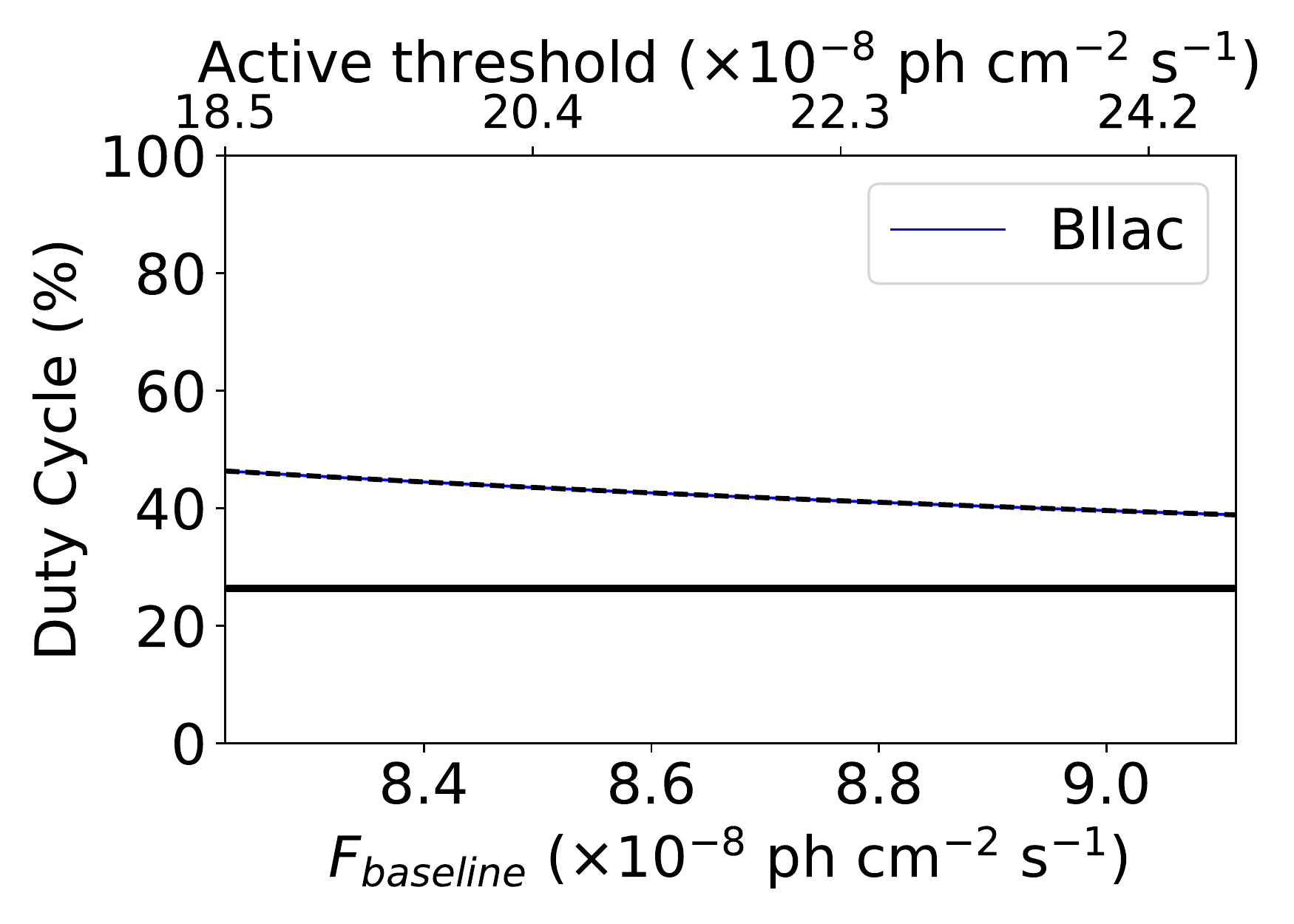}}
\qquad  %%%%%%%
\subfloat{
\includegraphics[width=0.29\textwidth]{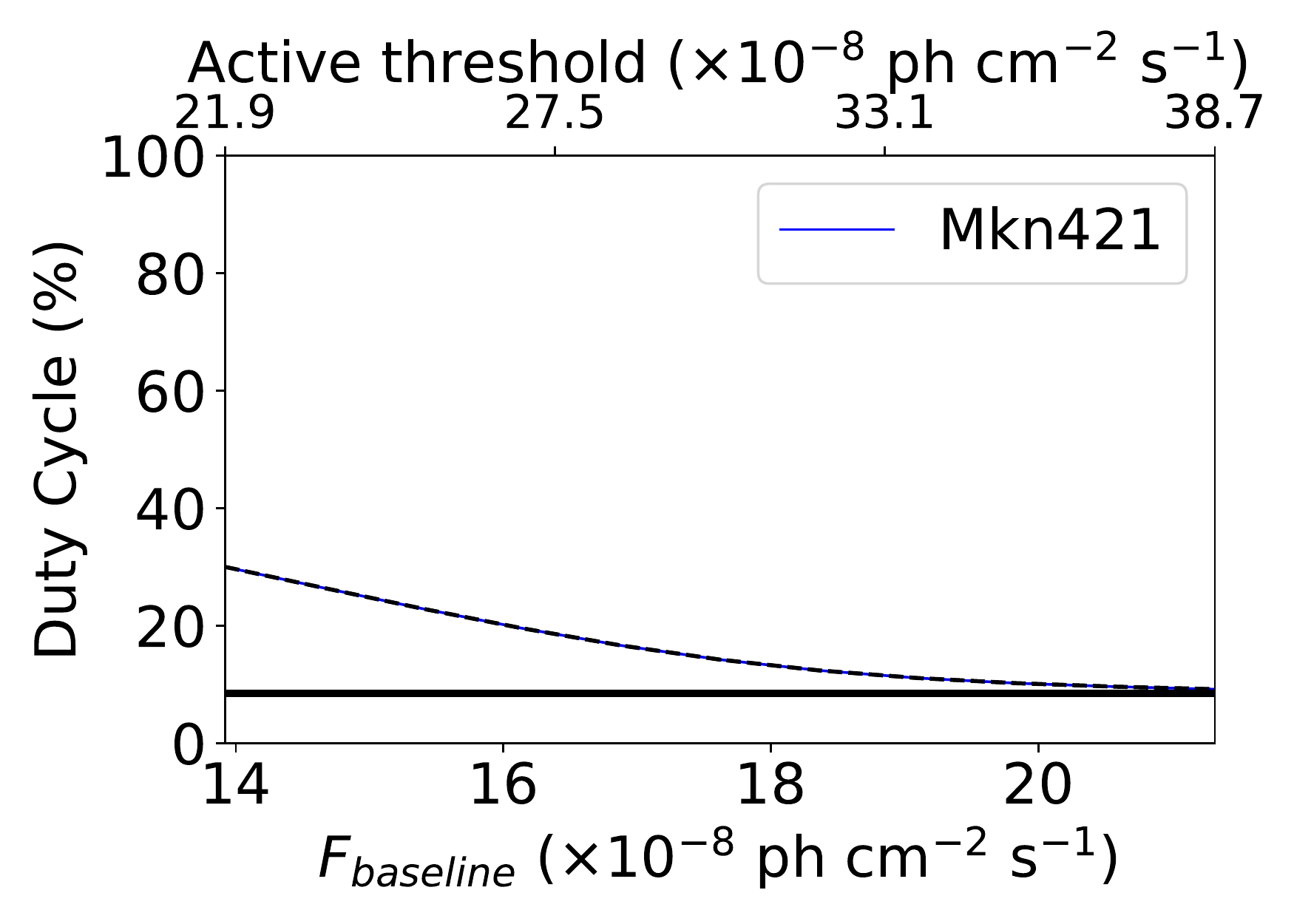}}
\subfloat{
\includegraphics[width=0.29\textwidth]{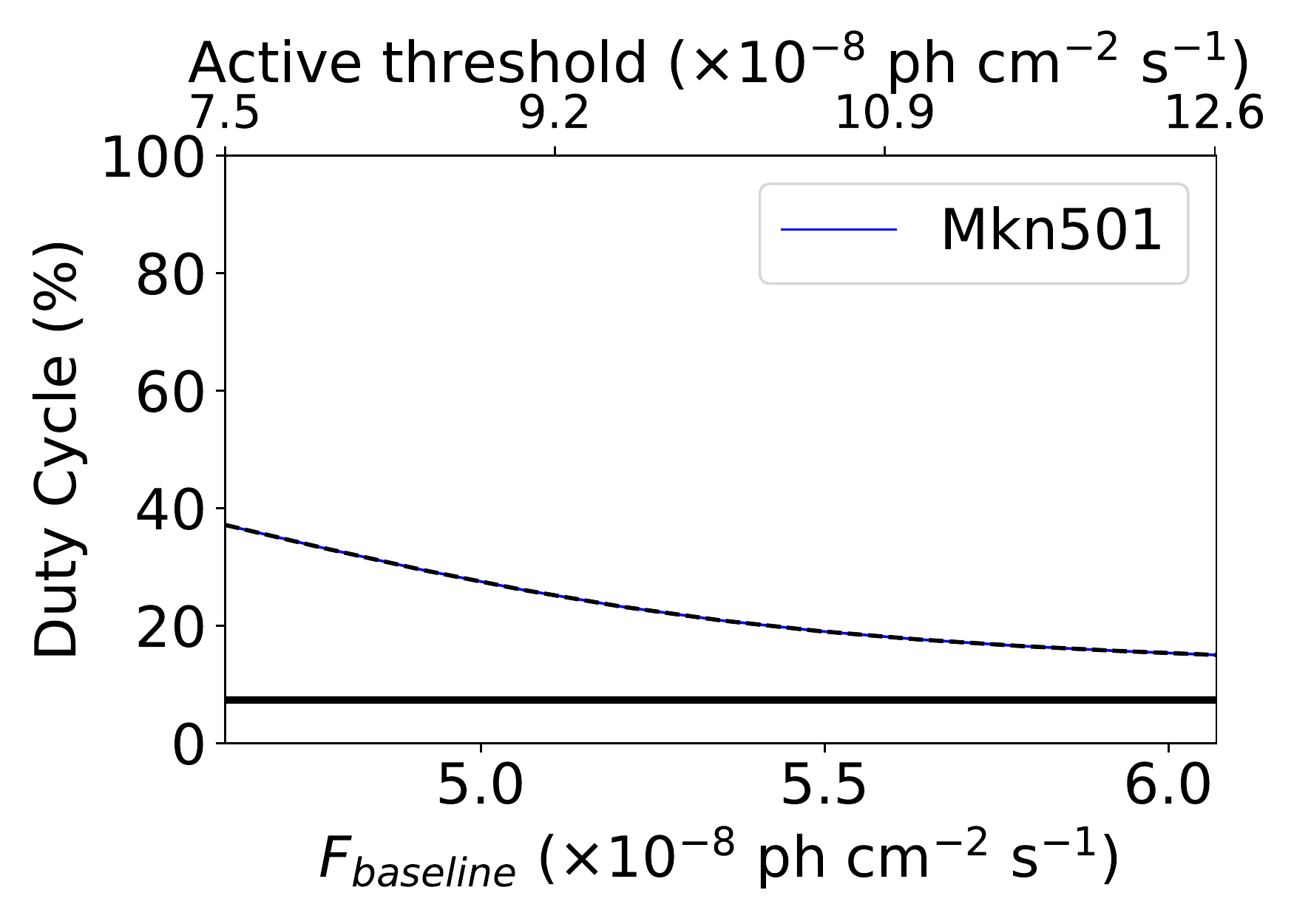}}
\subfloat{
\includegraphics[width=0.27\textwidth]{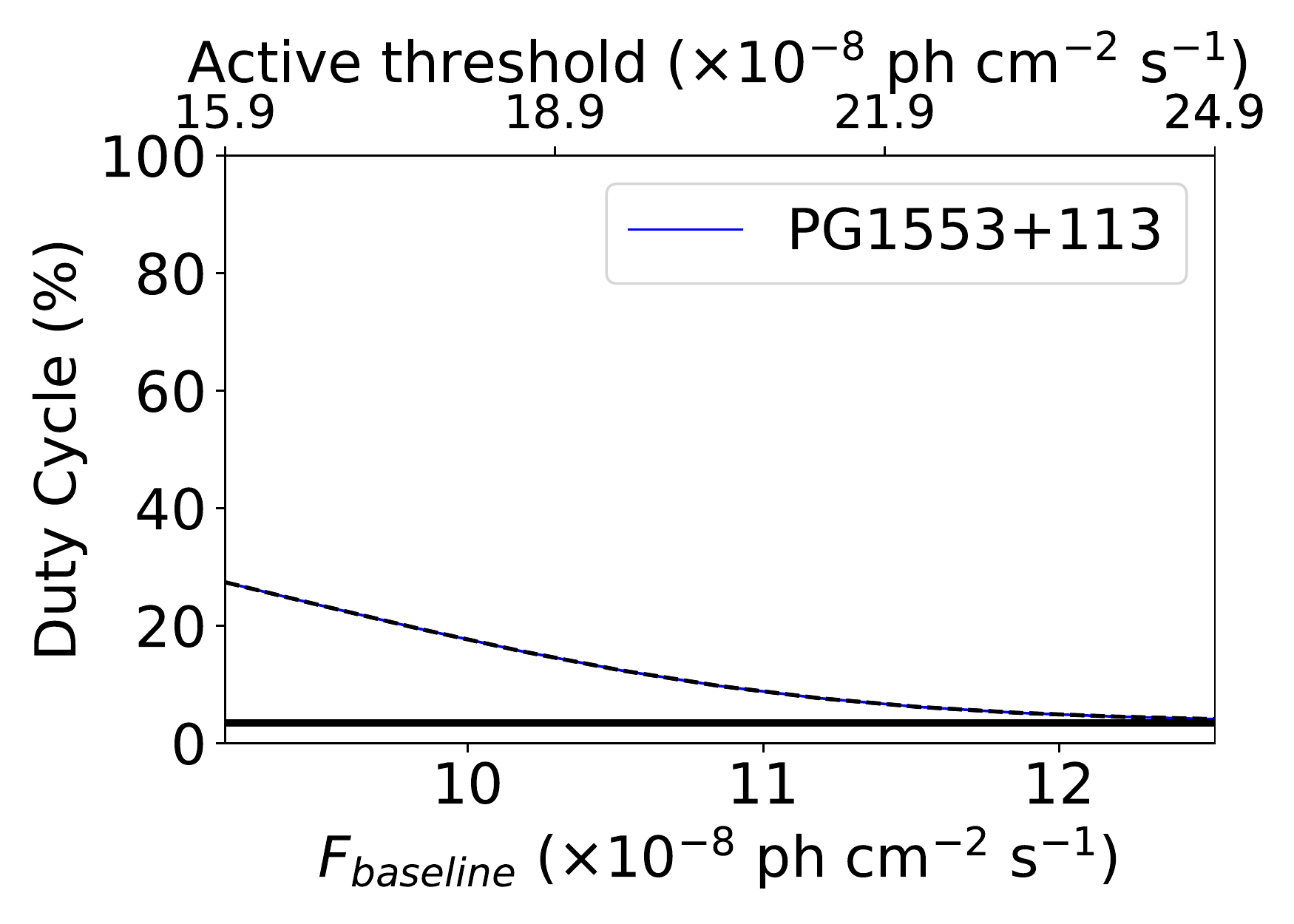}}
\qquad  %%%%%%%
\subfloat{
\includegraphics[width=0.29\textwidth]{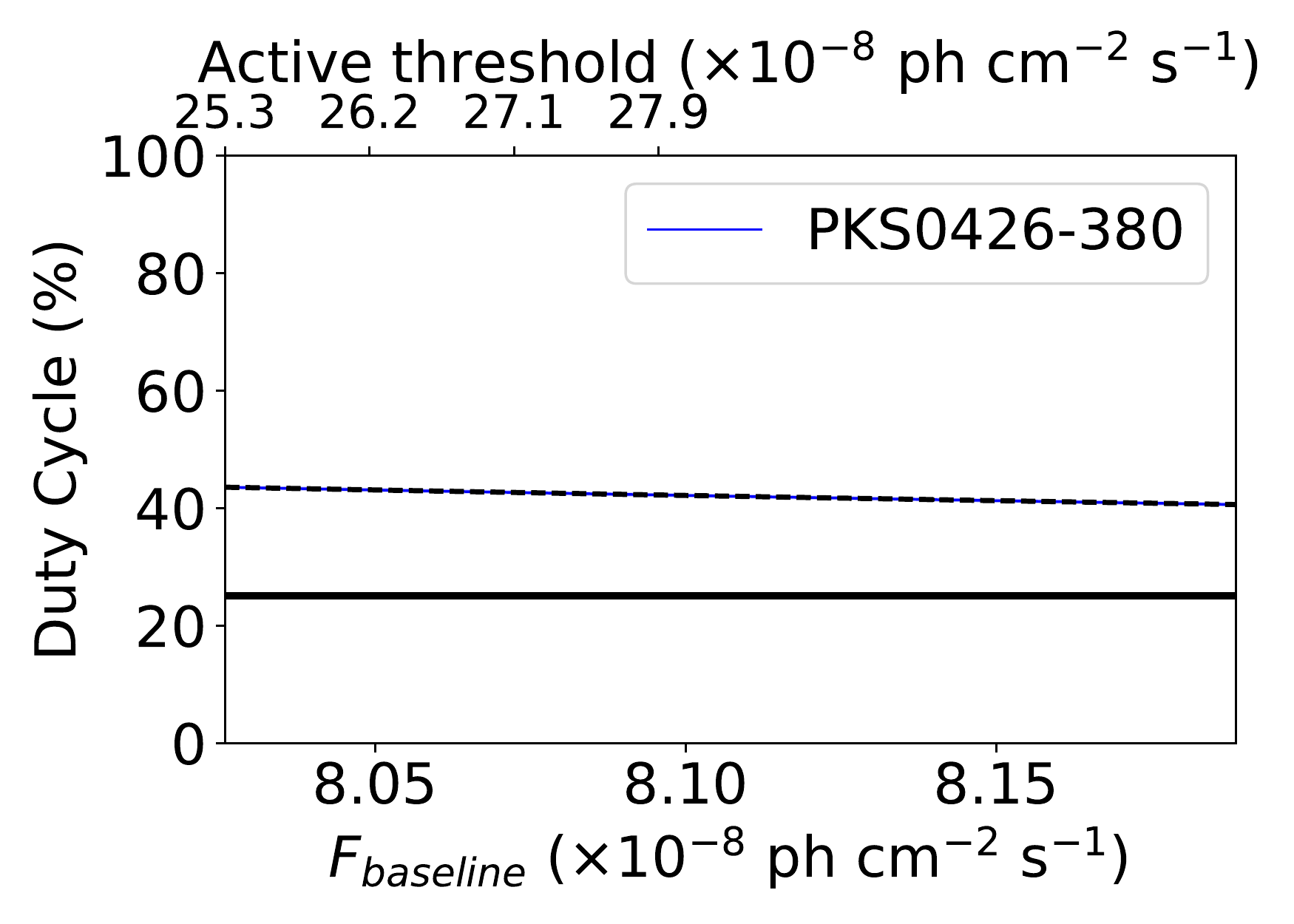}}
\subfloat{
\includegraphics[width=0.29\textwidth]{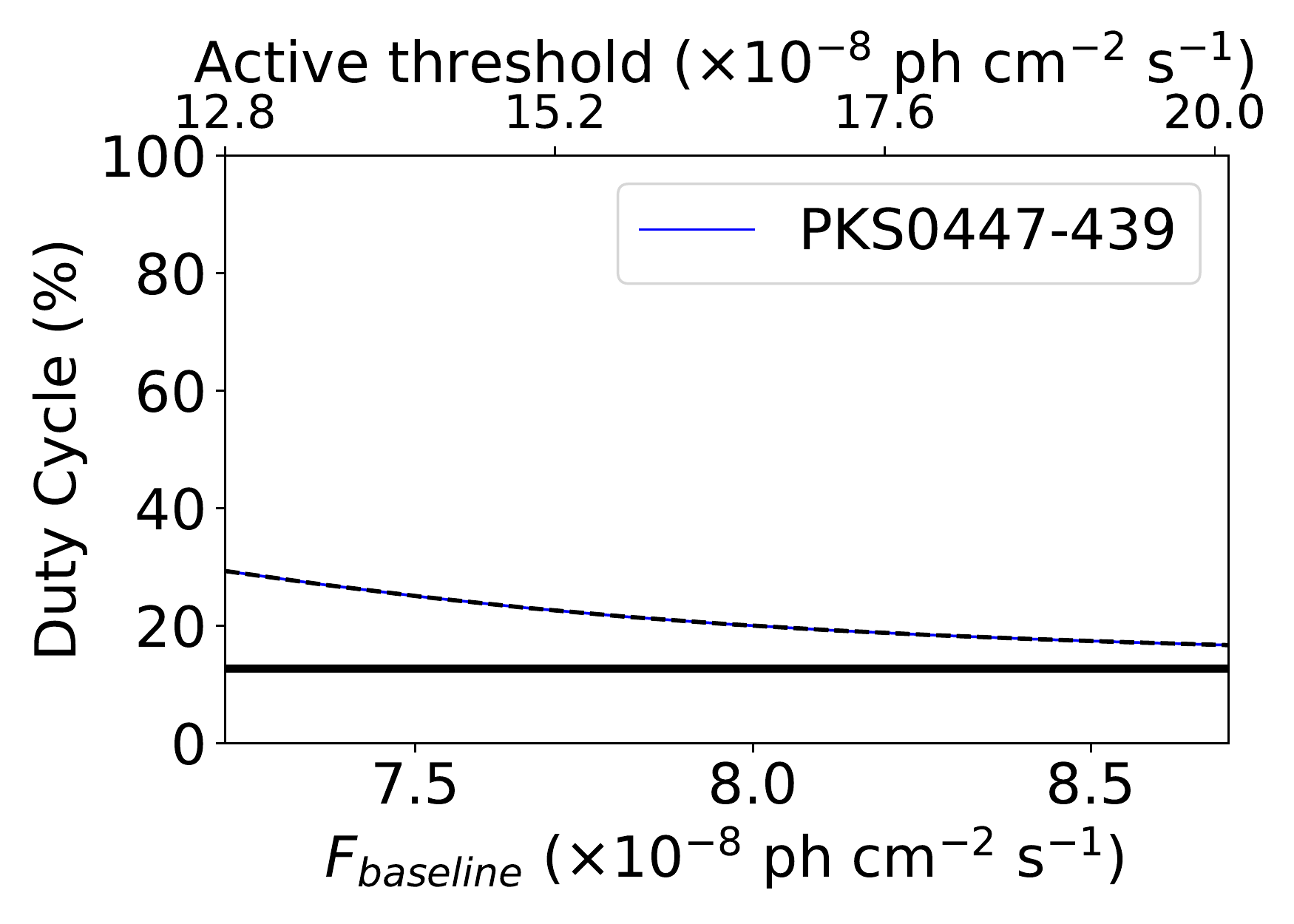}}
\subfloat{
\includegraphics[width=0.29\textwidth]{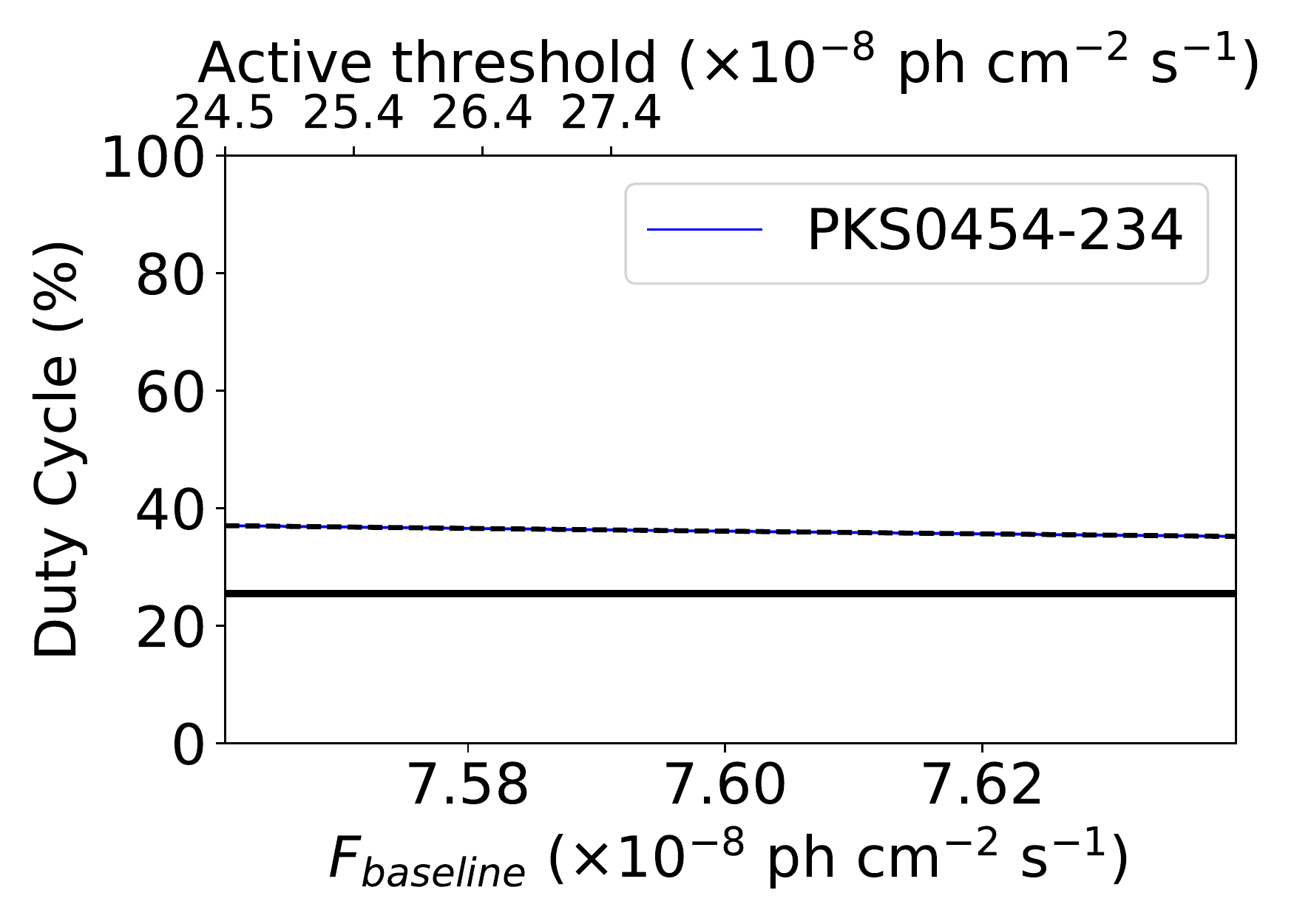}}
\caption{DC vs. flux conidering EBL absorption. The blue (black) lines represent the DC range inferred from Tluczykont (Vercellone's) criterion.}
\label{fig:DC_conEBL}
\end{figure*}

\begin{figure*}
\ContinuedFloat
\centering
\subfloat{
\includegraphics[width=0.29\textwidth]{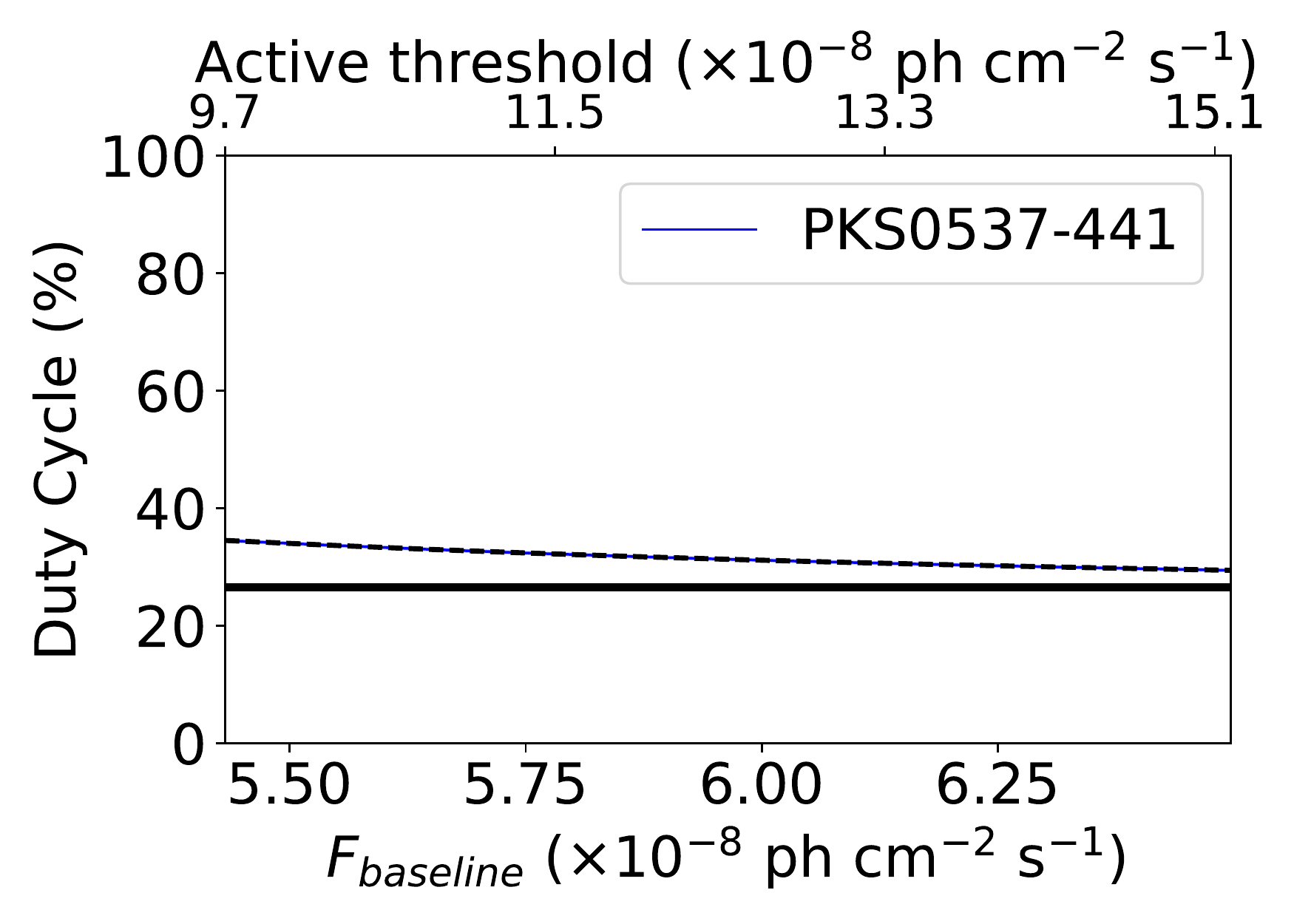}}
\subfloat{
\includegraphics[width=0.29\textwidth]{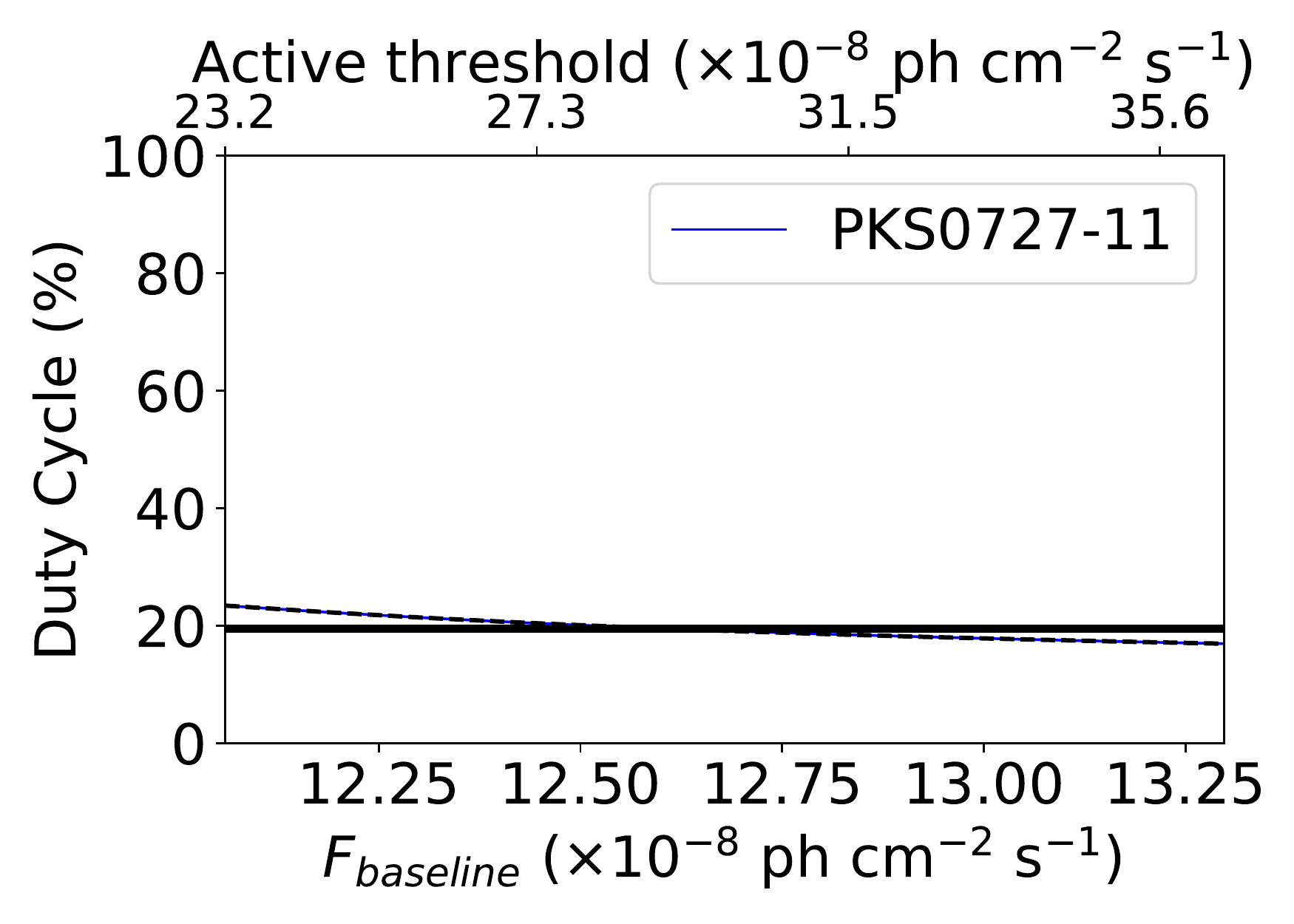}}
\subfloat{
\includegraphics[width=0.29\textwidth]{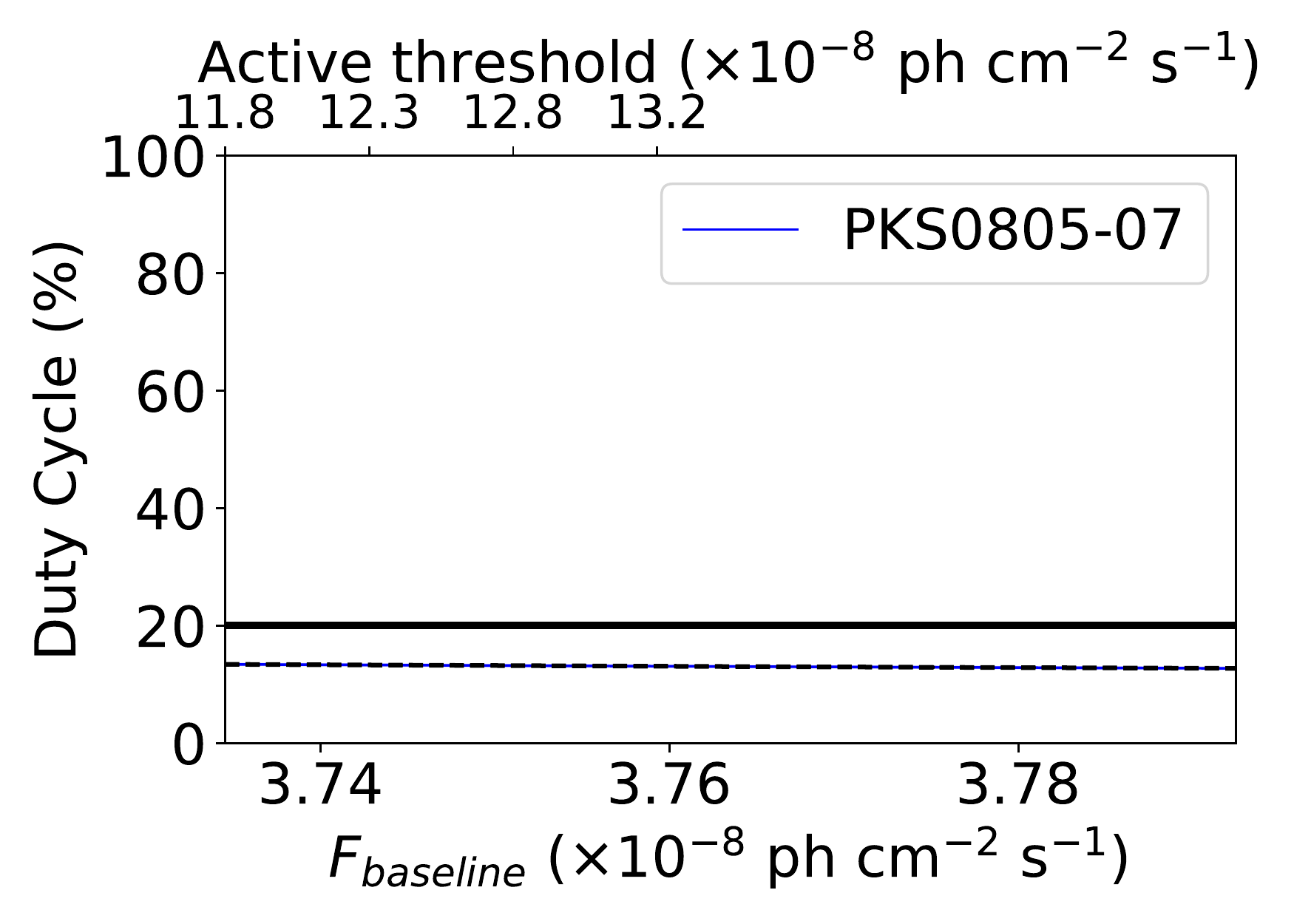}}
\qquad  %%%%%%%
\subfloat{
\includegraphics[width=0.29\textwidth]{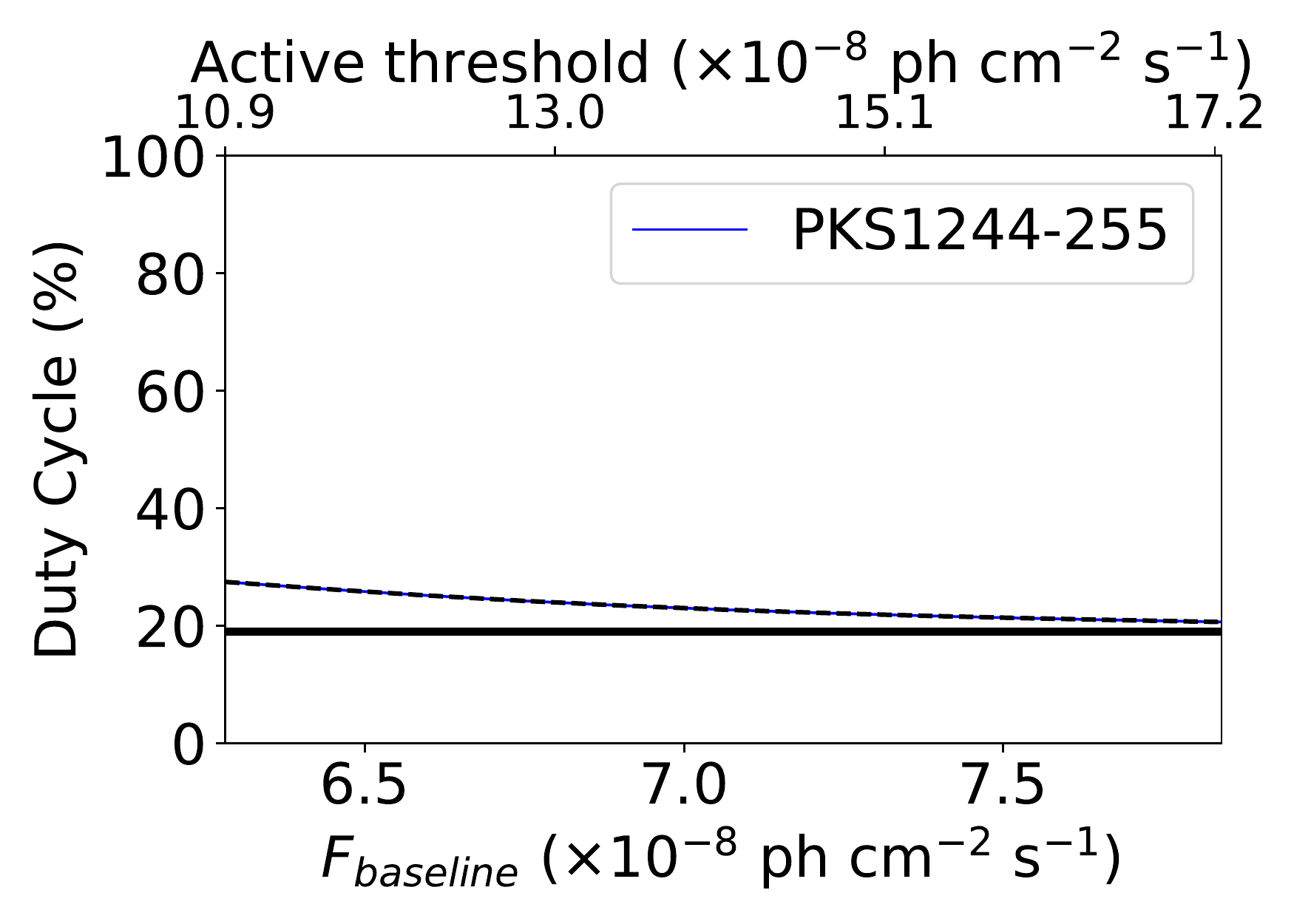}}
\subfloat{
\includegraphics[width=0.29\textwidth]{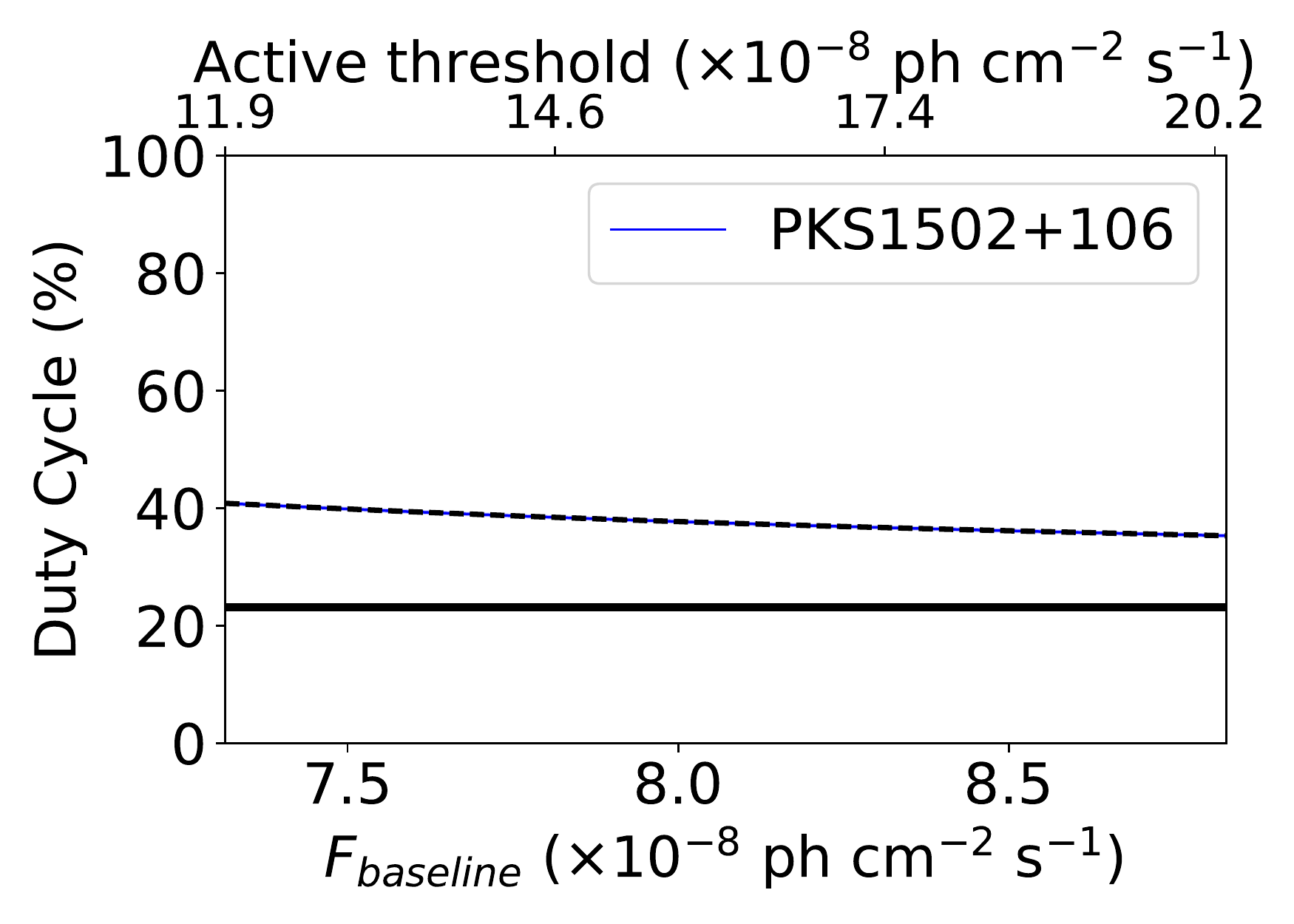}}
\subfloat{
\includegraphics[width=0.29\textwidth]{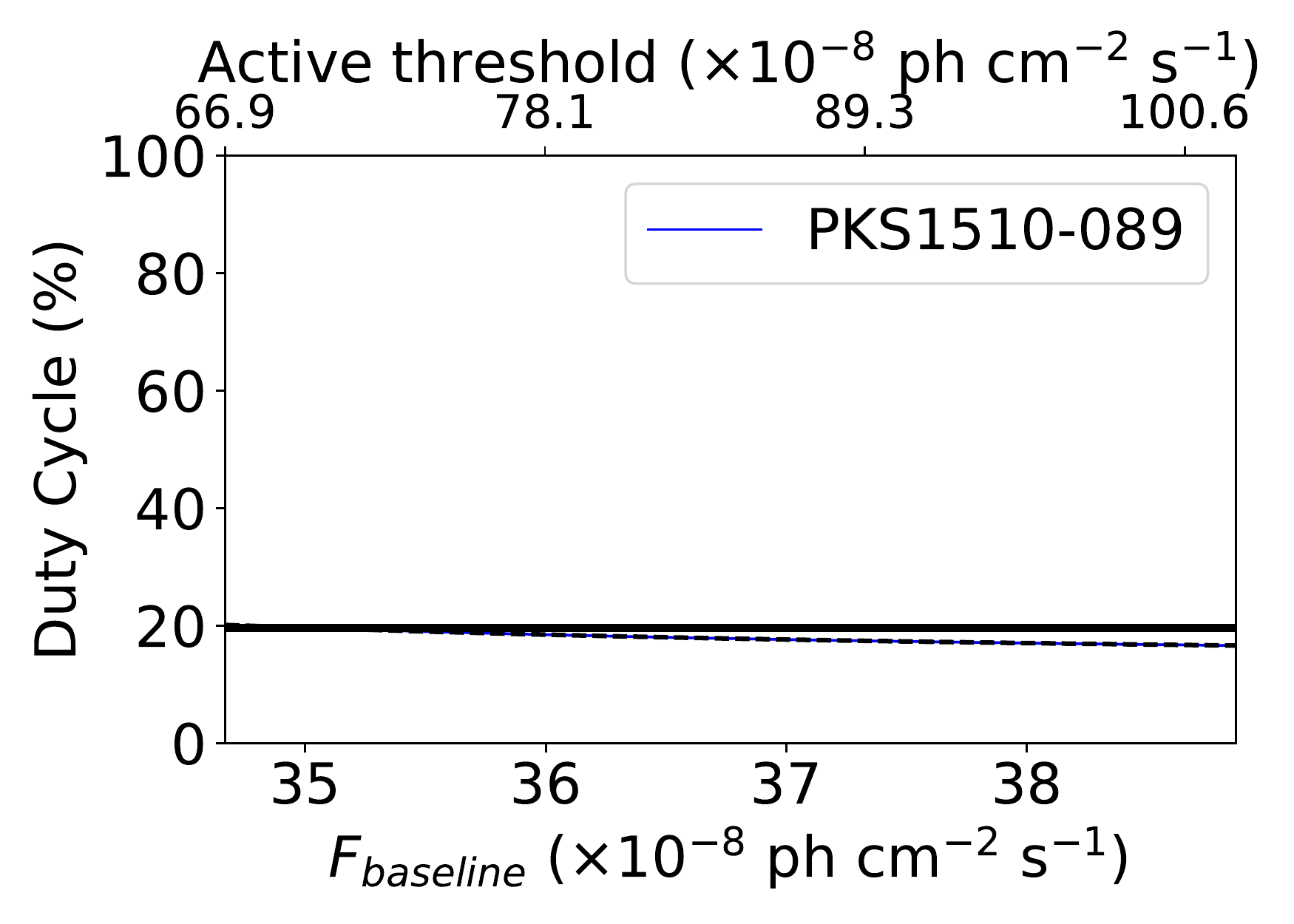}}

\qquad  %%%%%%%
\subfloat{
\includegraphics[width=0.29\textwidth]{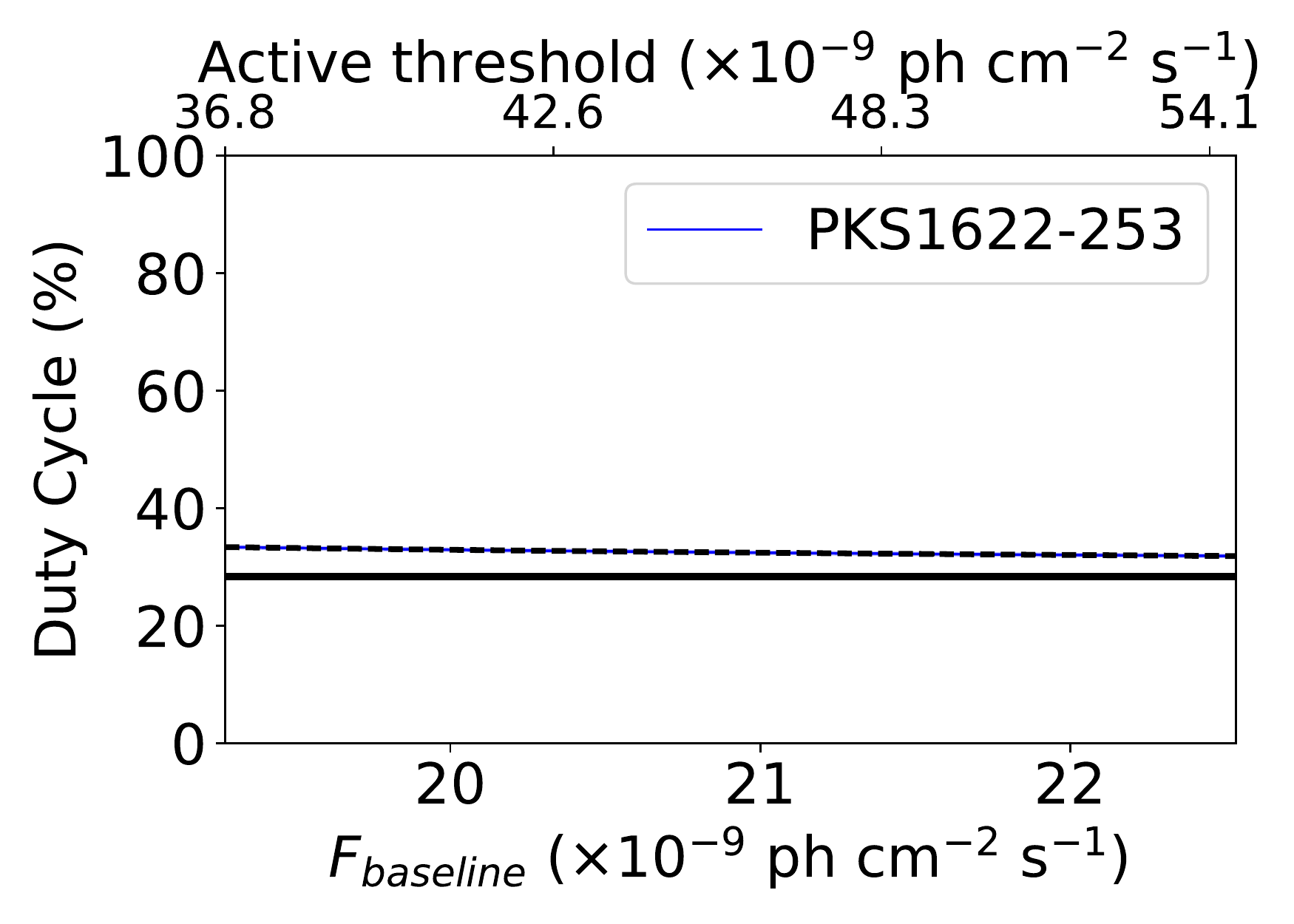}}
\subfloat{
\includegraphics[width=0.29\textwidth]{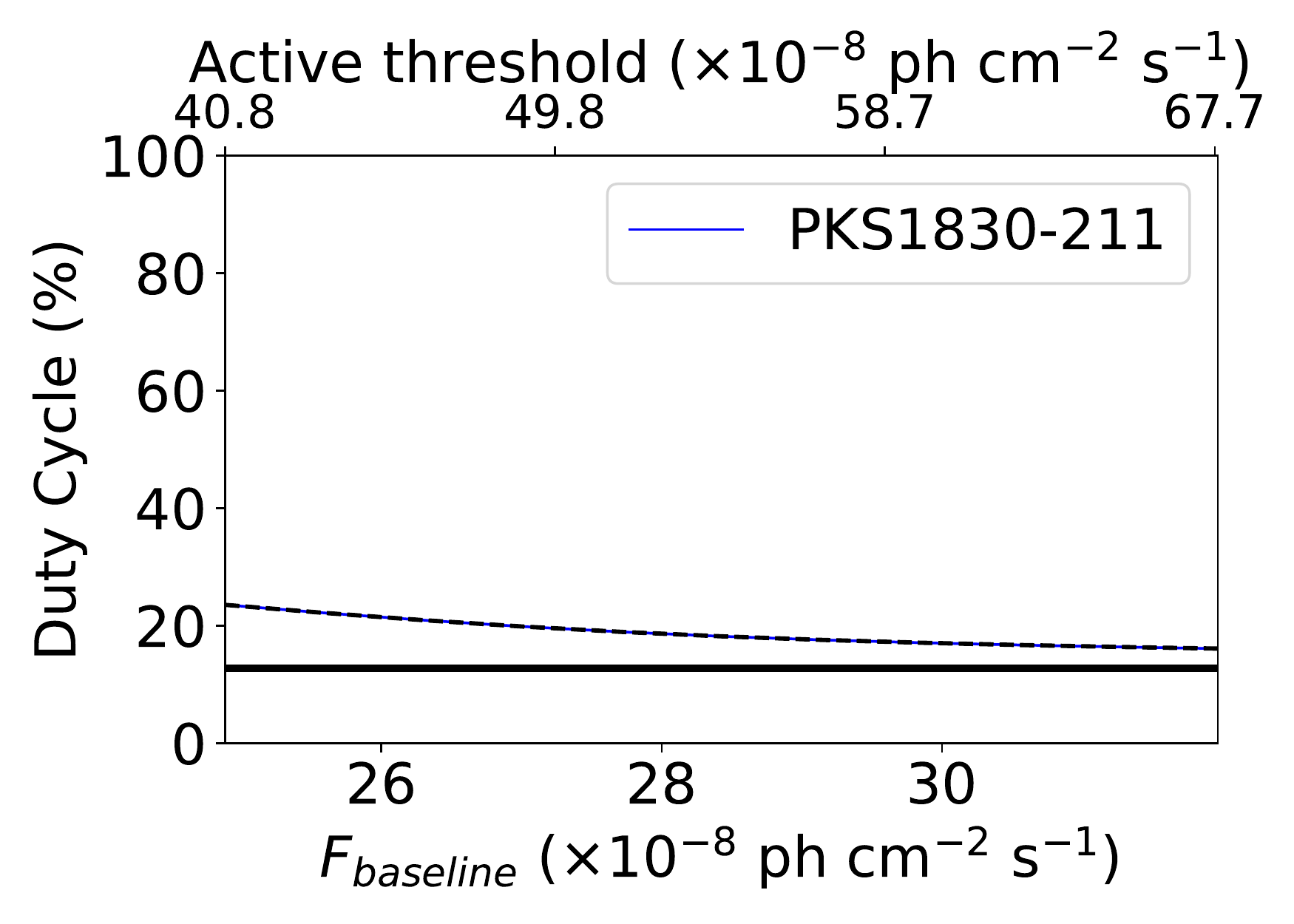}}
\subfloat{
\includegraphics[width=0.29\textwidth]{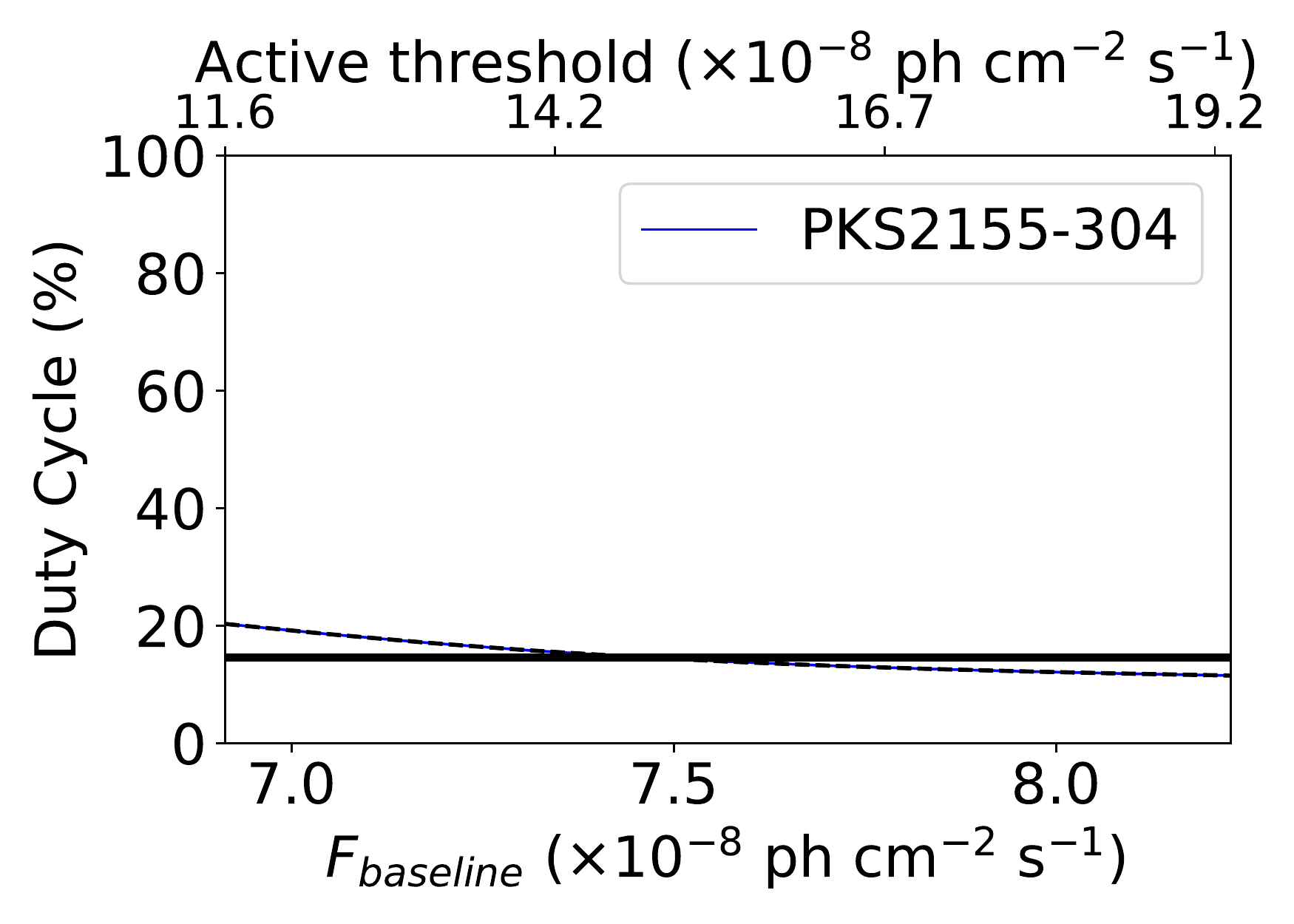}}
\qquad  %%%%%%%
\subfloat{
\includegraphics[width=0.29\textwidth]{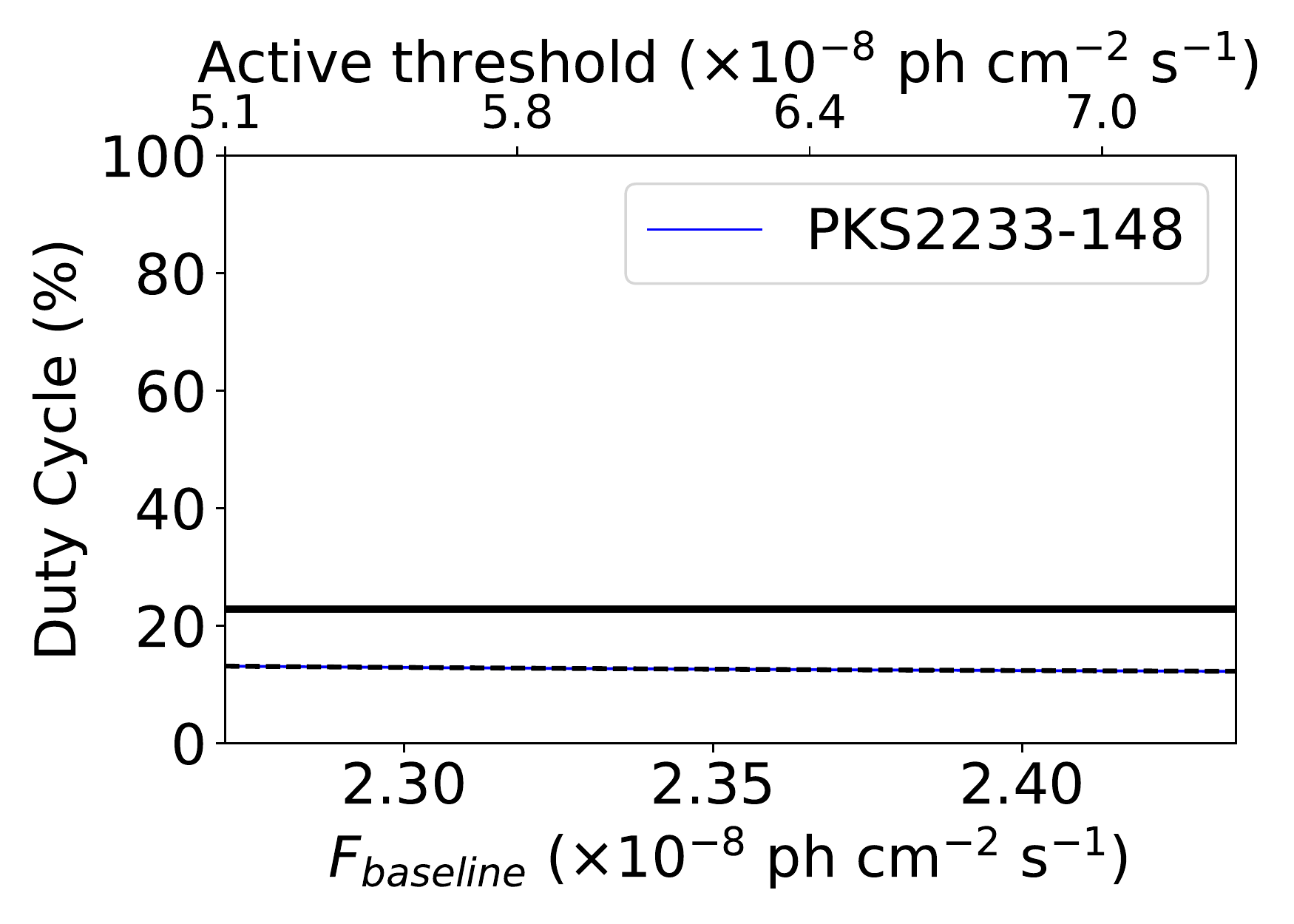}}
\subfloat{
\includegraphics[width=0.29\textwidth]{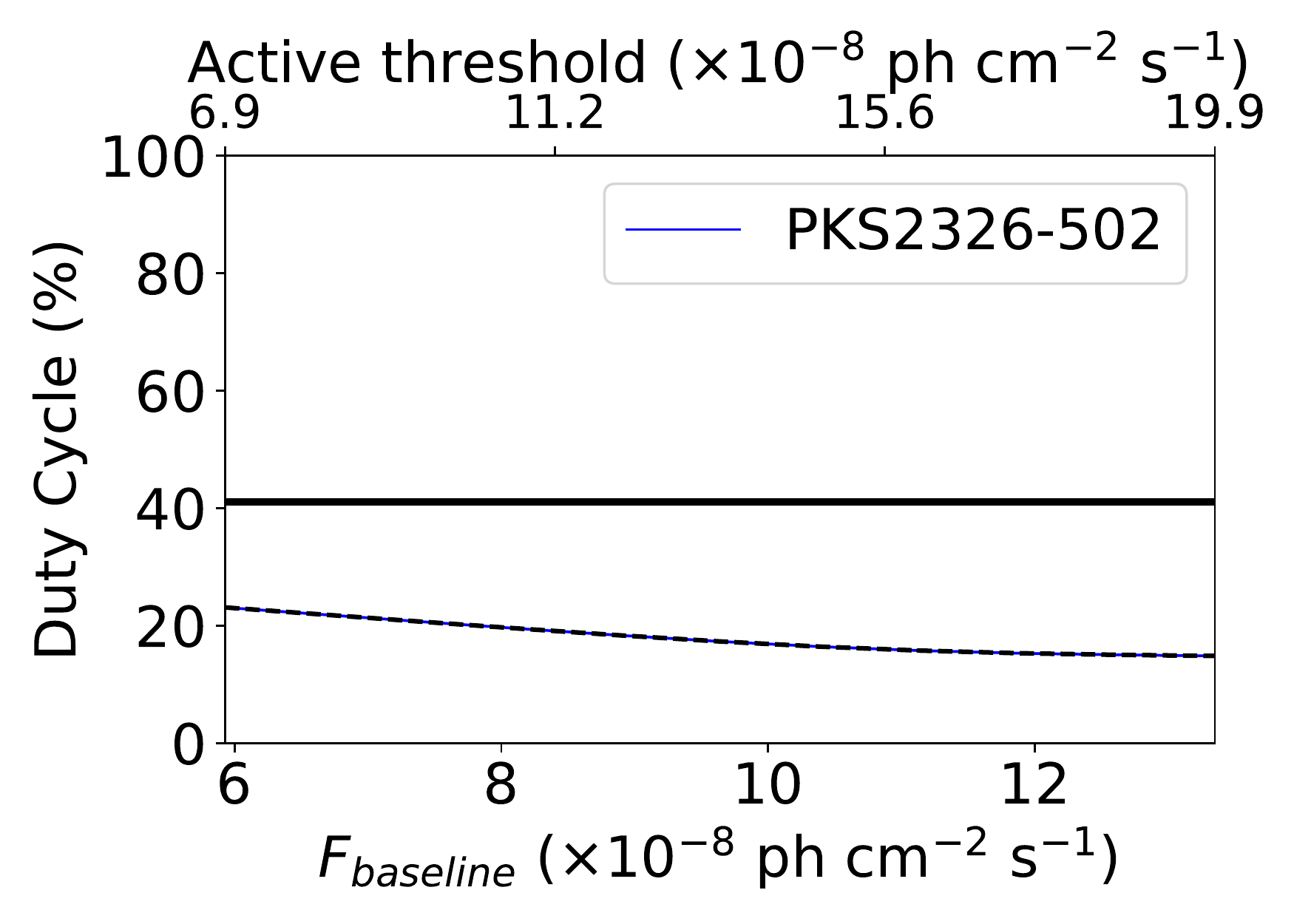}}
\subfloat{
\includegraphics[width=0.29\textwidth]{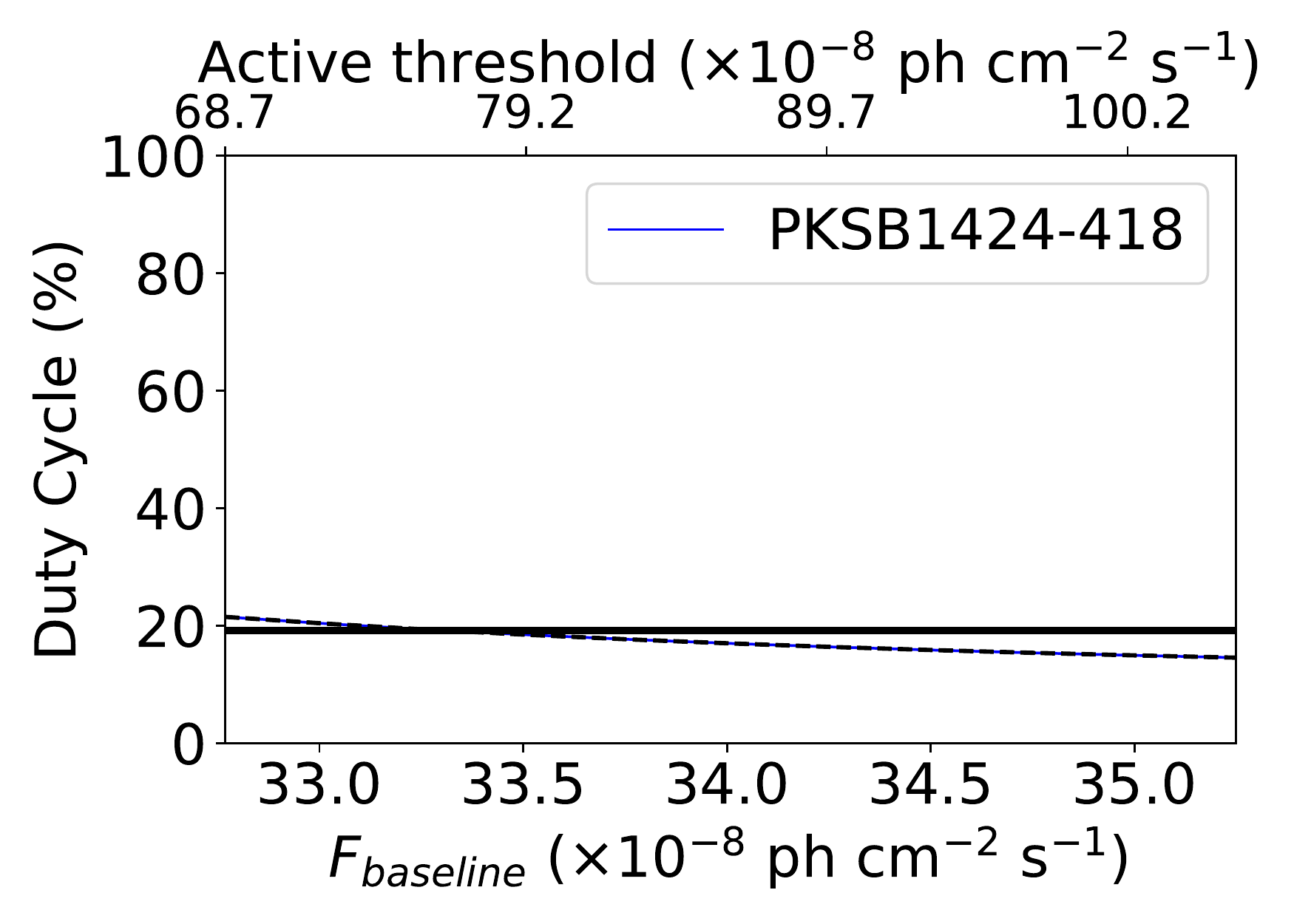}}

\qquad  %%%%%%%

\subfloat{
\includegraphics[width=0.29\textwidth]{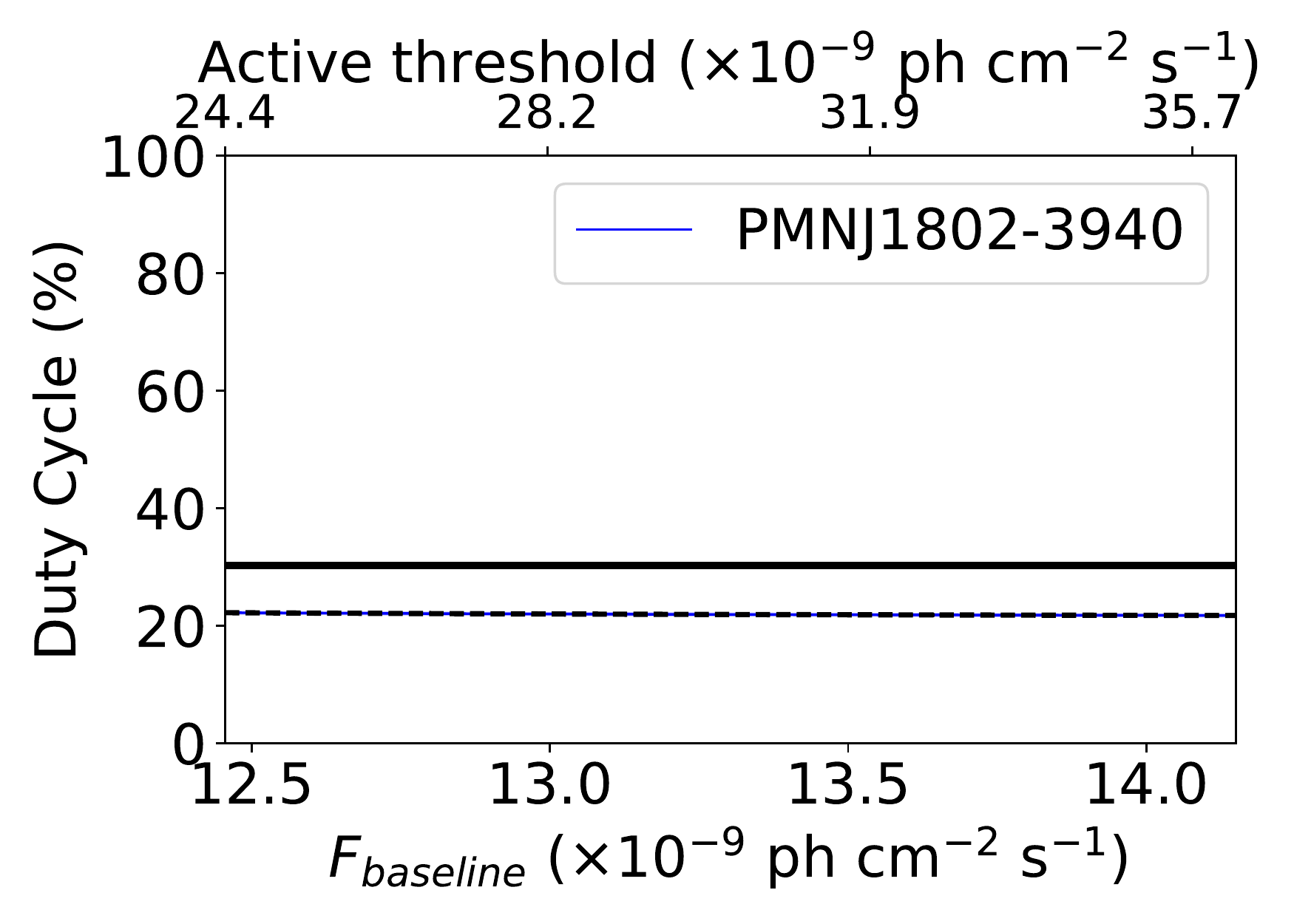}}
\subfloat{
\includegraphics[width=0.29\textwidth]{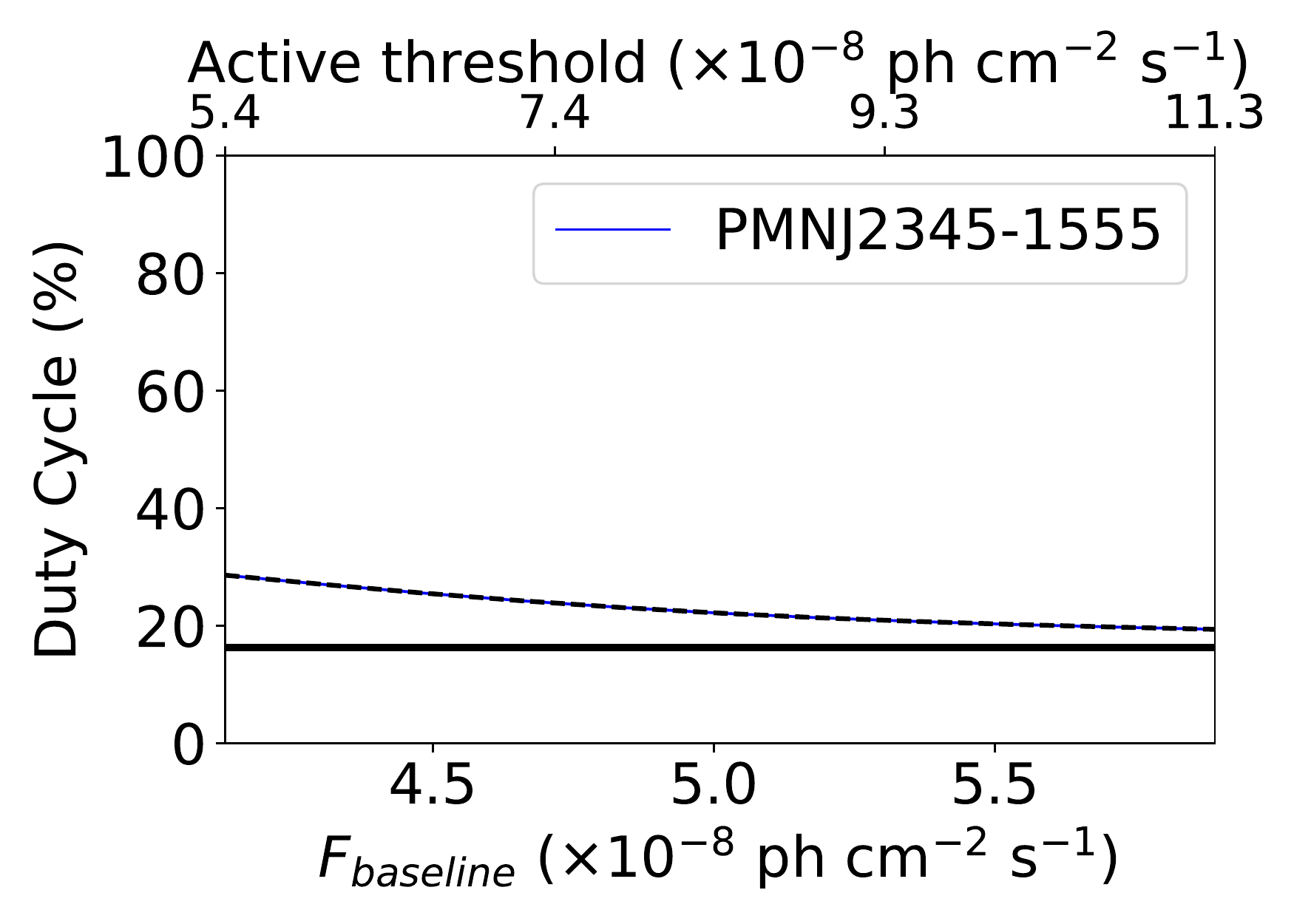}}
\subfloat{
\includegraphics[width=0.29\textwidth]{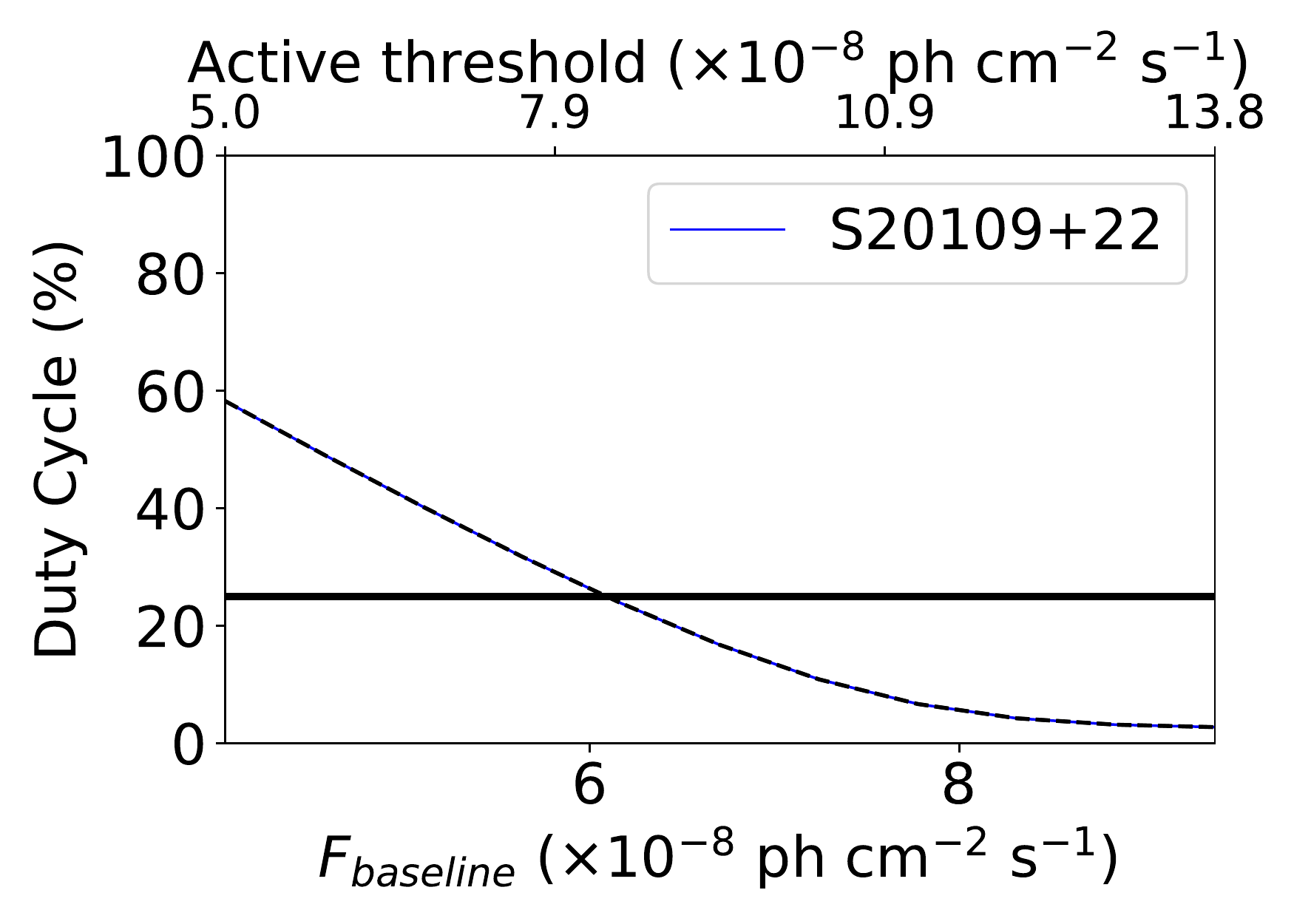}}

\qquad  %%%%%%%
\subfloat{
\includegraphics[width=0.29\textwidth]{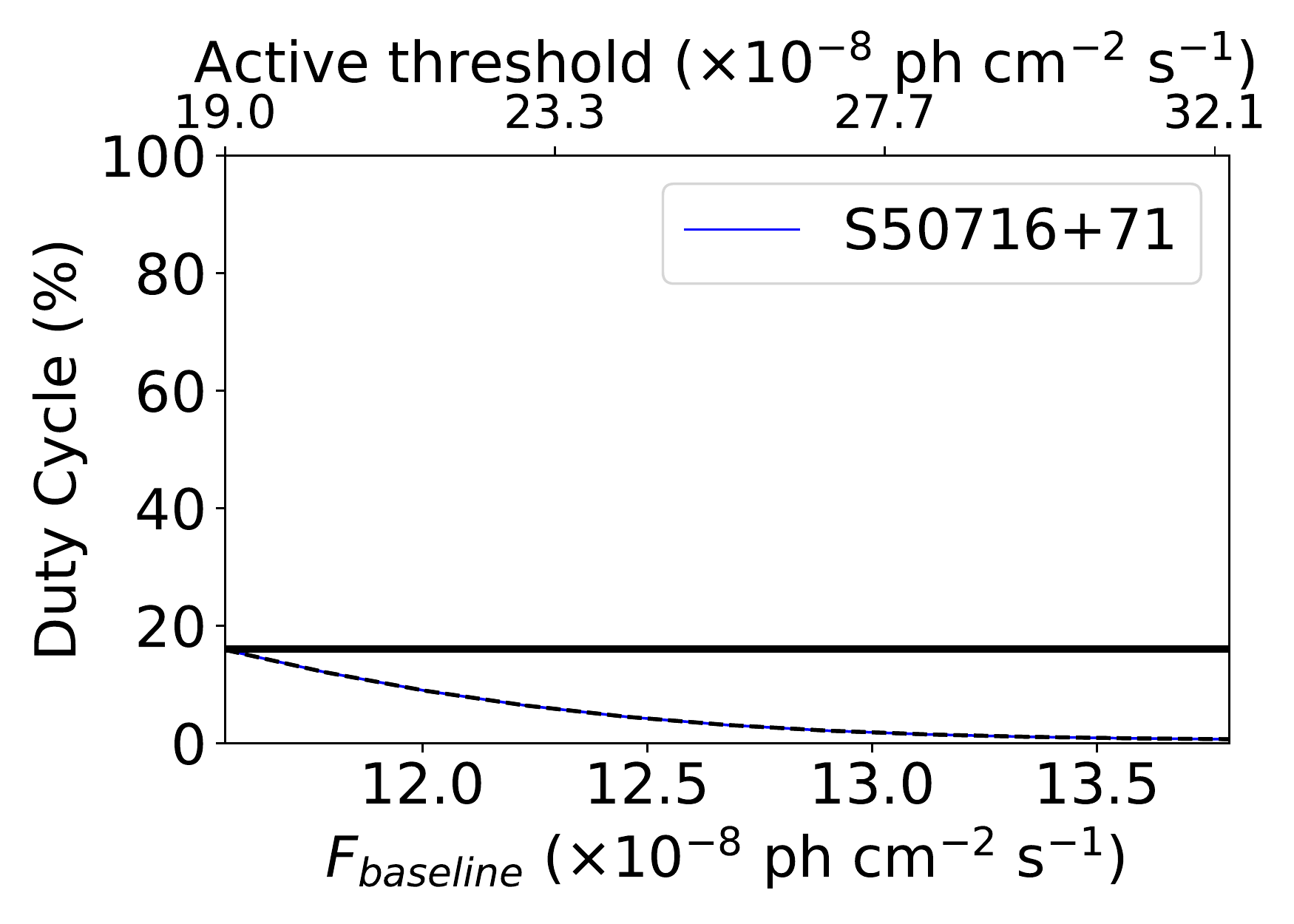}}
\subfloat{
\includegraphics[width=0.29\textwidth]{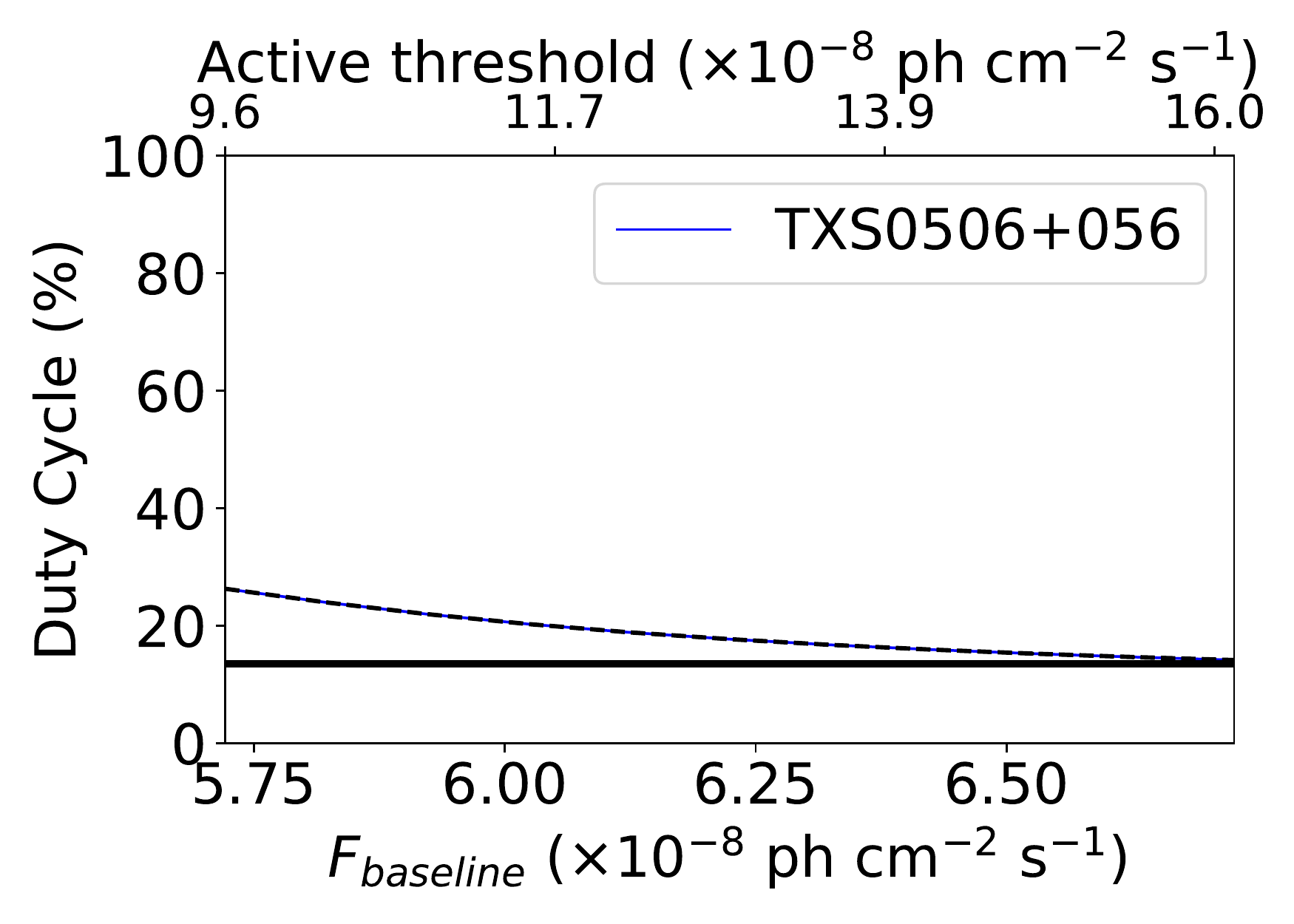}}
\subfloat{
\includegraphics[width=0.29\textwidth]{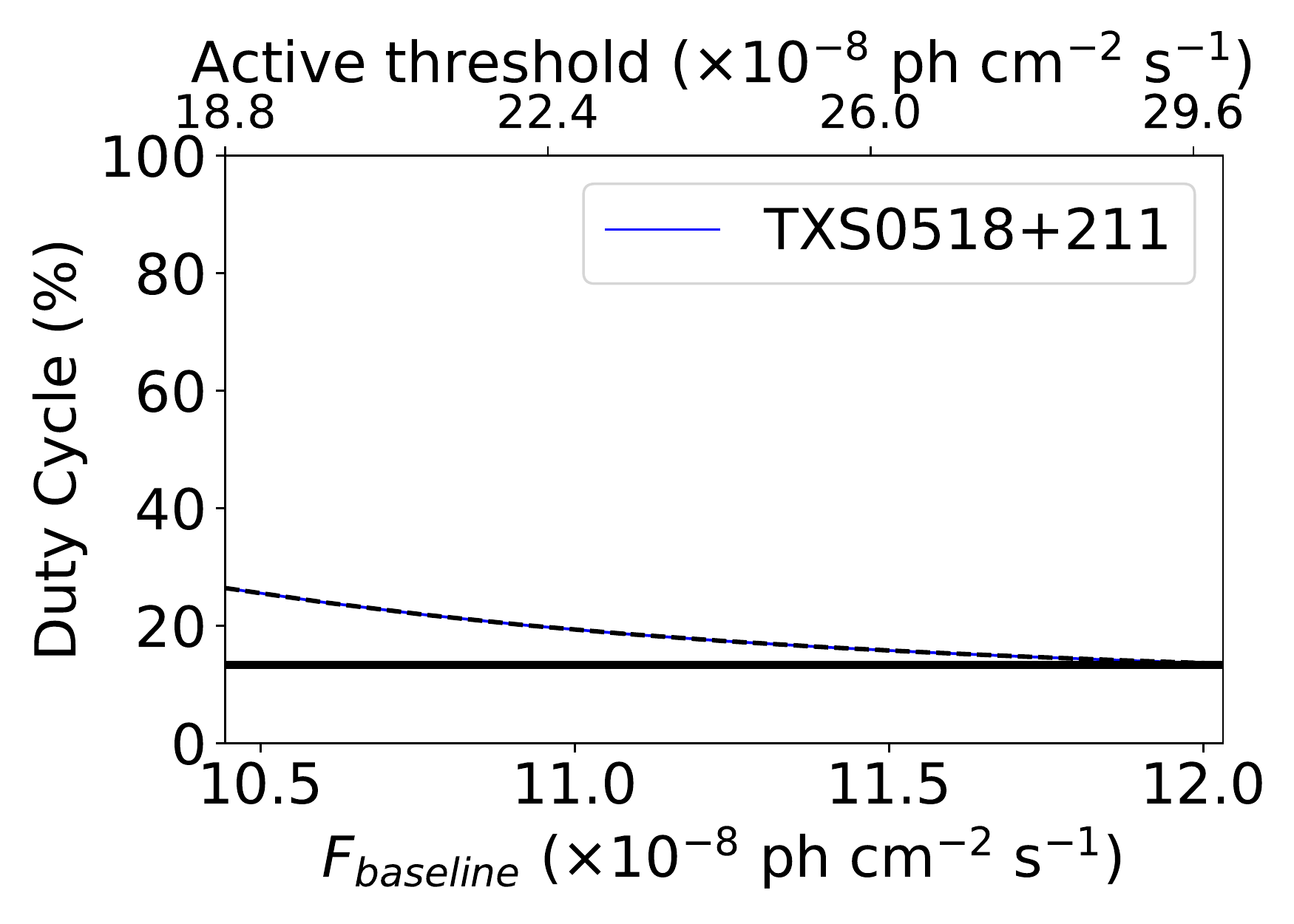}}
\qquad  %%%%%%%
\caption{(continued)}
\end{figure*}

%%%%%%%%%%%%%%%%%%%%%%%%%%%%%%%%%%%%%%%%%%%%%%%%%%FIG SIN EBL
\begin{figure*}
\centering
\subfloat{
\includegraphics[width=0.29\textwidth]{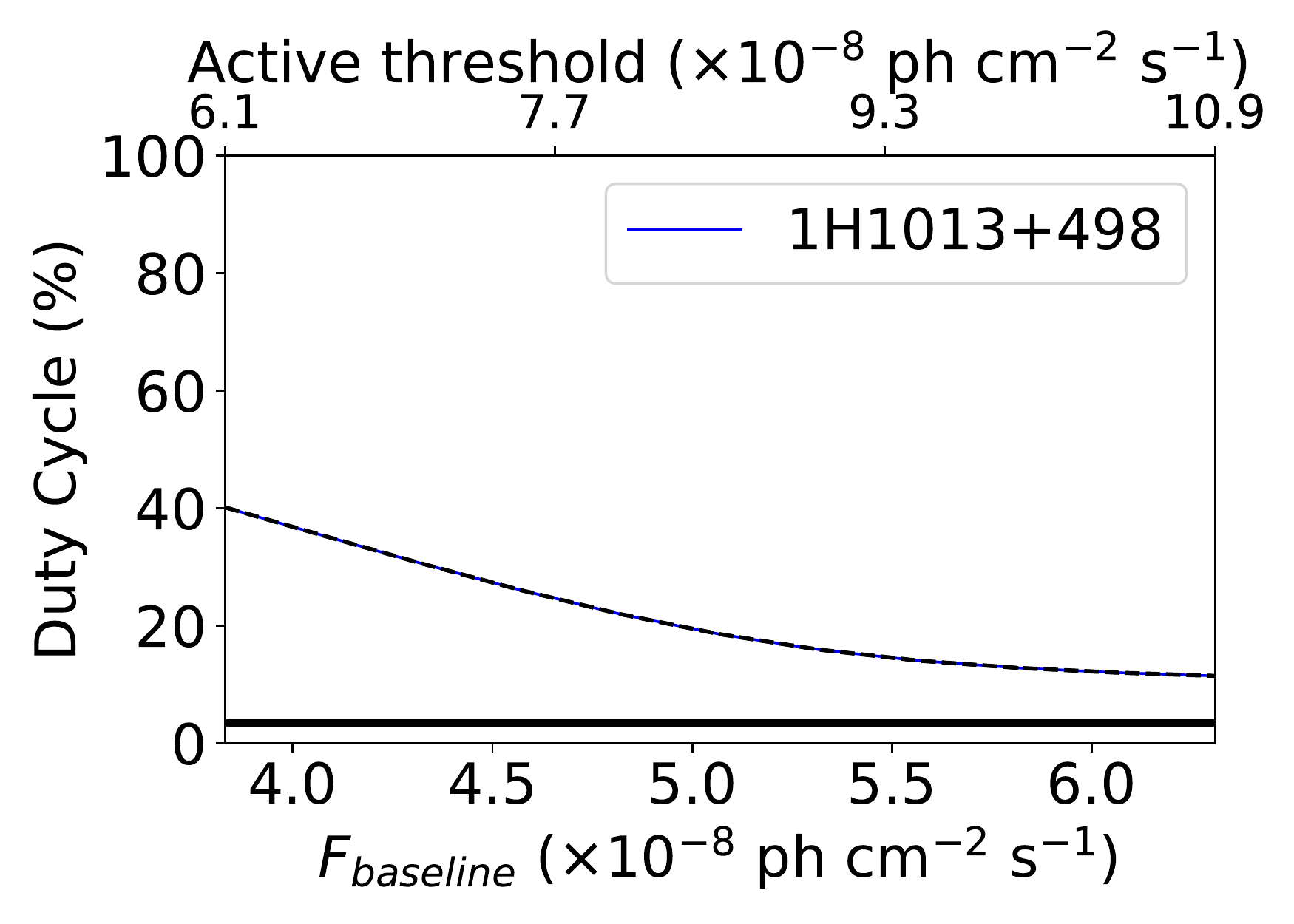}}
\subfloat{
\includegraphics[width=0.29\textwidth]{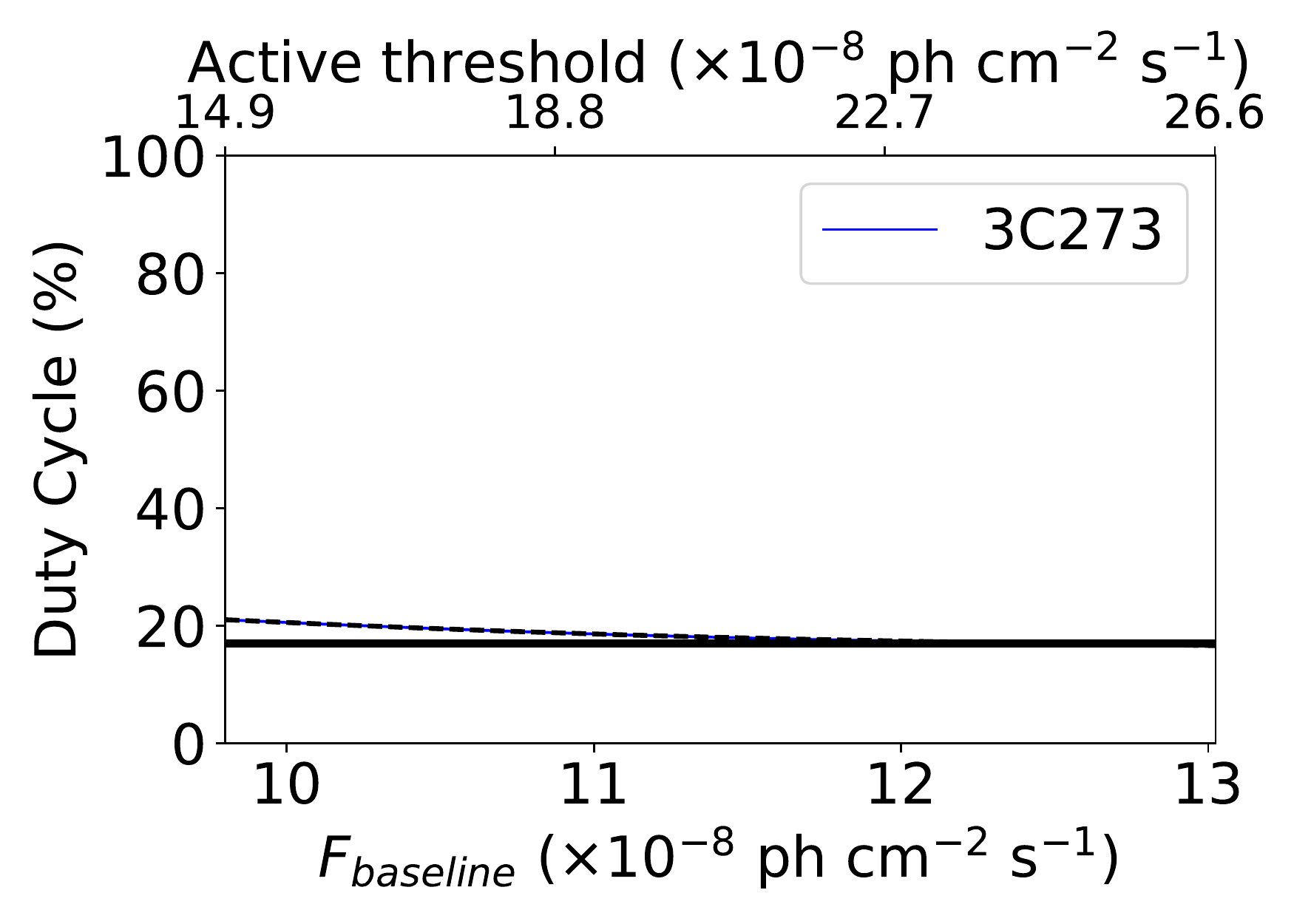}
}
\subfloat{
\includegraphics[width=0.29\textwidth]{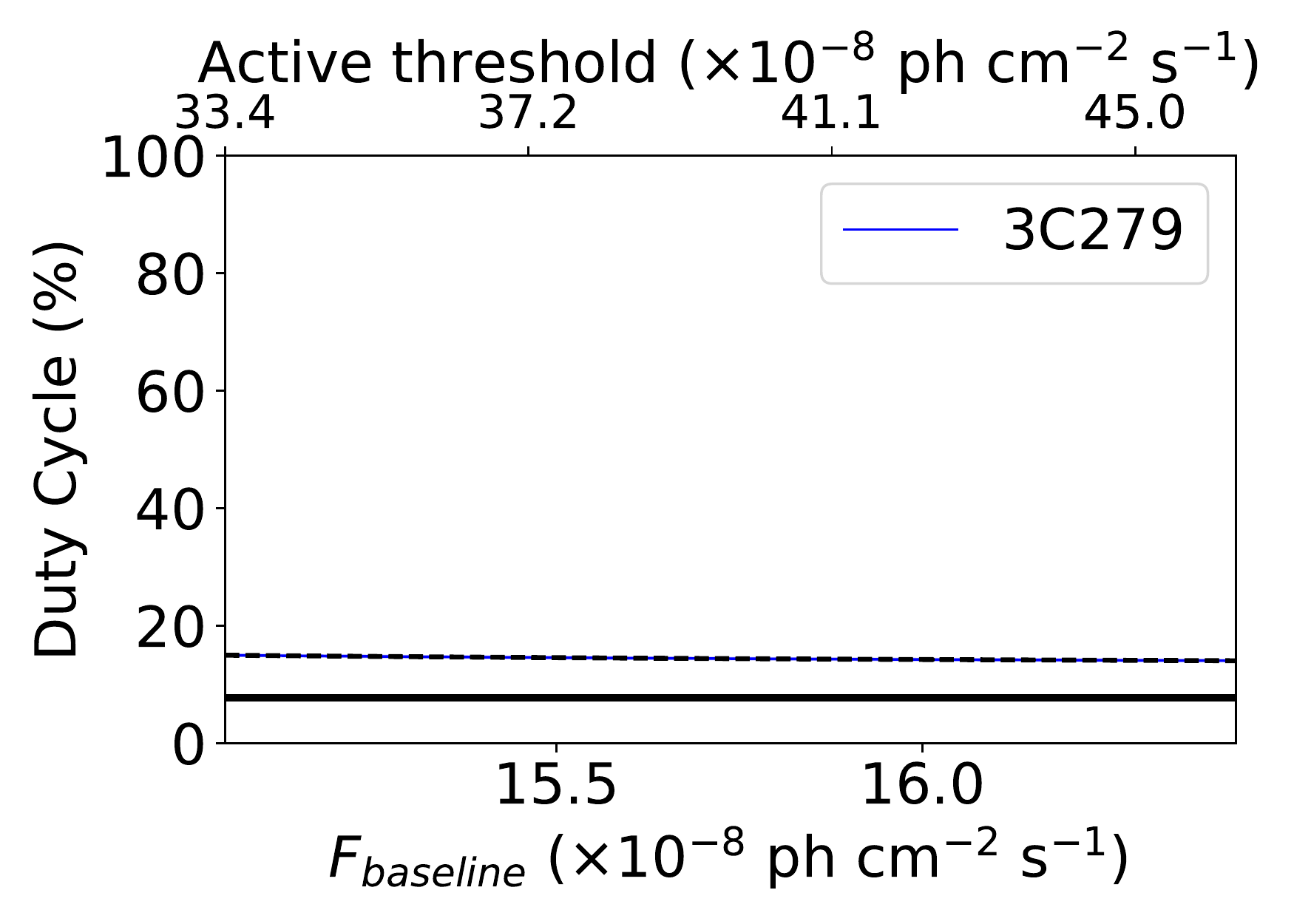}}
\qquad  %%%%%%%
\subfloat{
\includegraphics[width=0.29\textwidth]{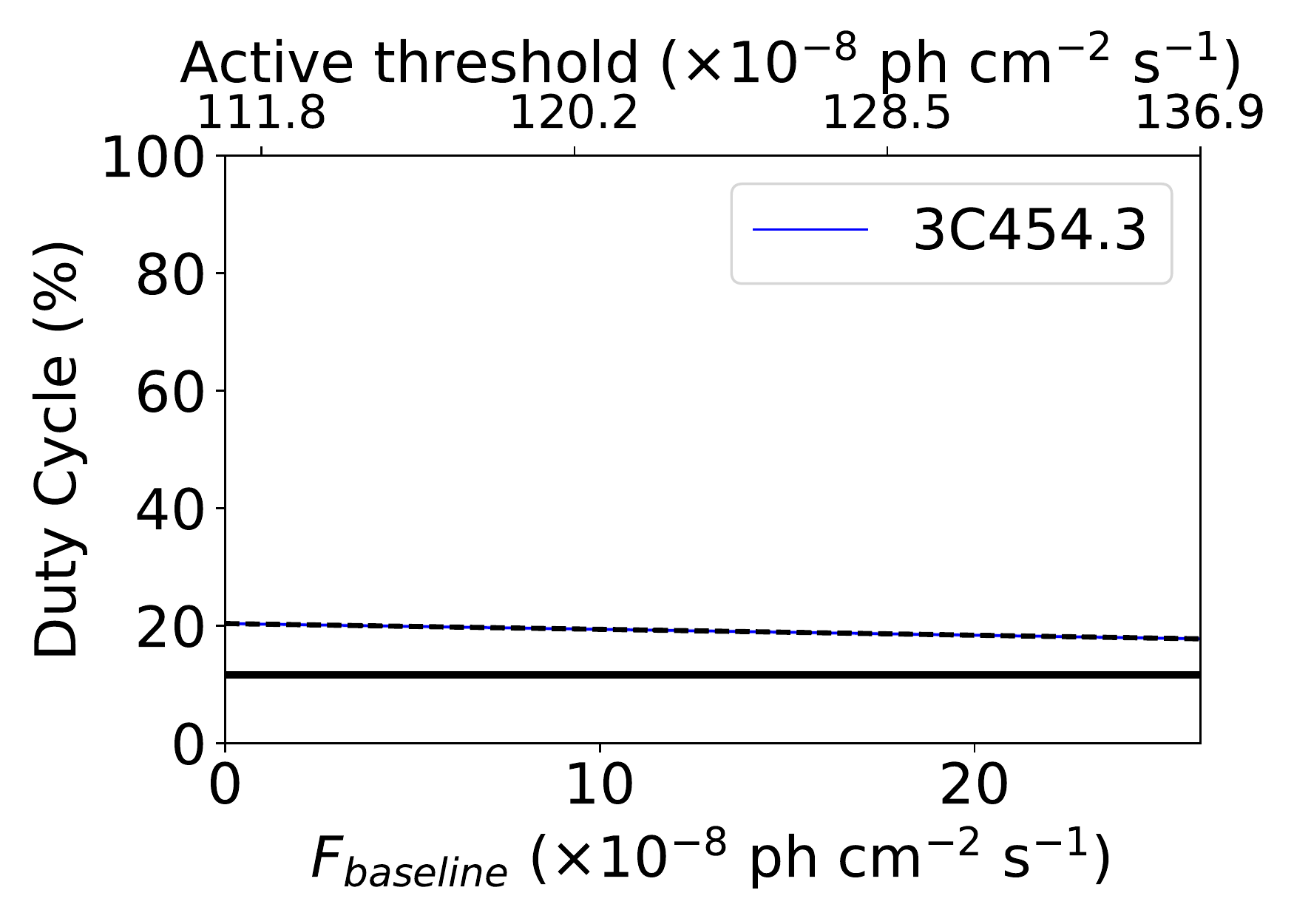}}
\subfloat{
\includegraphics[width=0.29\textwidth]{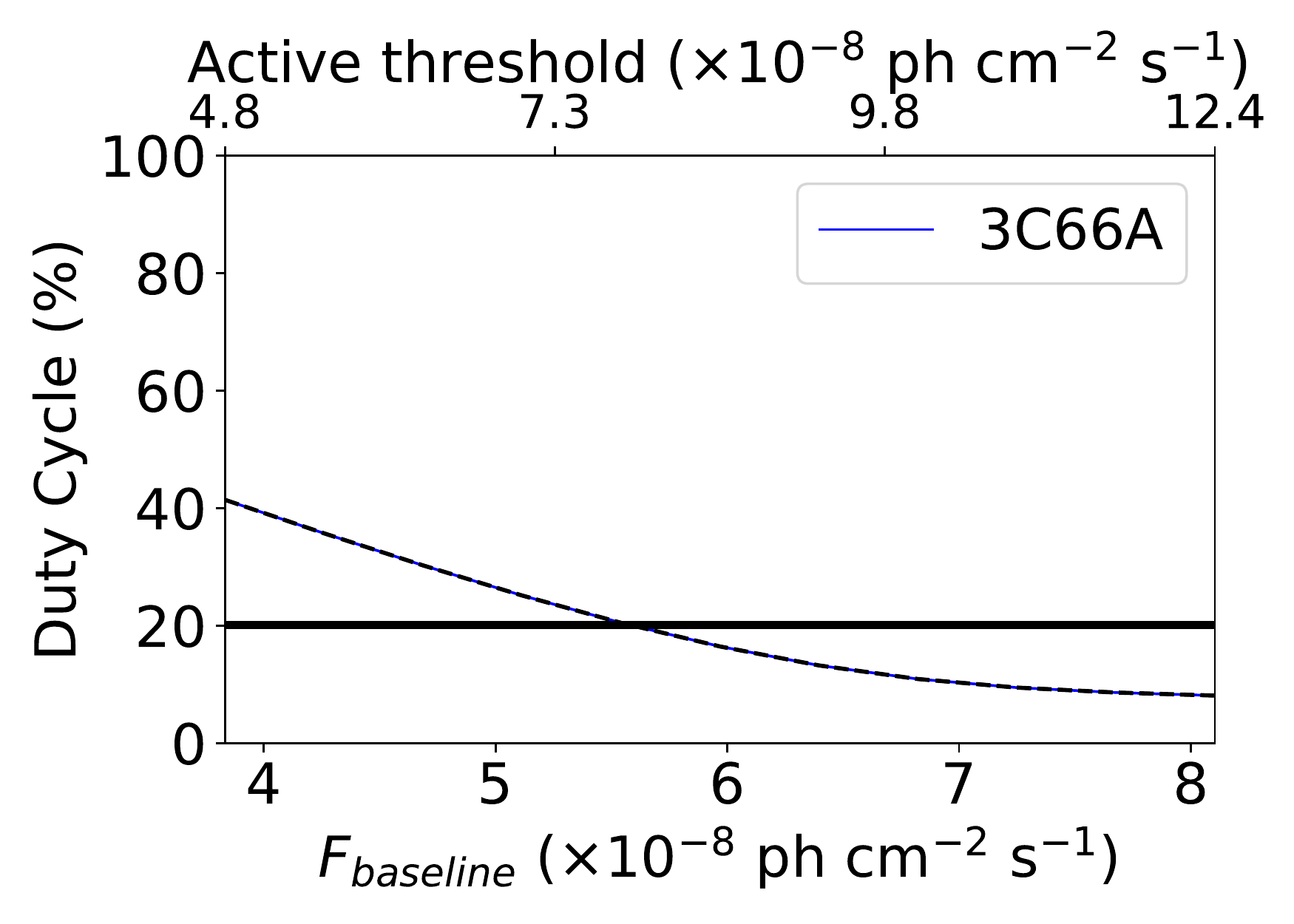}
}
\subfloat{
\includegraphics[width=0.29\textwidth]{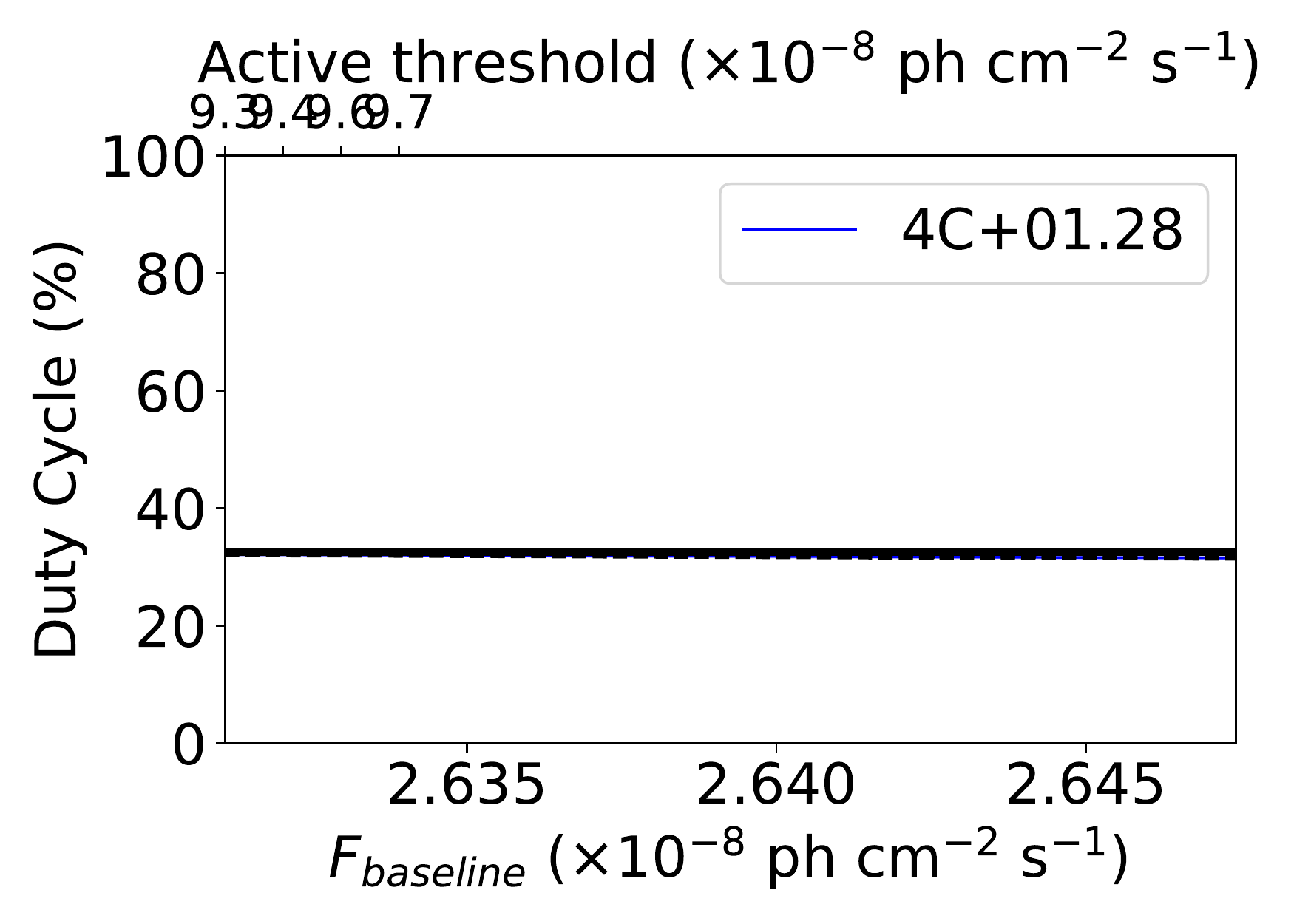}}
\qquad  %%%%%%%
\subfloat{
\includegraphics[width=0.29\textwidth]{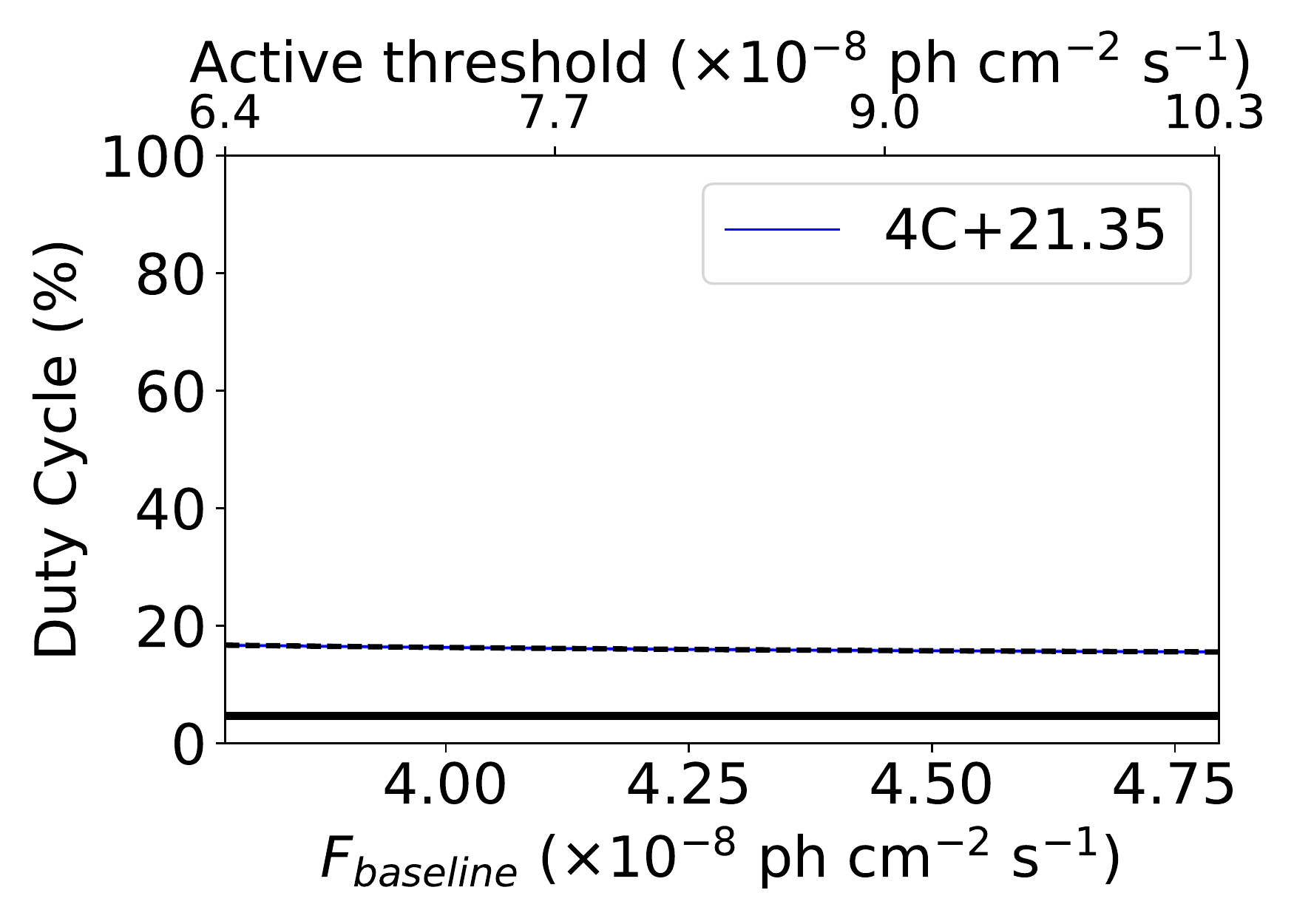}}
\subfloat{
\includegraphics[width=0.29\textwidth]{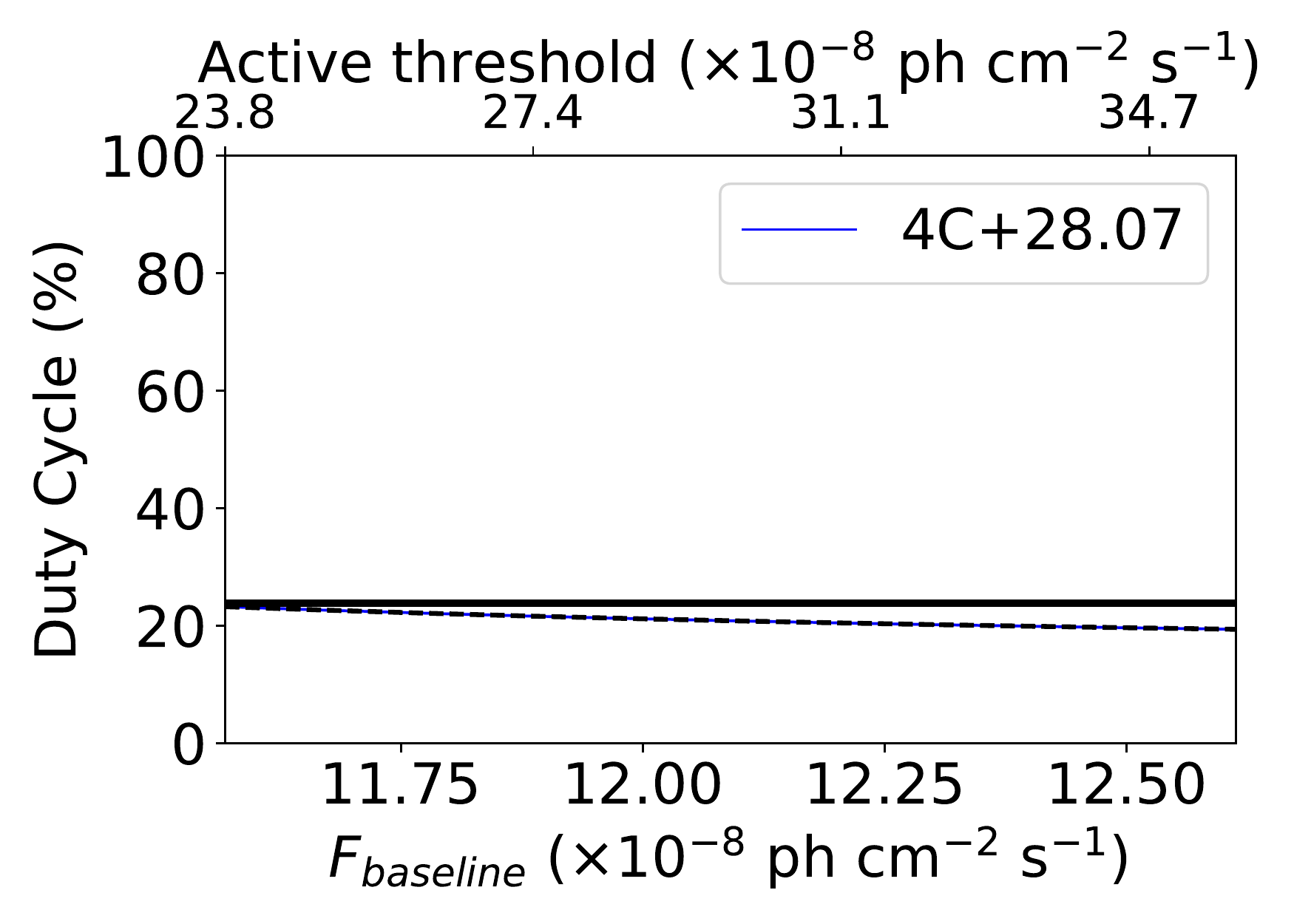}}
\subfloat{
\includegraphics[width=0.29\textwidth]{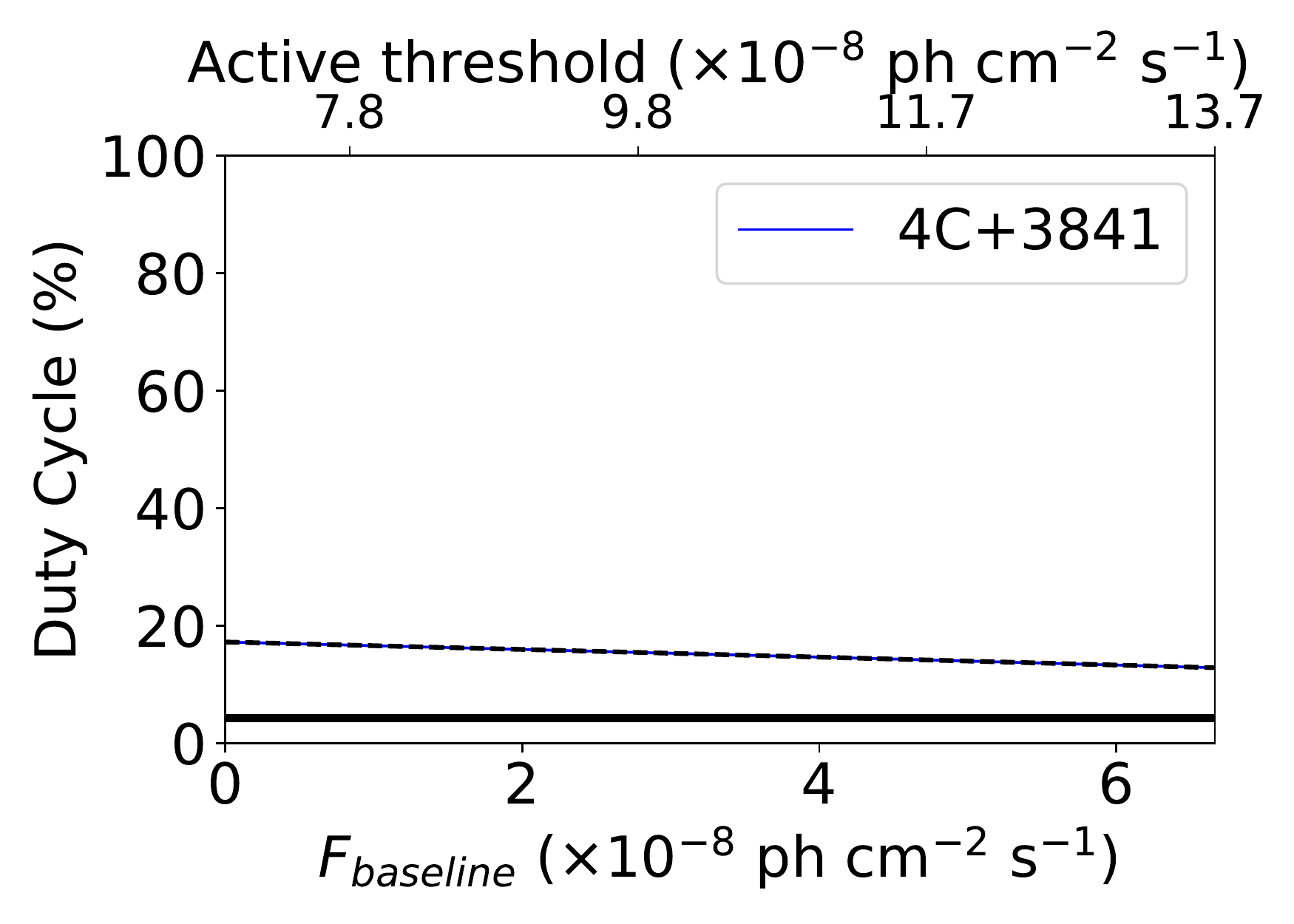}}
\qquad  %%%%%%%
\subfloat{
\includegraphics[width=0.29\textwidth]{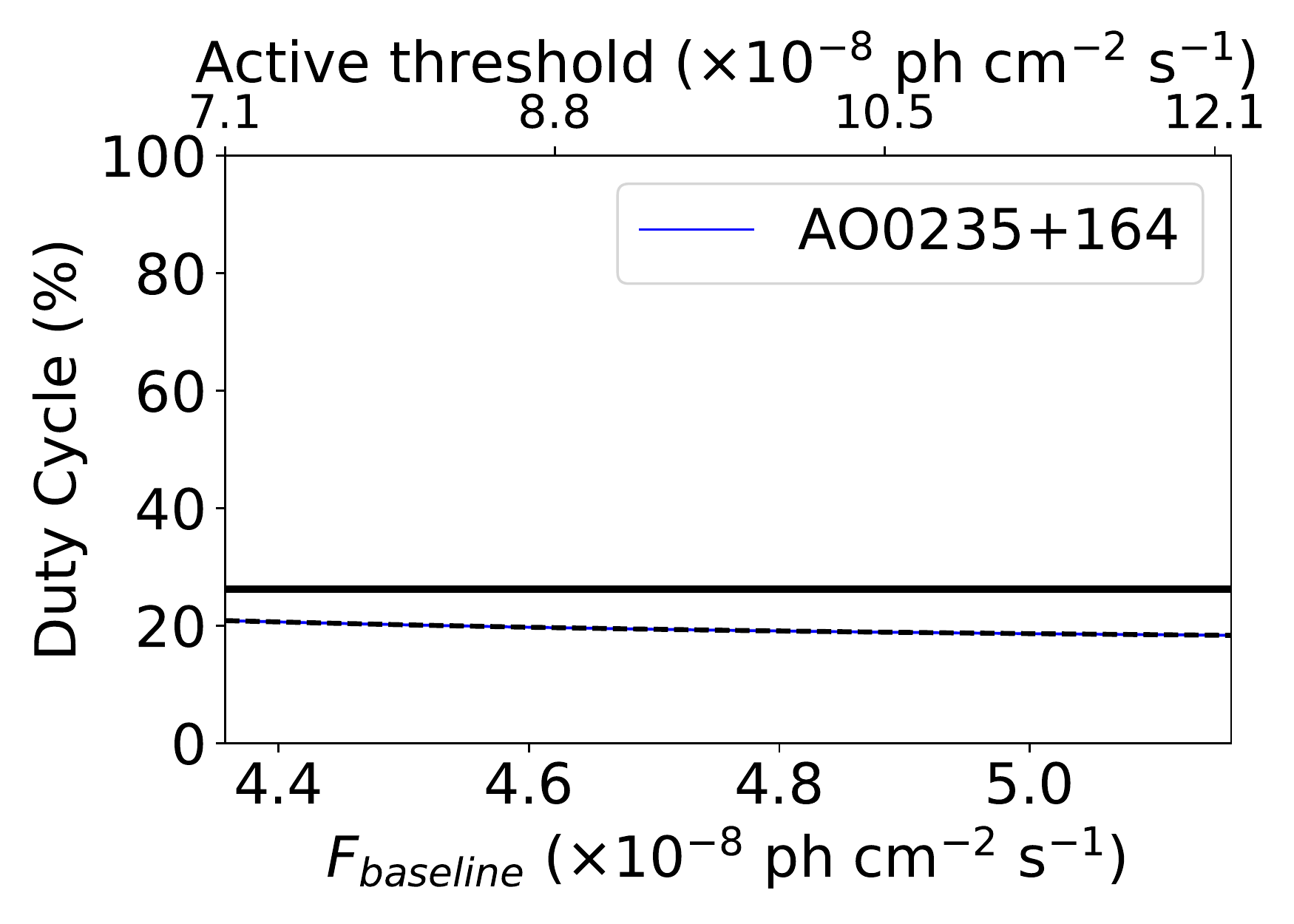}}
\subfloat{
\includegraphics[width=0.29\textwidth]{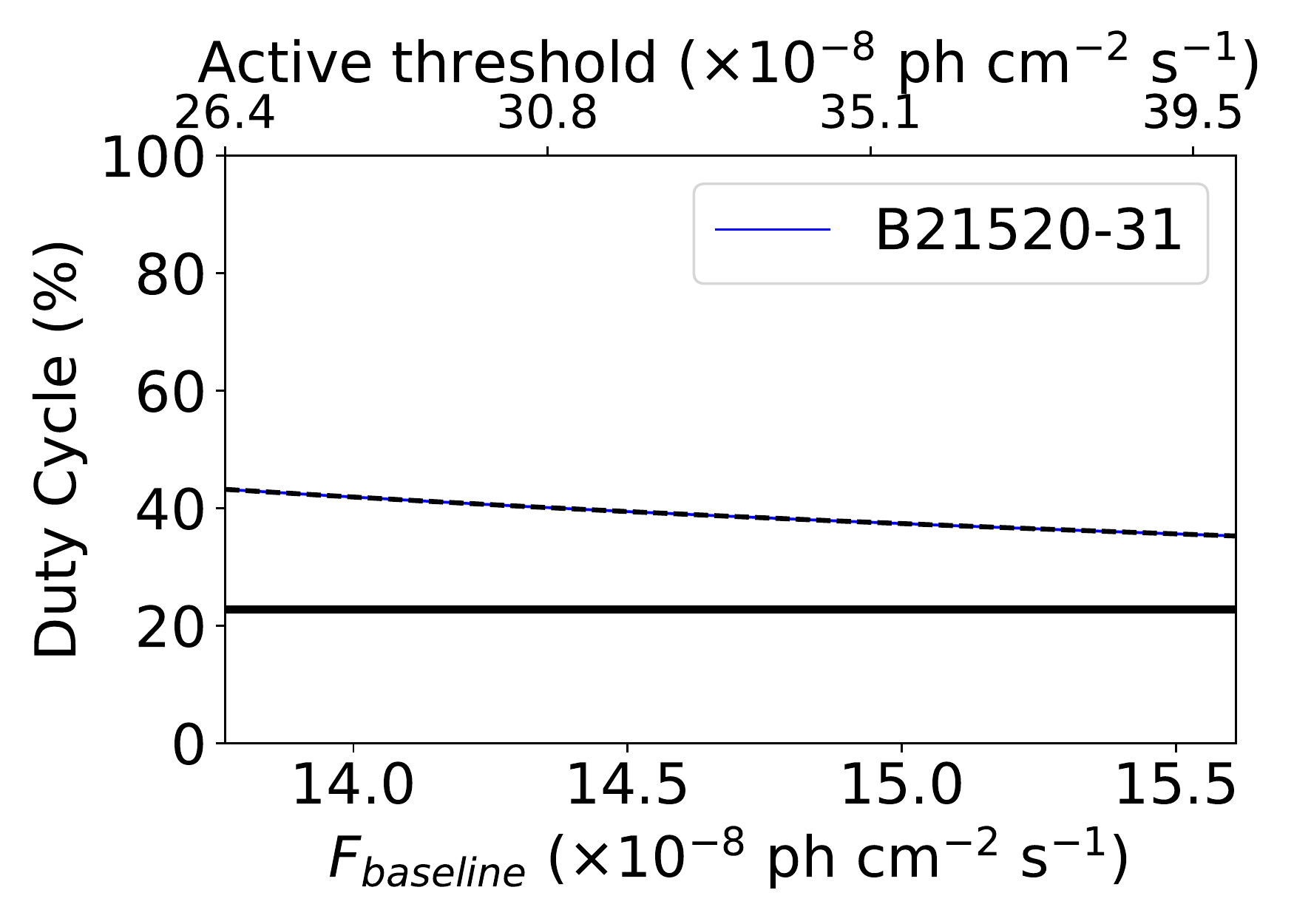}}
\subfloat{
\includegraphics[width=0.29\textwidth]{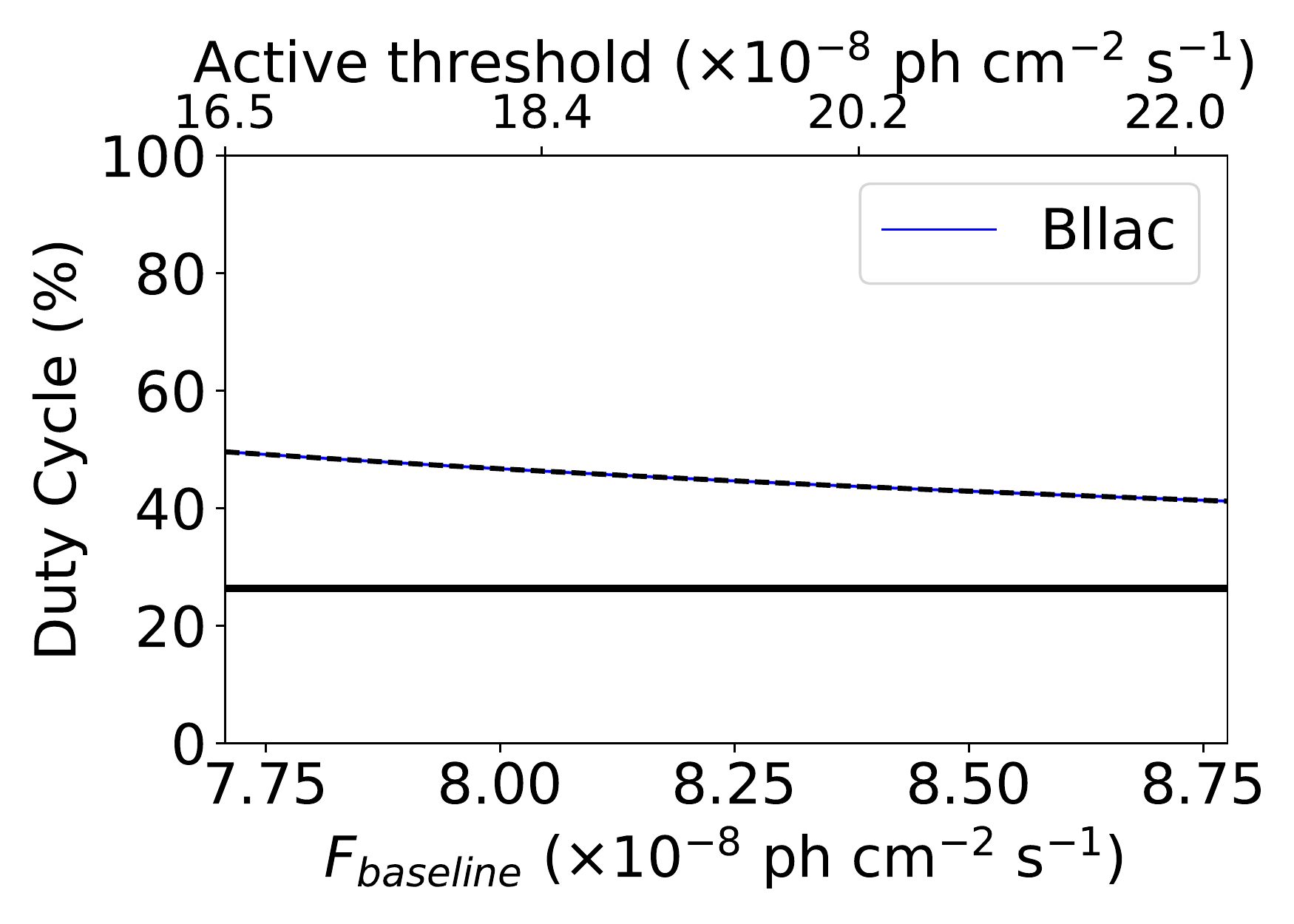}}
\qquad  %%%%%%%
\subfloat{
\includegraphics[width=0.29\textwidth]{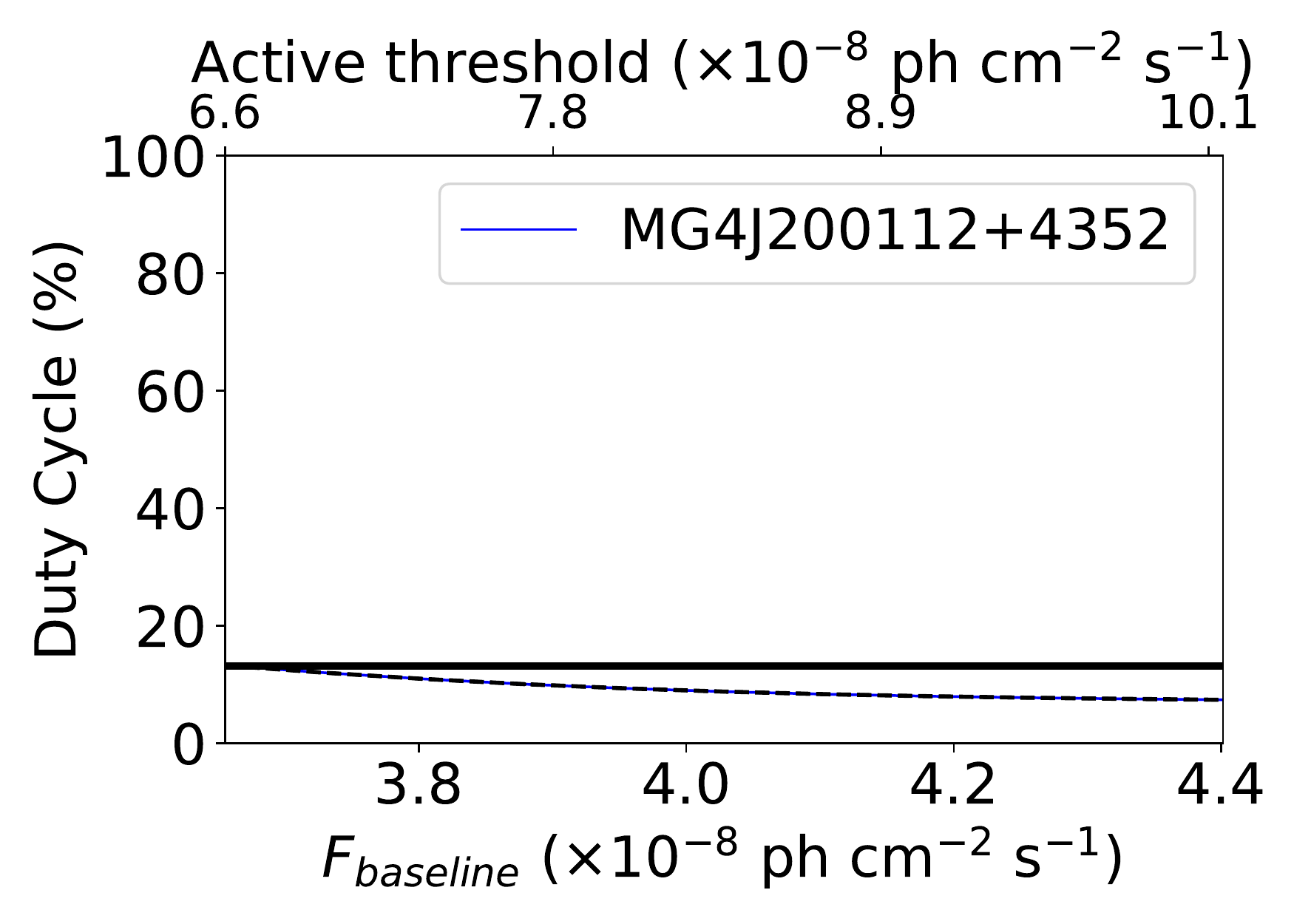}}
\subfloat{
\includegraphics[width=0.29\textwidth]{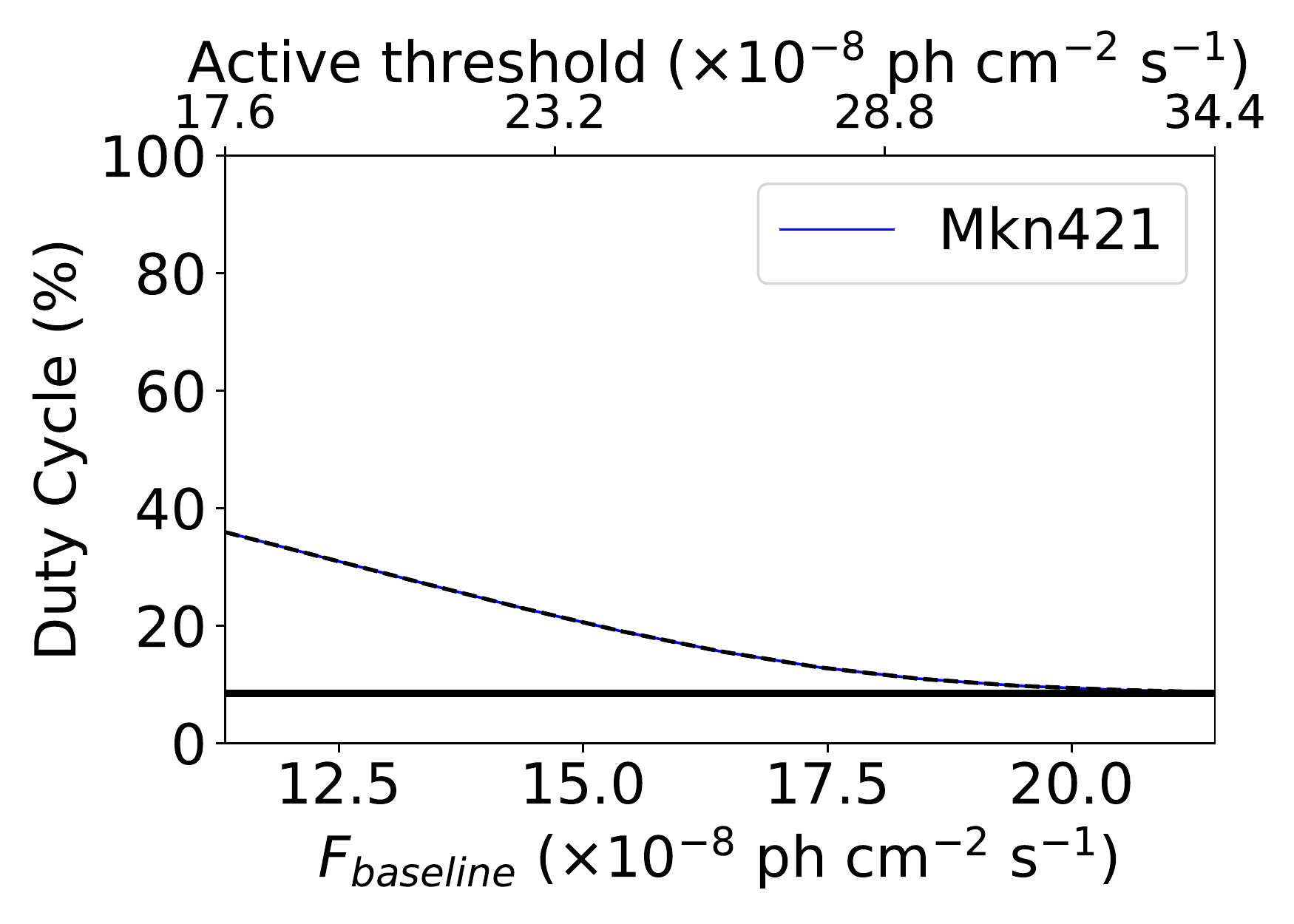}}
\subfloat{
\includegraphics[width=0.29\textwidth]{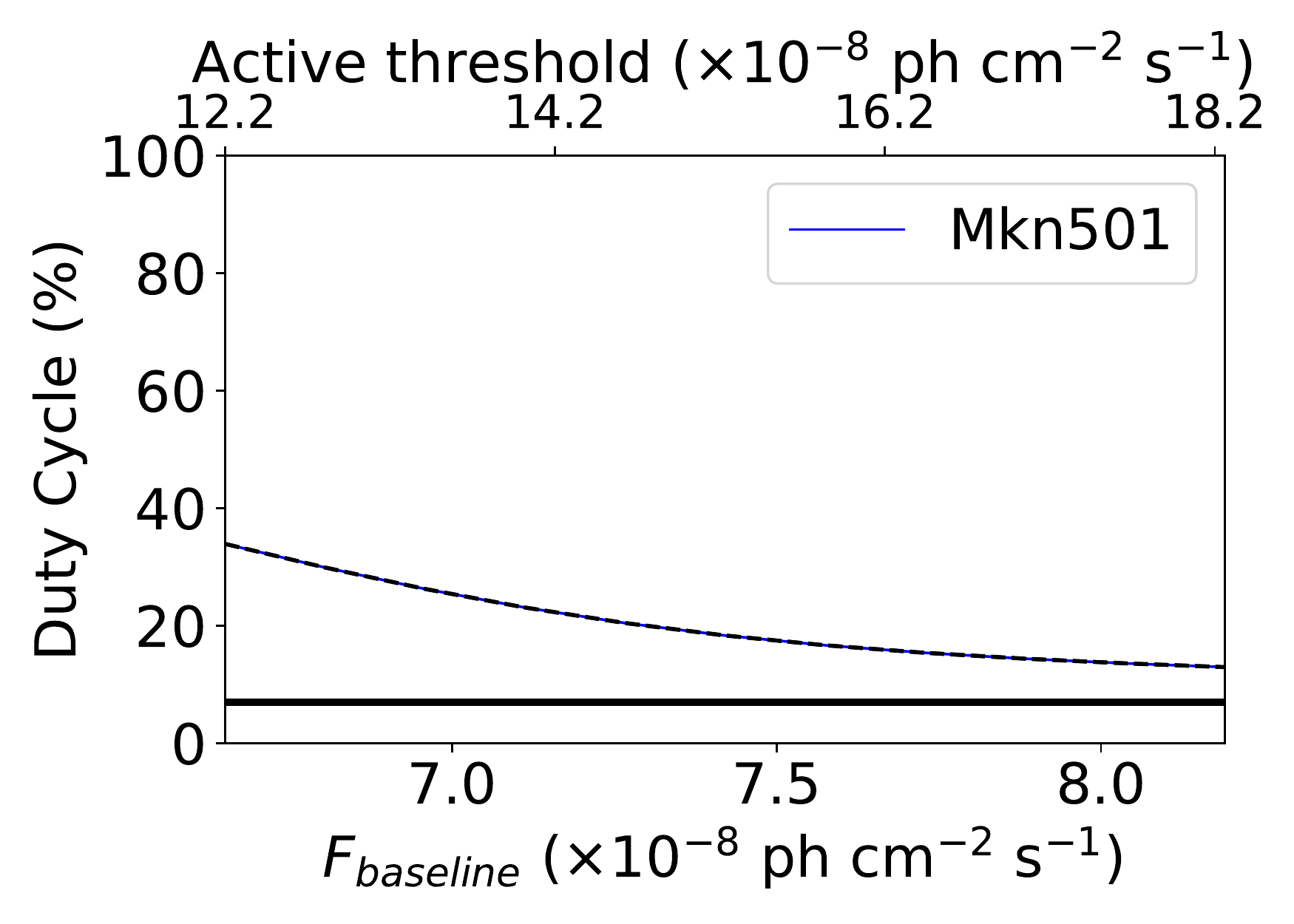}}
\qquad  %%%%%%%
\subfloat{
\includegraphics[width=0.27\textwidth]{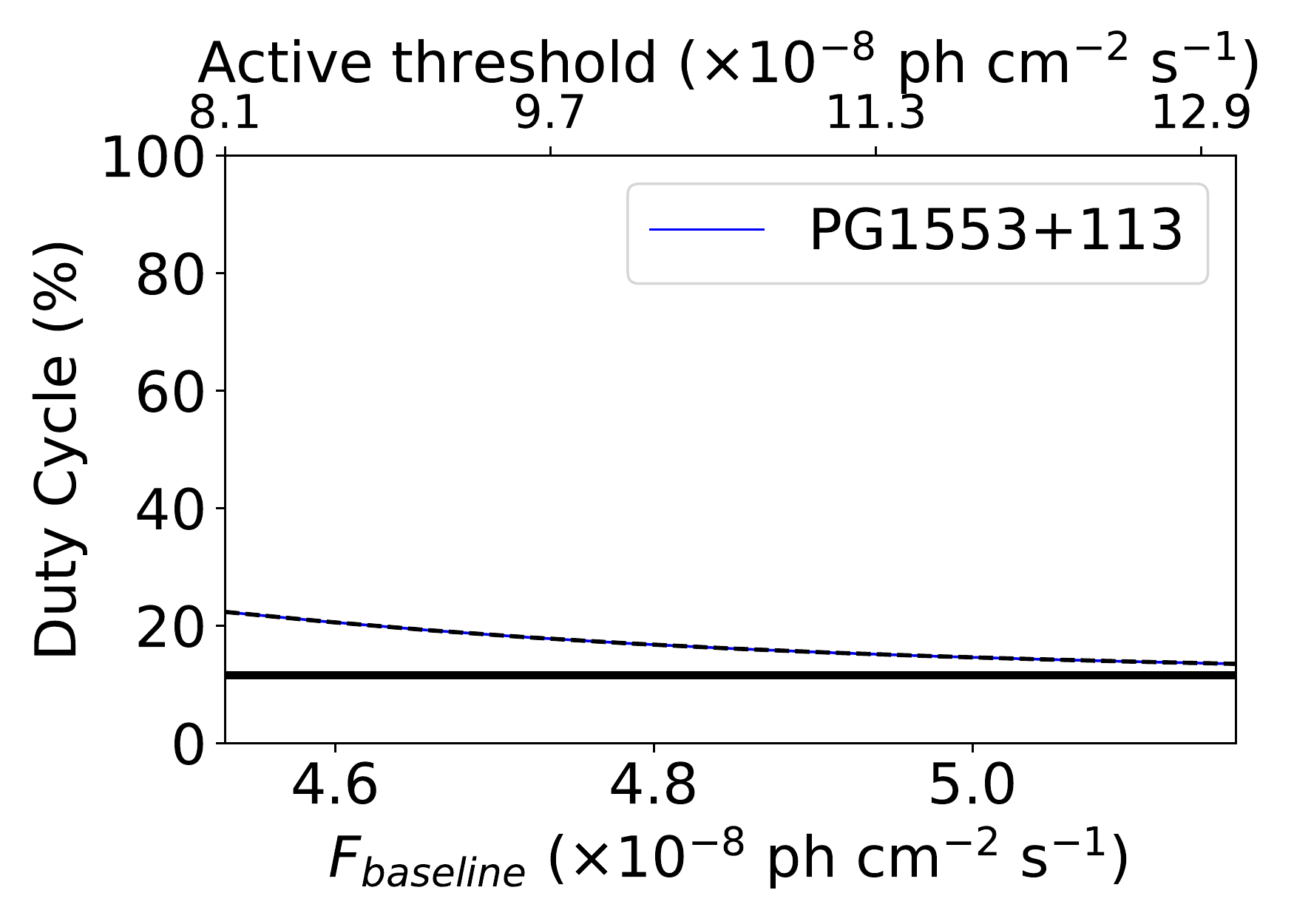}}
\subfloat{
\includegraphics[width=0.29\textwidth]{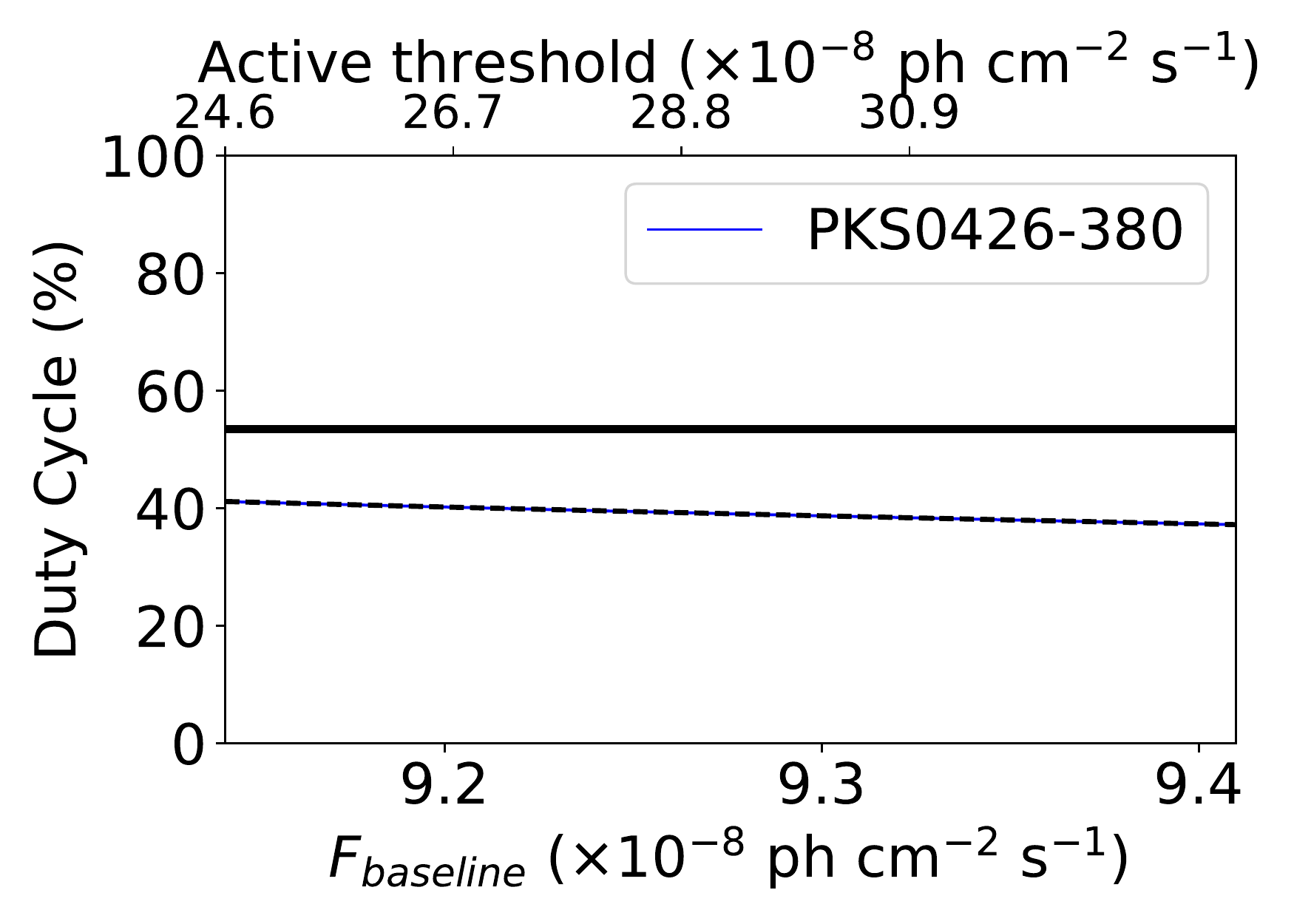}}
\subfloat{
\includegraphics[width=0.29\textwidth]{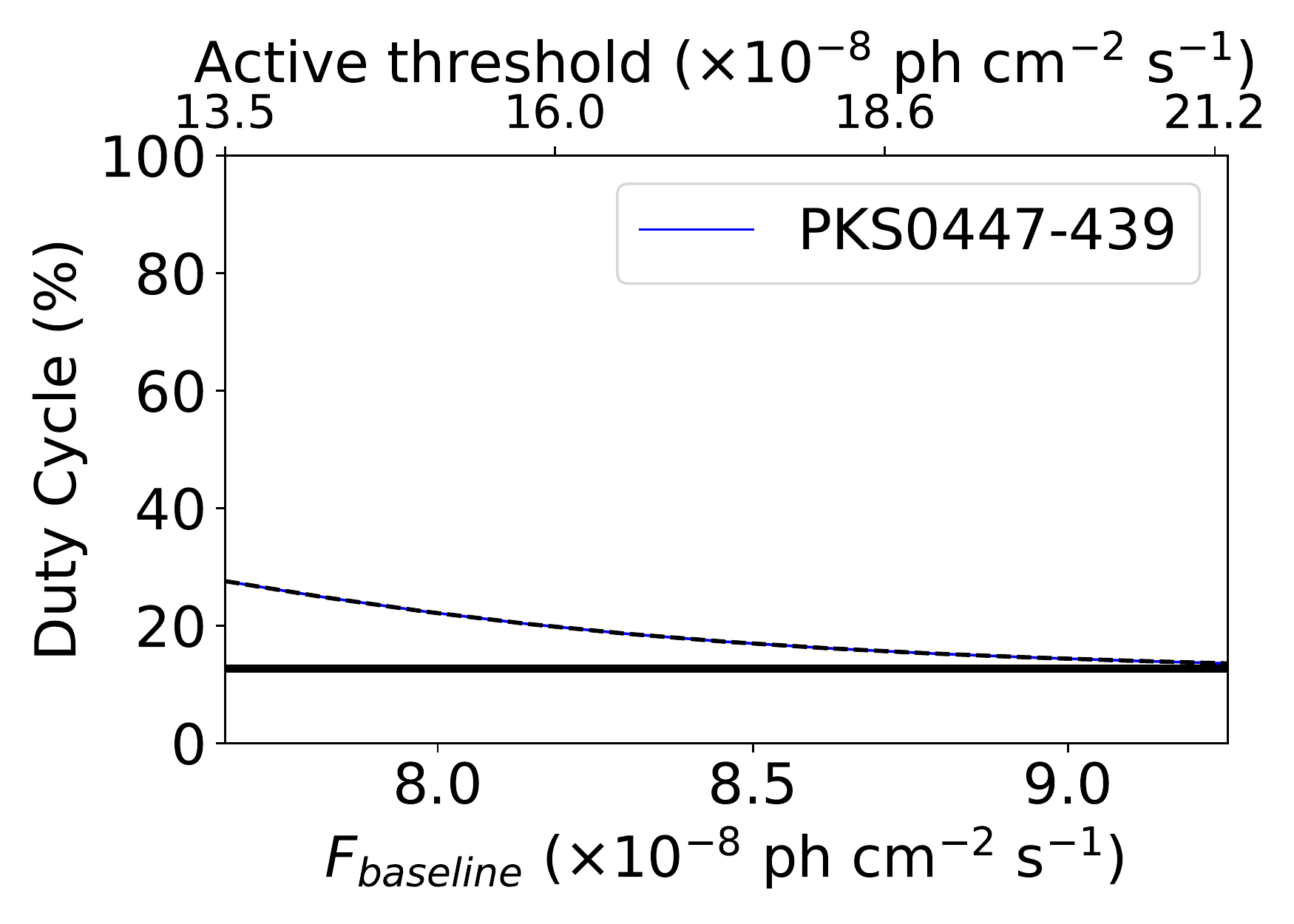}}
\qquad  %%%%%%%
\subfloat{
\includegraphics[width=0.29\textwidth]{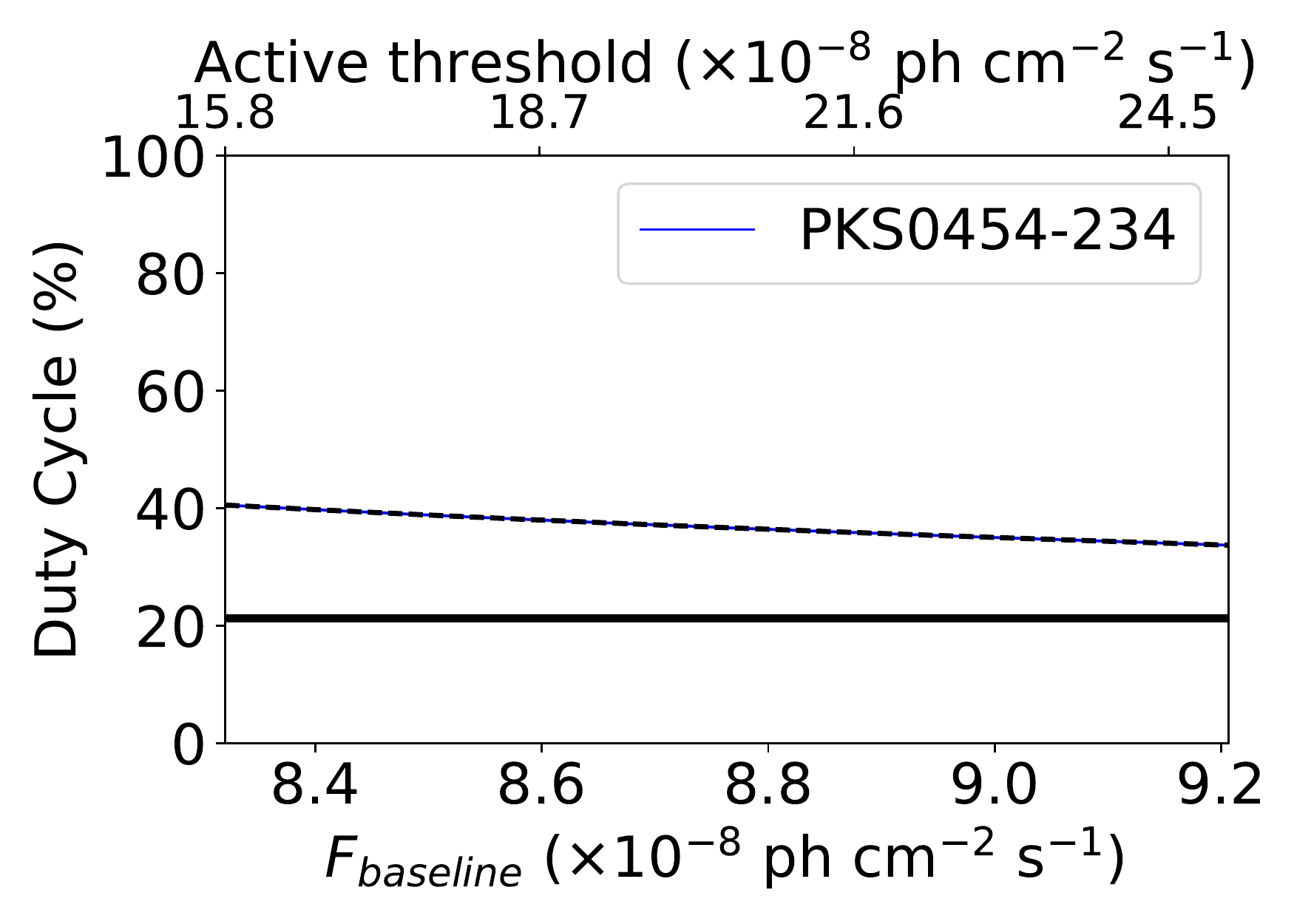}}

\caption{DC vs. flux for each source of the sample, with no EBL absorption. The blue (black) lines represent the DC range inferred from Tluczykont (Vercellone's) criterion.}
\label{fig:fitG-LN}
\end{figure*}

\begin{figure*}
\ContinuedFloat
\centering
\subfloat{
\includegraphics[width=0.29\textwidth]{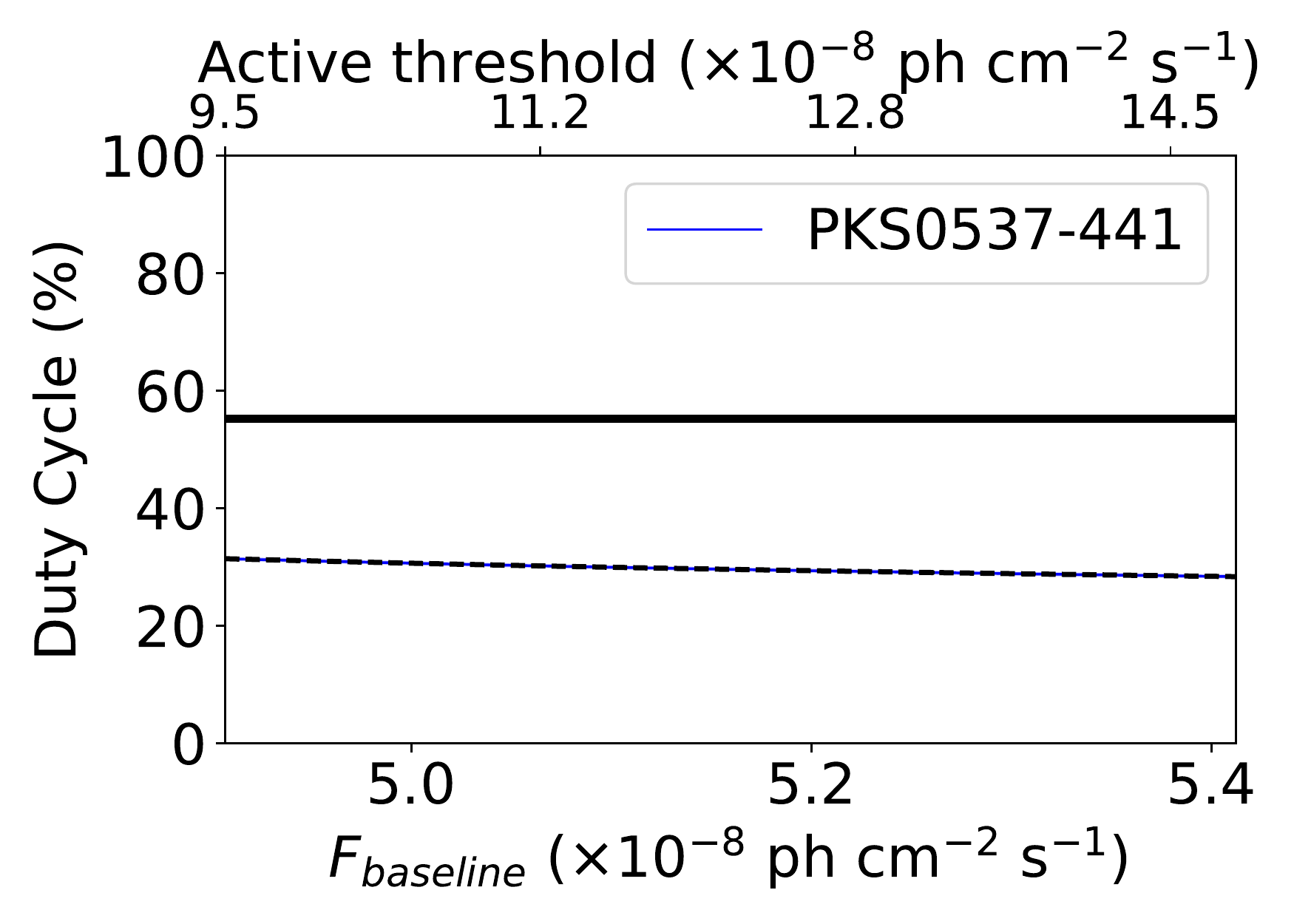}}
\subfloat{
\includegraphics[width=0.29\textwidth]{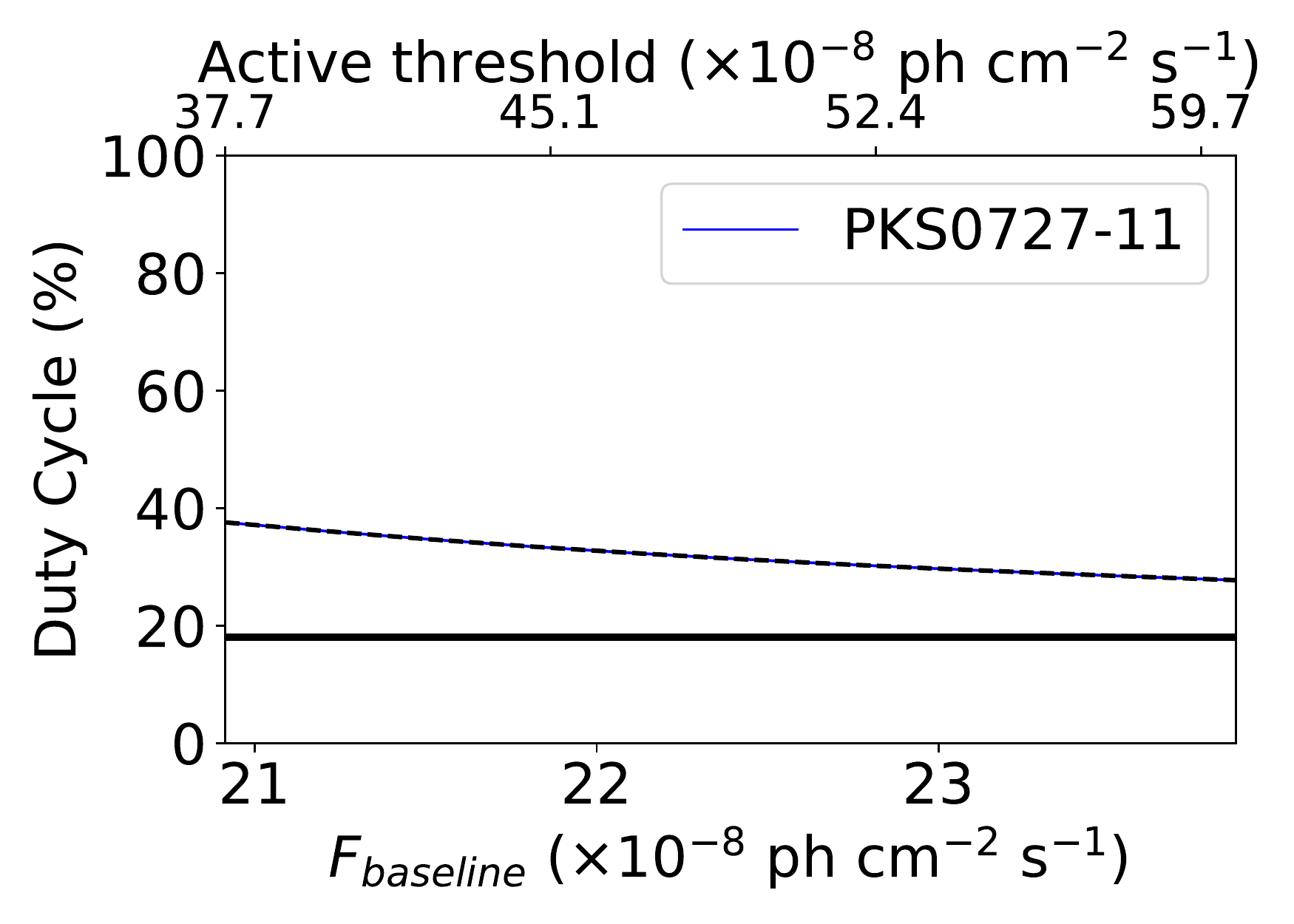}}
\subfloat{
\includegraphics[width=0.29\textwidth]{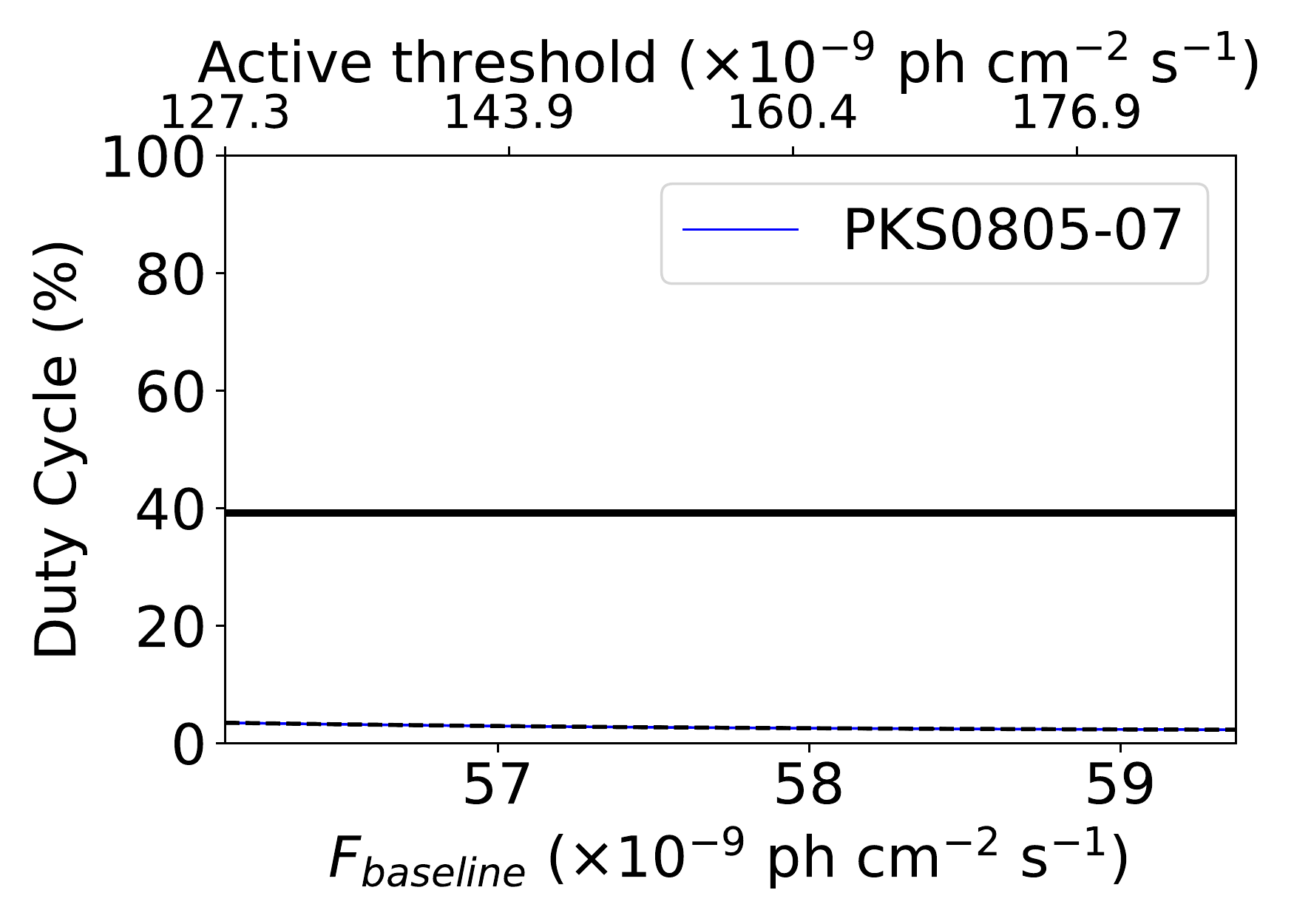}}
\qquad  %%%%%%%
\subfloat{
\includegraphics[width=0.29\textwidth]{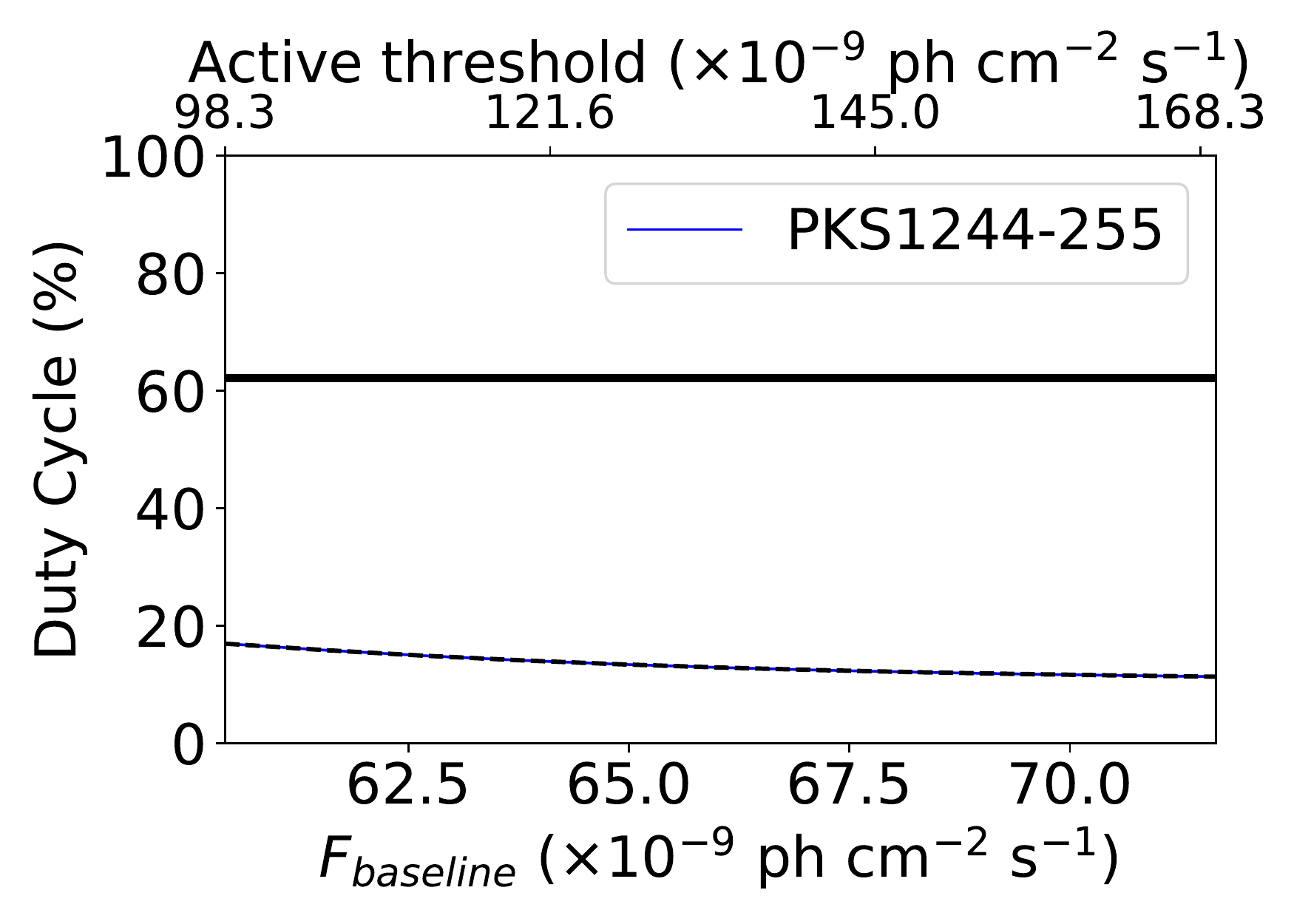}}
\subfloat{
\includegraphics[width=0.29\textwidth]{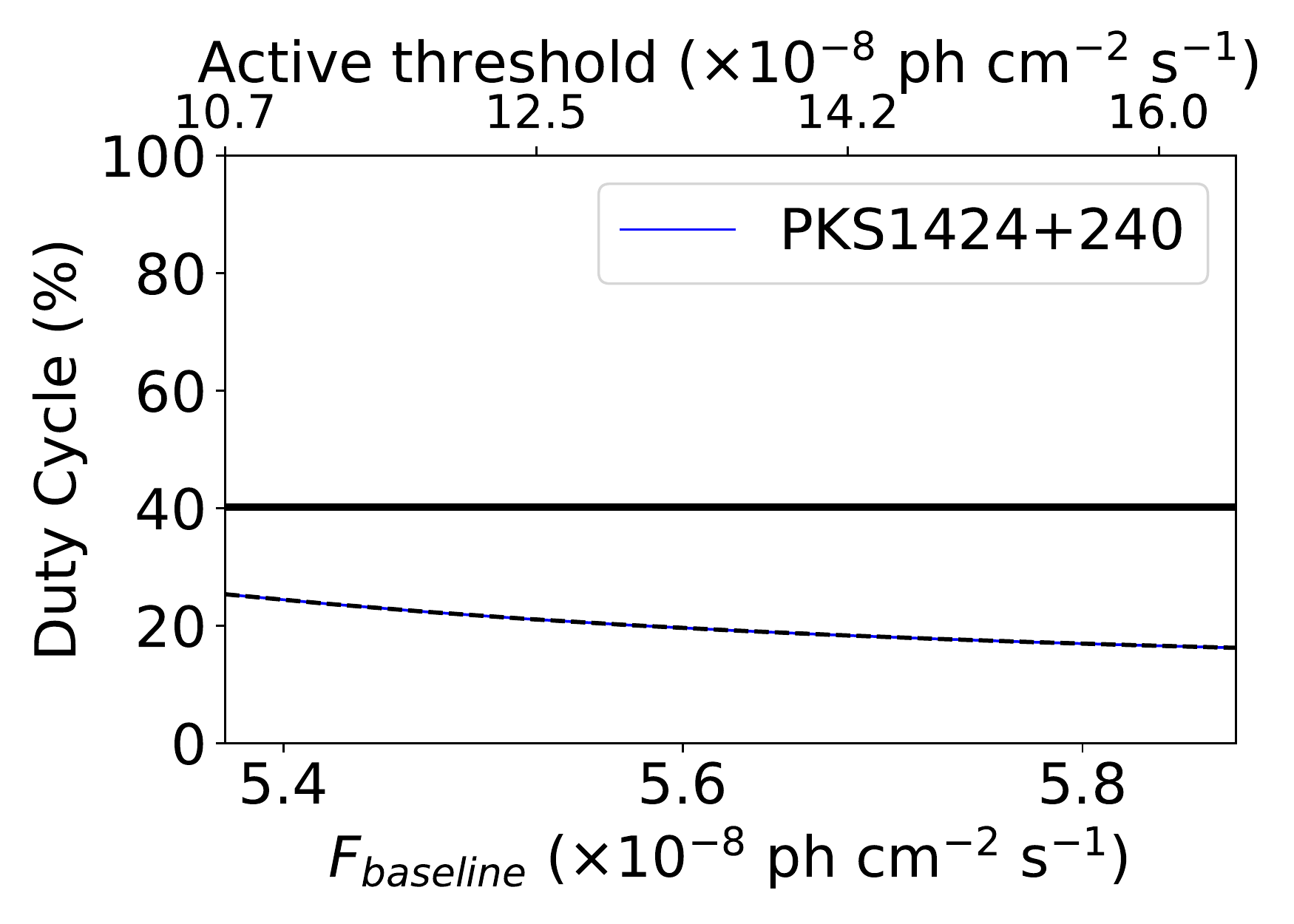}}
\subfloat{
\includegraphics[width=0.29\textwidth]{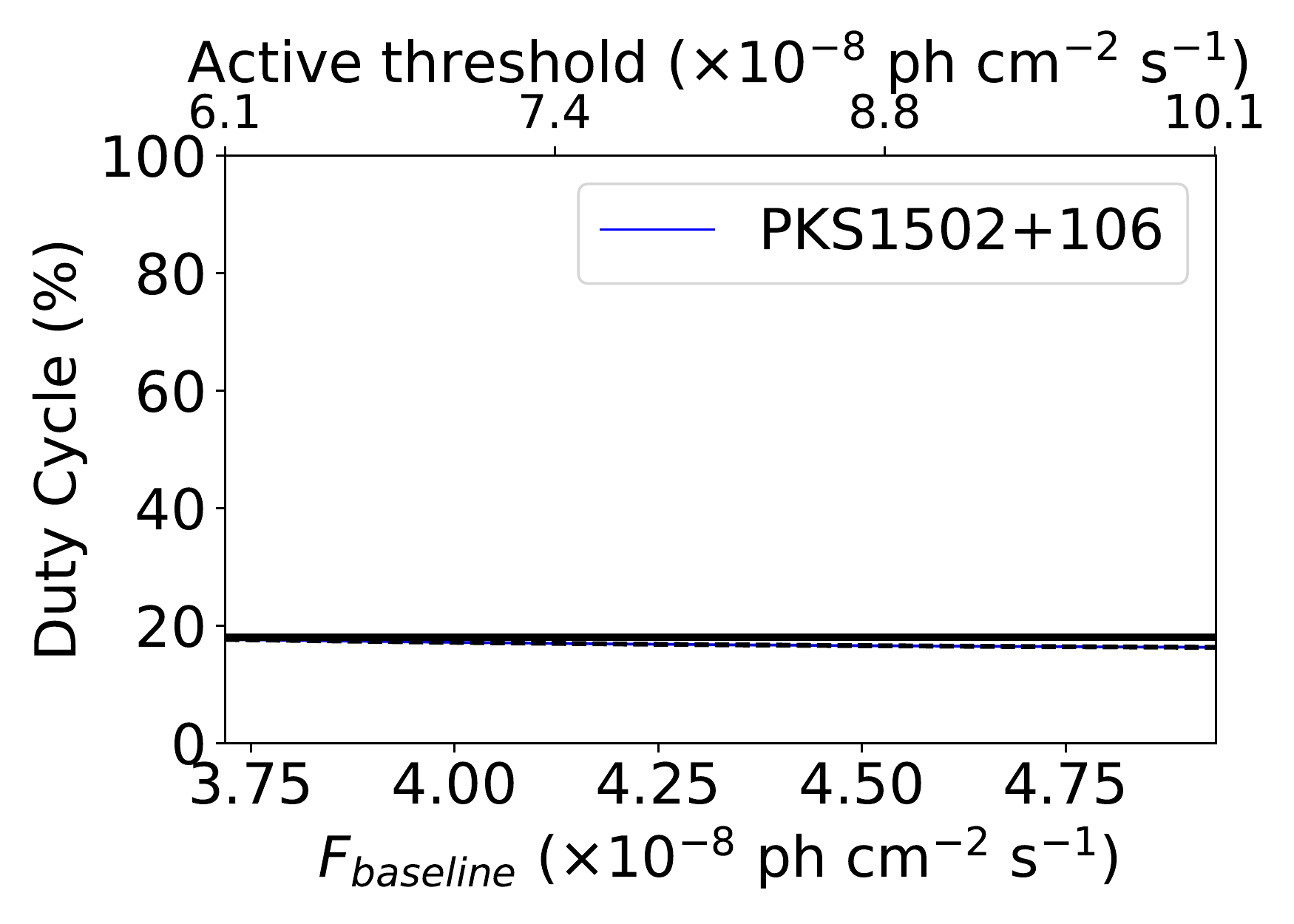}}
\qquad  %%%%%%%
\subfloat{
\includegraphics[width=0.29\textwidth]{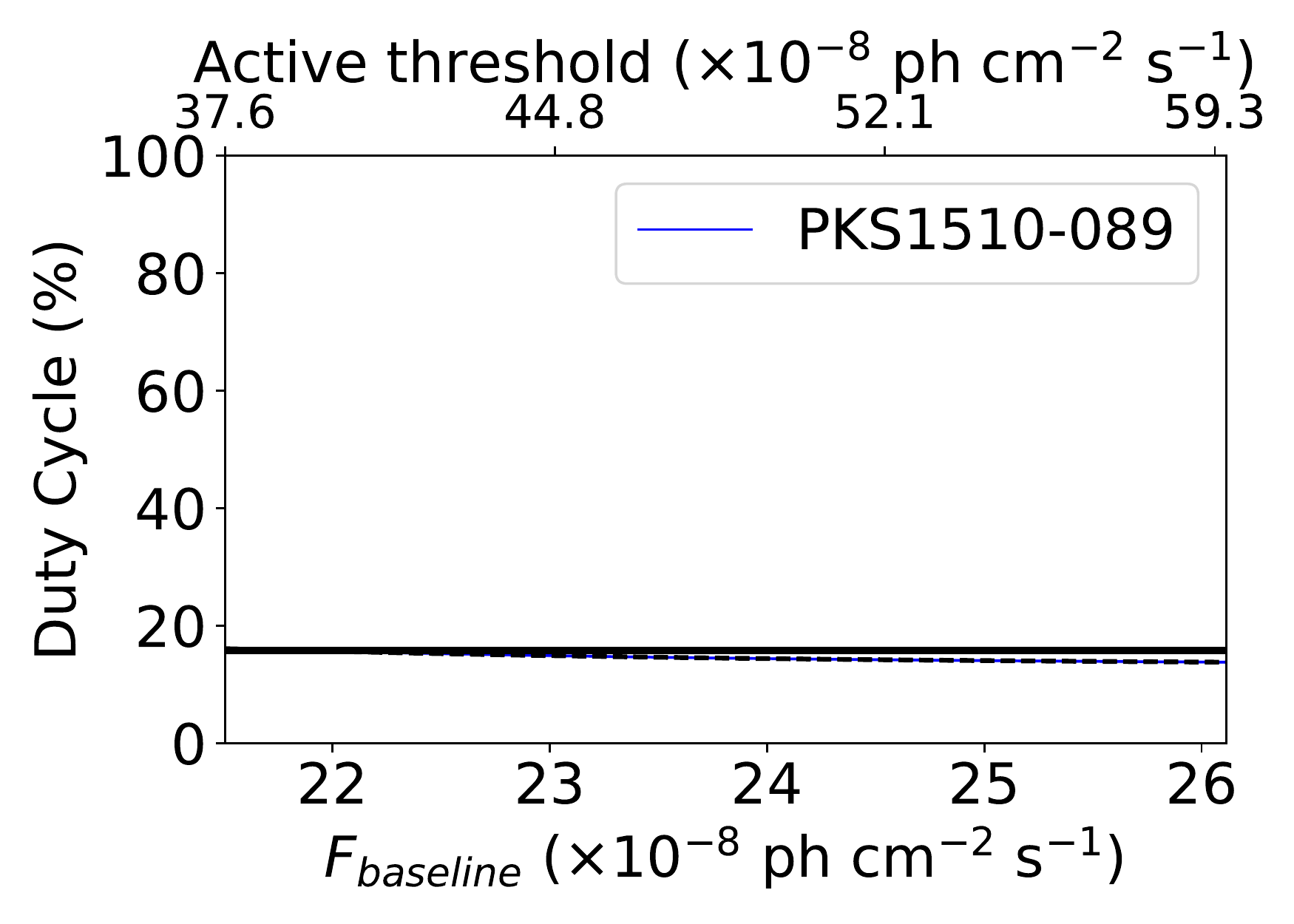}}
\subfloat{
\includegraphics[width=0.29\textwidth]{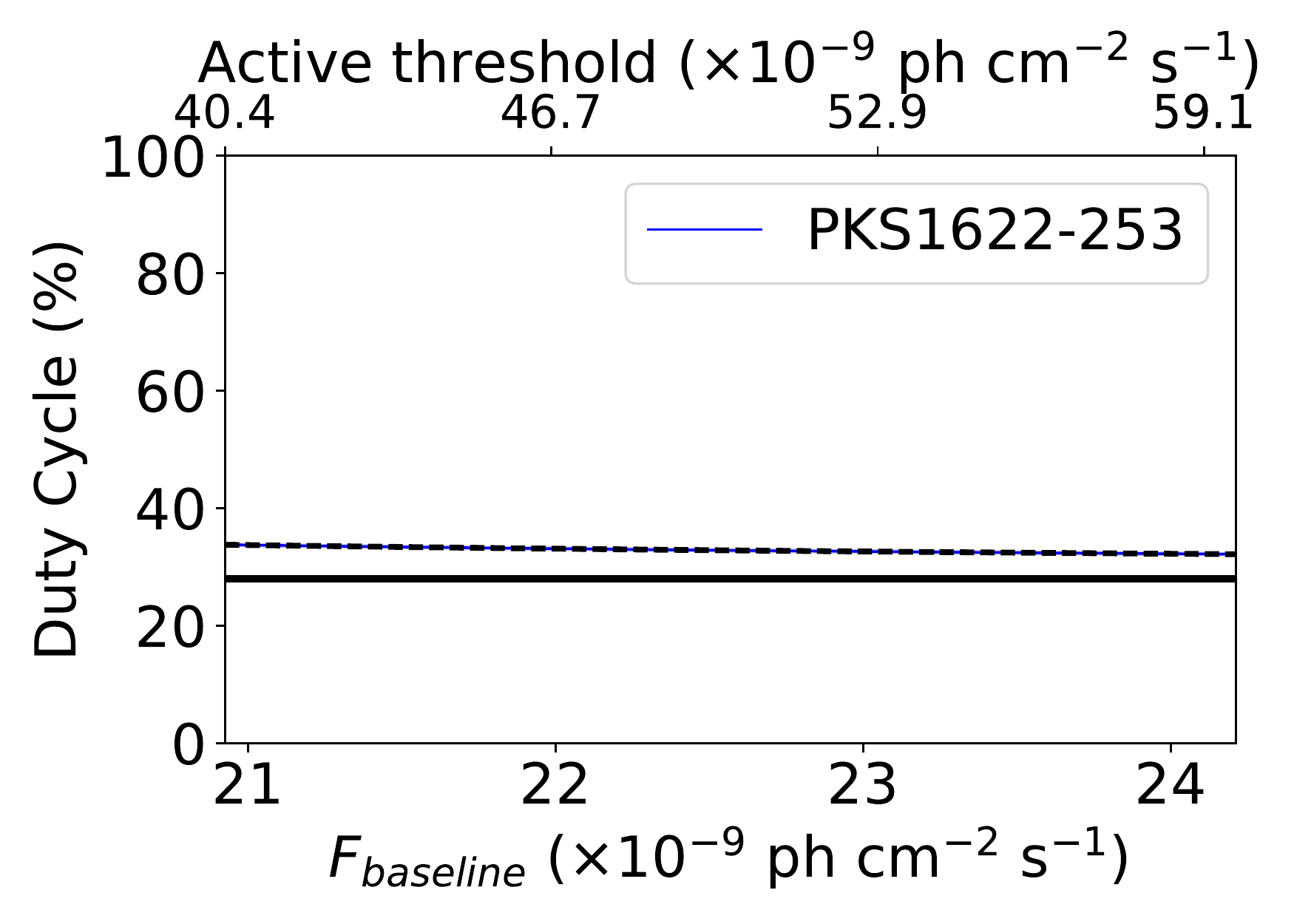}}
\subfloat{
\includegraphics[width=0.29\textwidth]{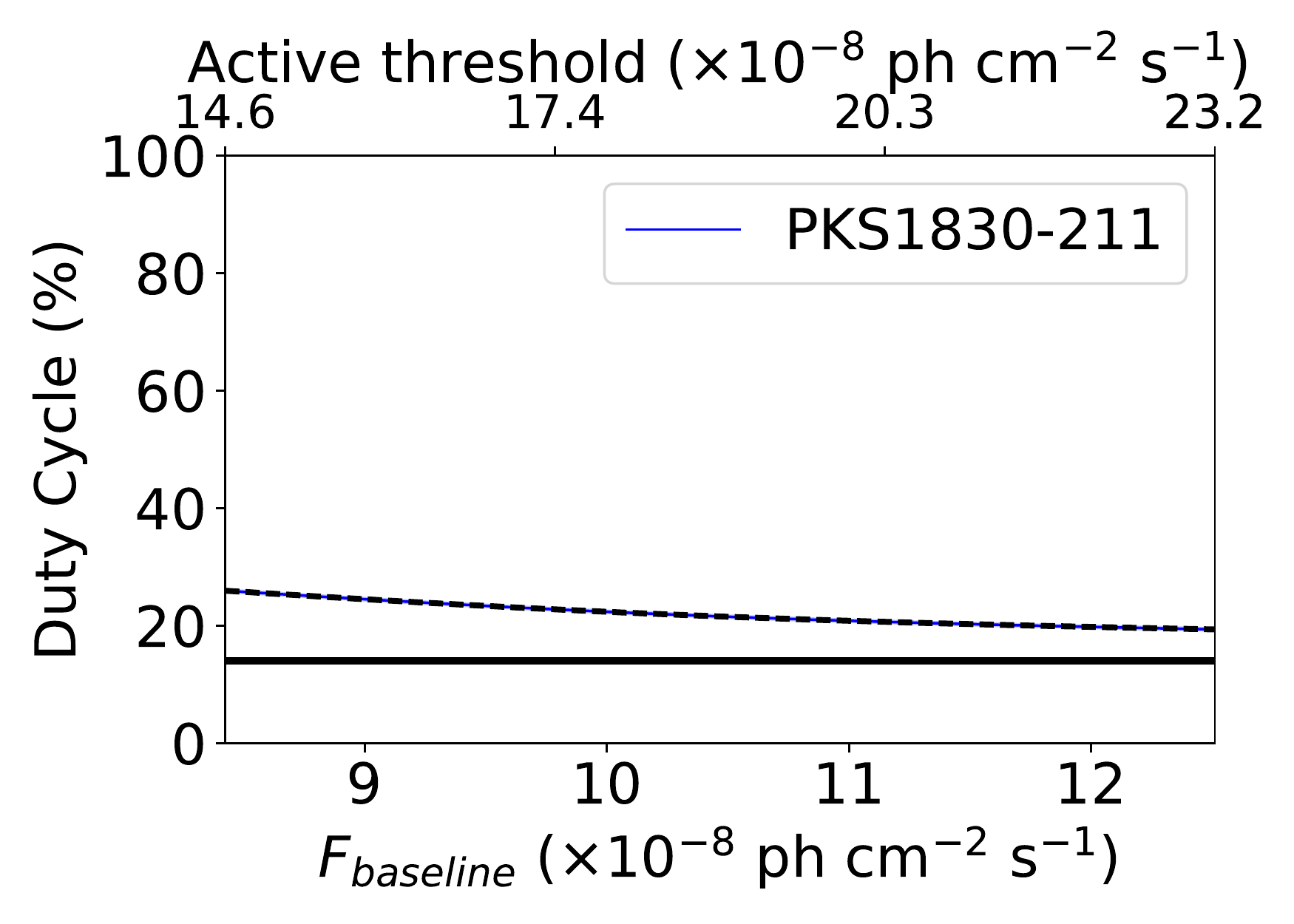}}
\qquad  %%%%%%%
\subfloat{
\includegraphics[width=0.29\textwidth]{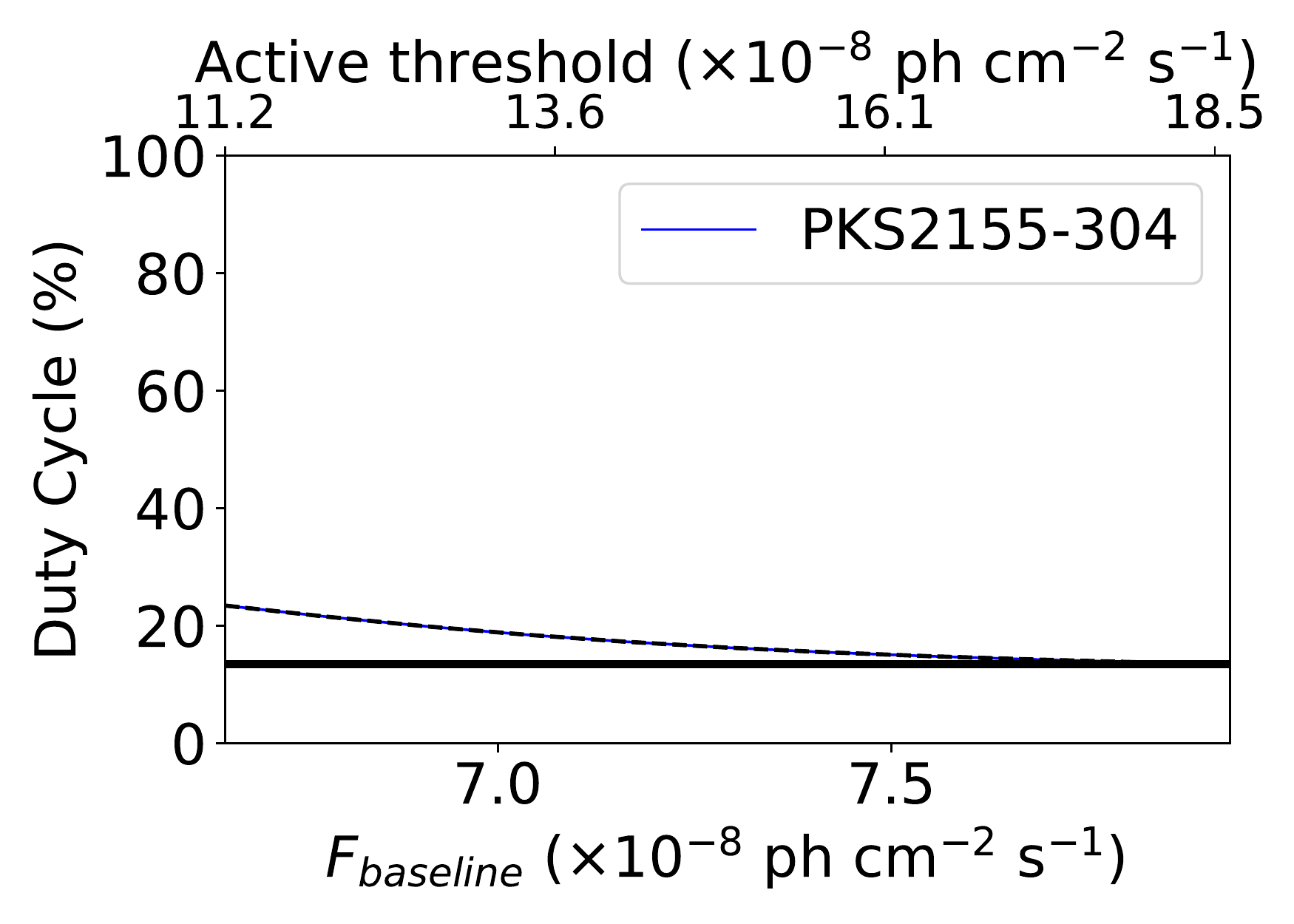}}
\subfloat{
\includegraphics[width=0.29\textwidth]{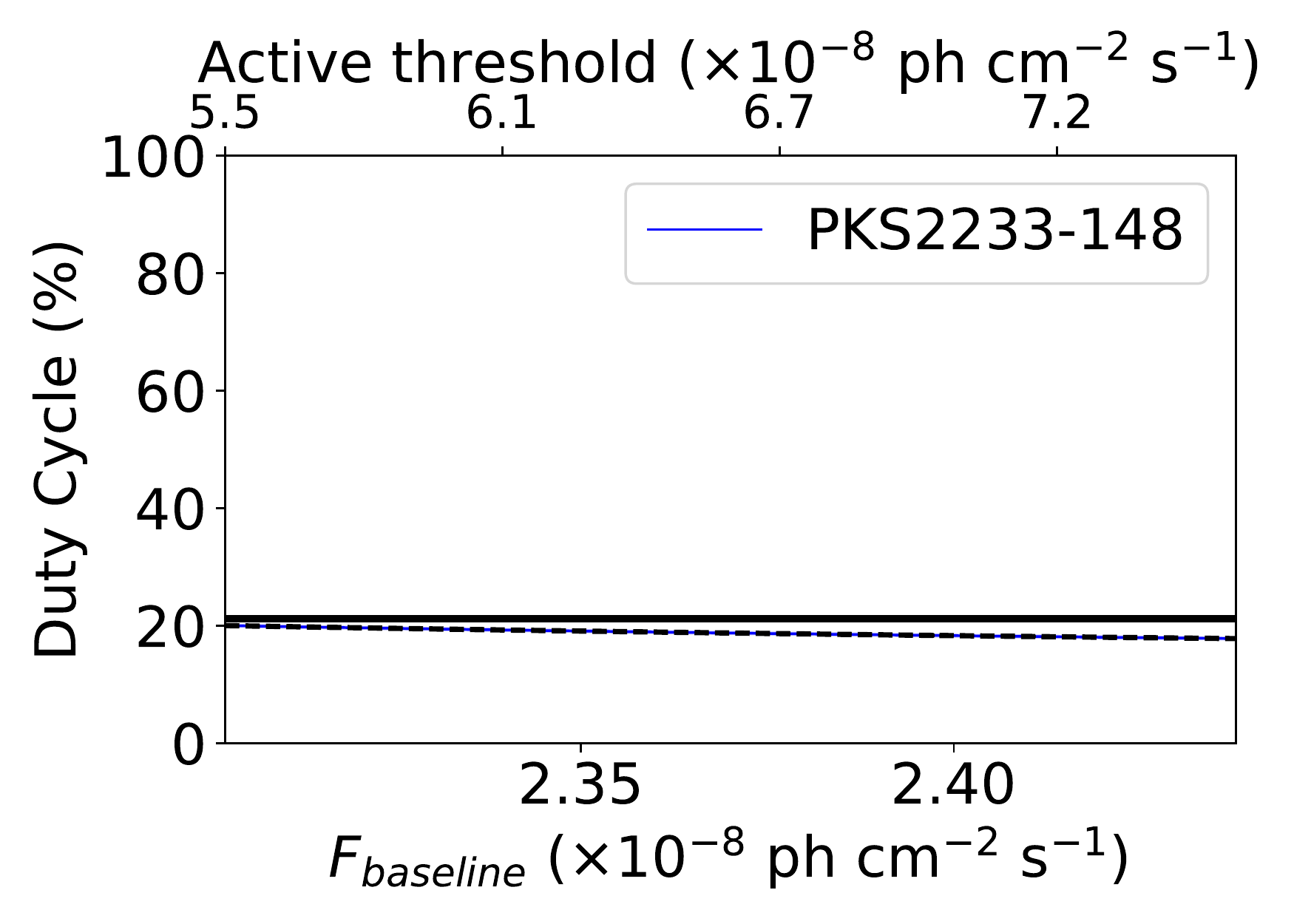}}
\subfloat{
\includegraphics[width=0.29\textwidth]{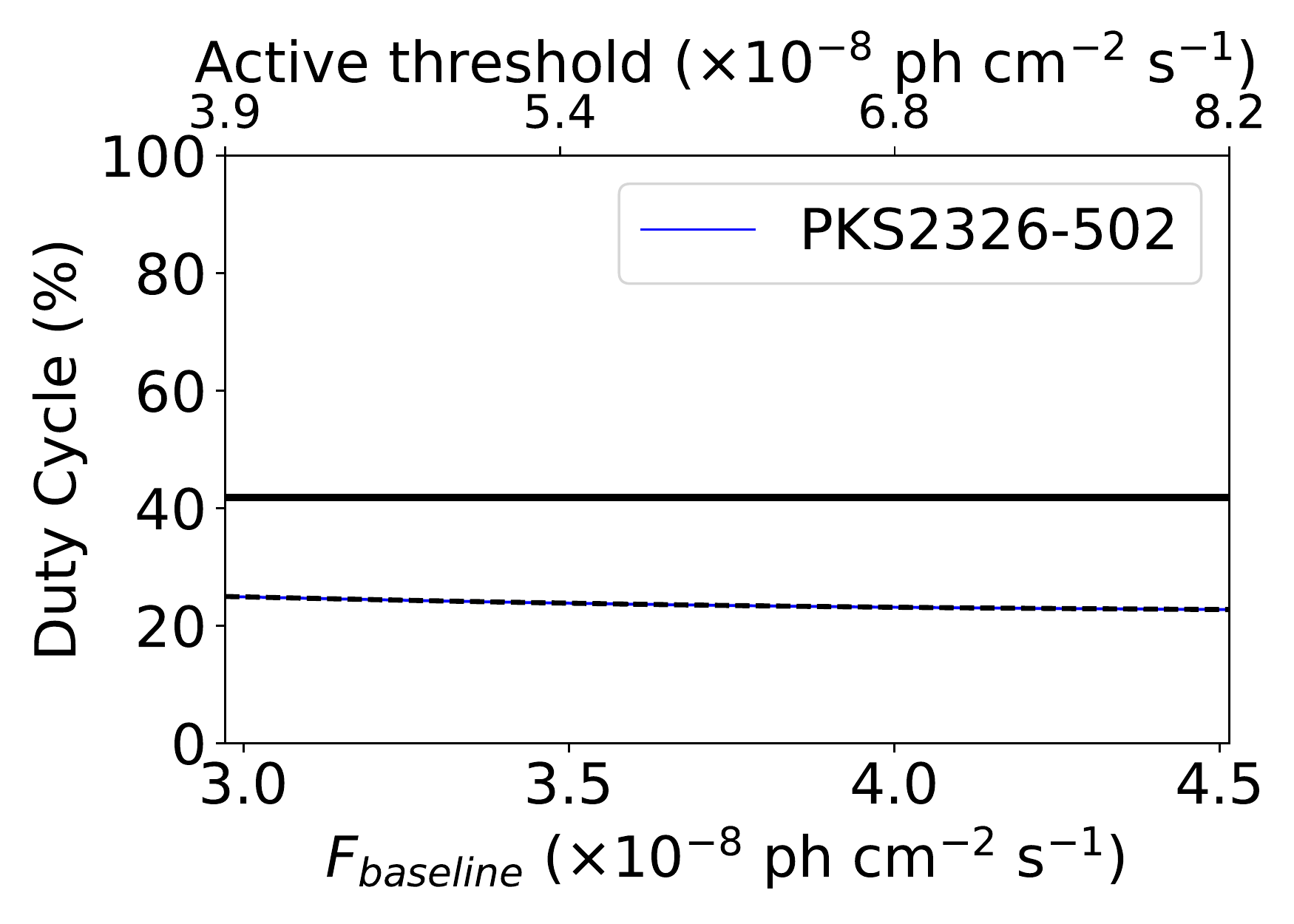}}
\qquad  %%%%%%%
\subfloat{
\includegraphics[width=0.29\textwidth]{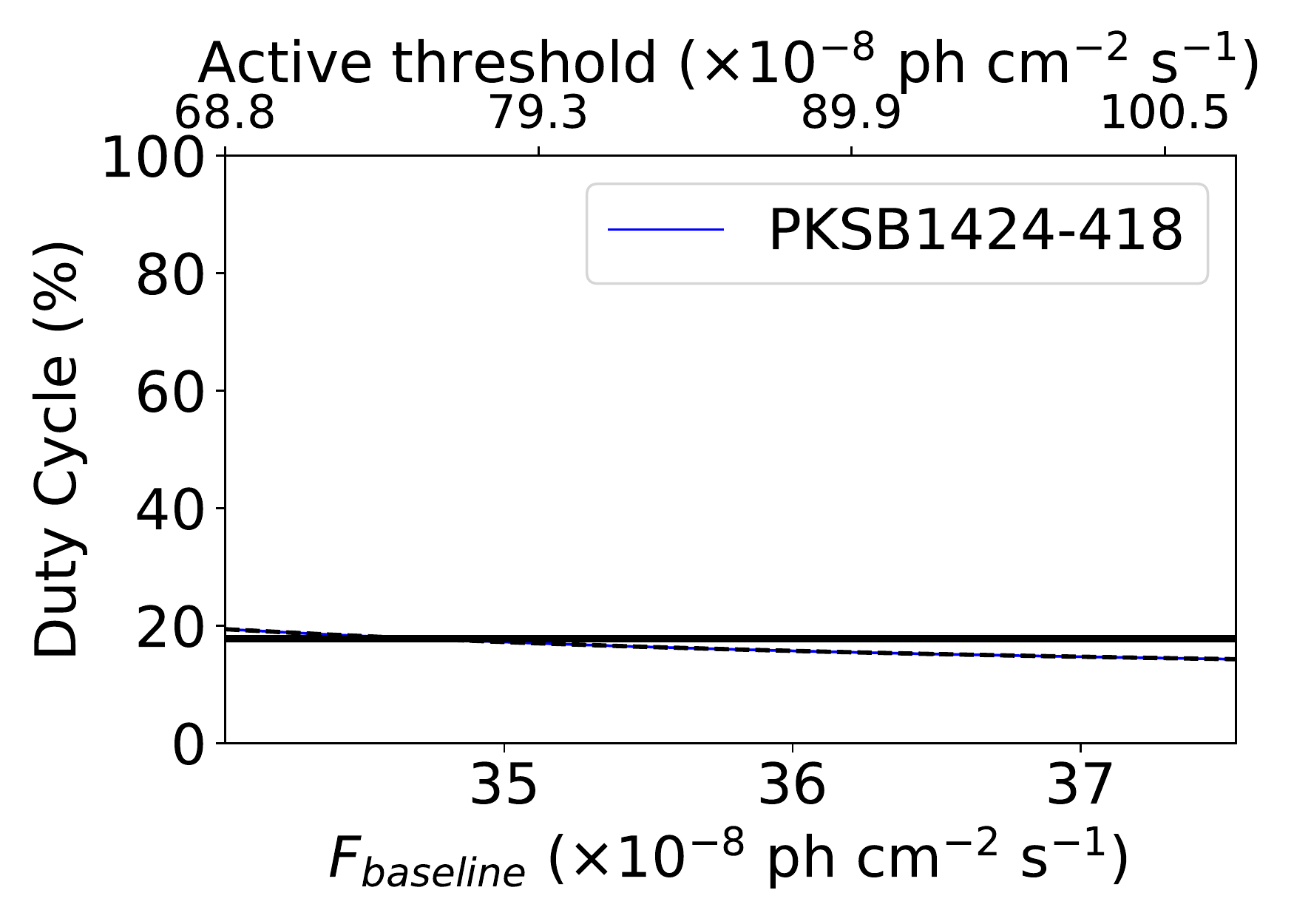}}
\subfloat{
\includegraphics[width=0.29\textwidth]{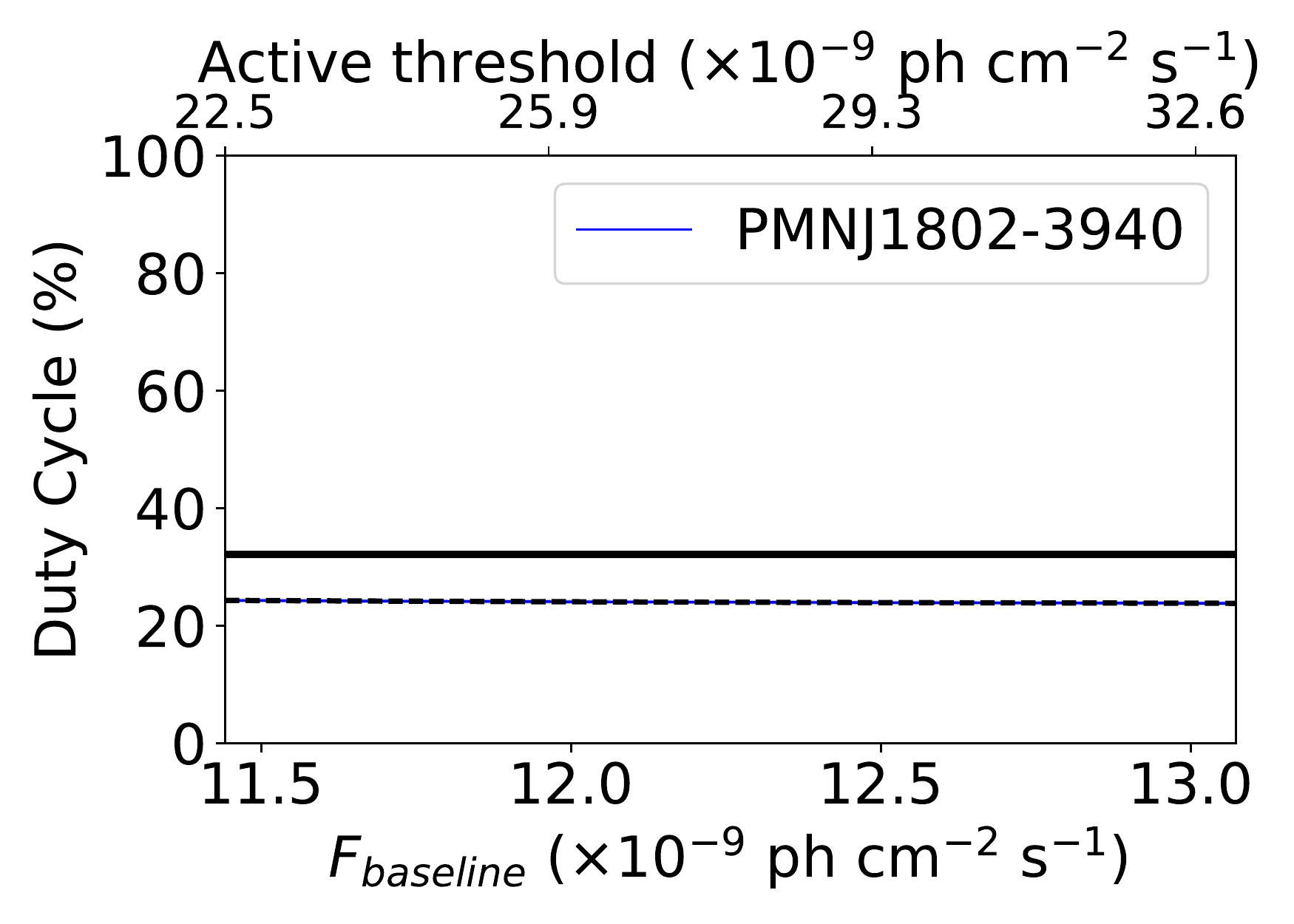}}
\subfloat{
\includegraphics[width=0.29\textwidth]{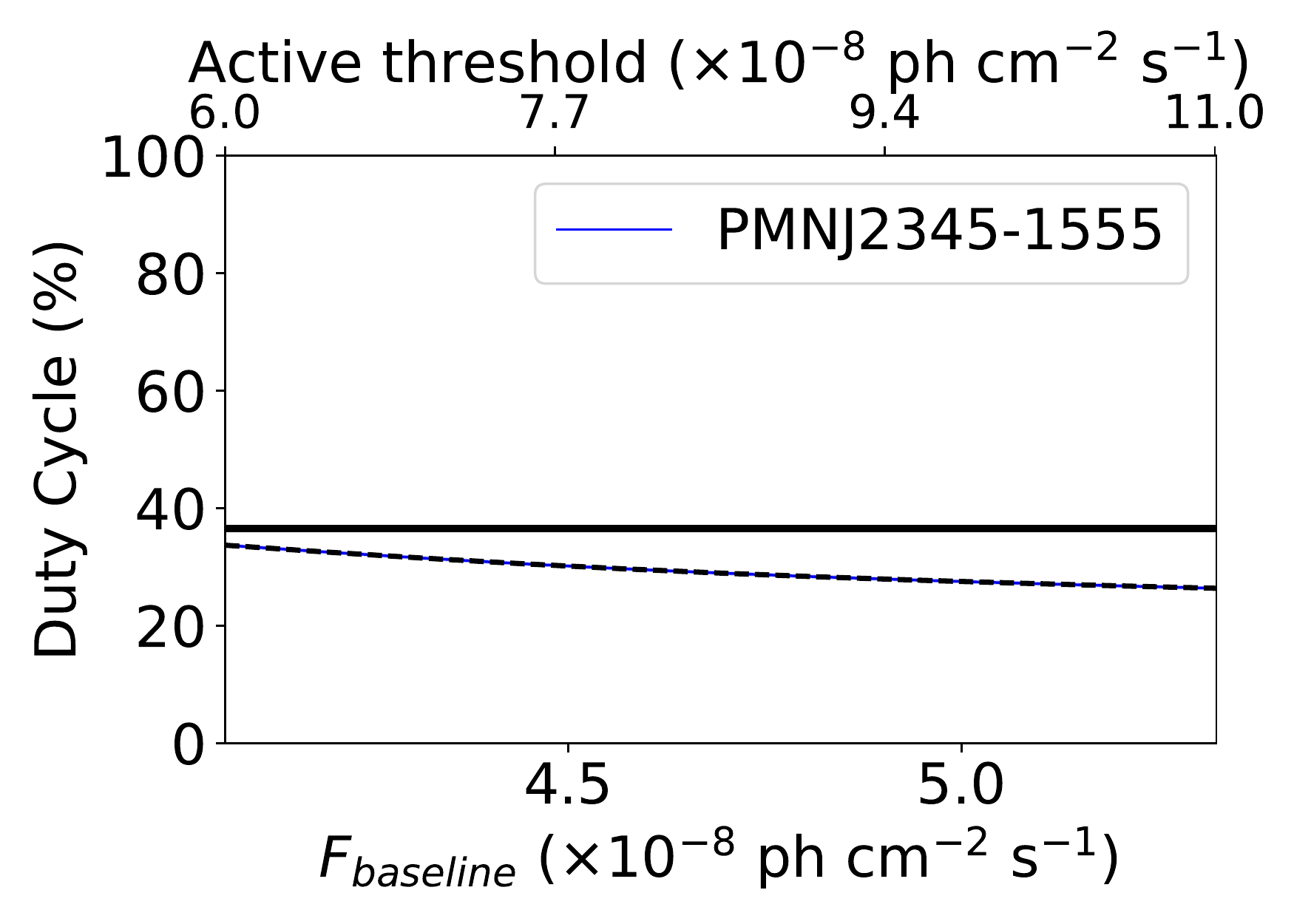}}
\qquad  %%%%%%%
\subfloat{
\includegraphics[width=0.29\textwidth]{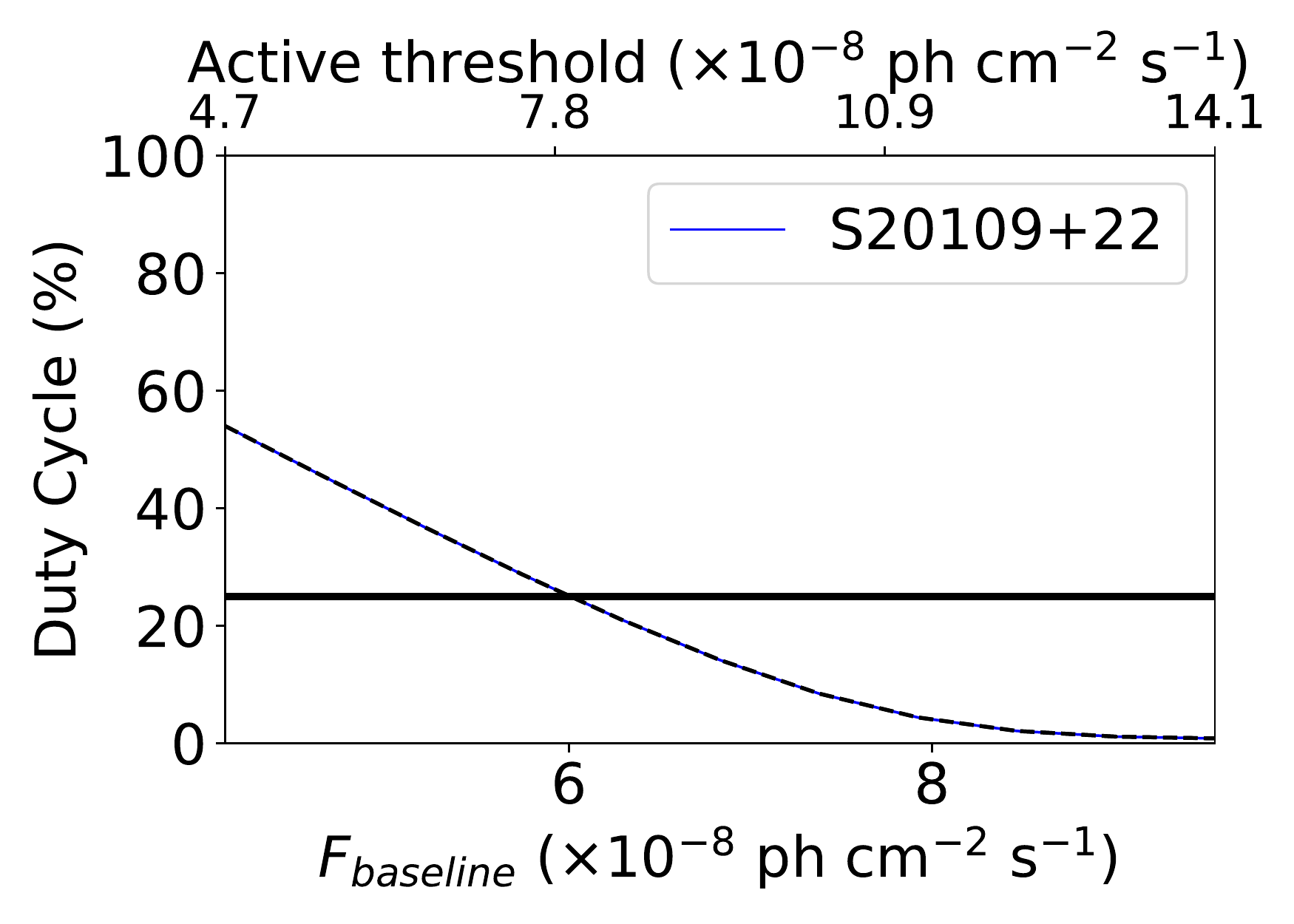}}
\subfloat{
\includegraphics[width=0.29\textwidth]{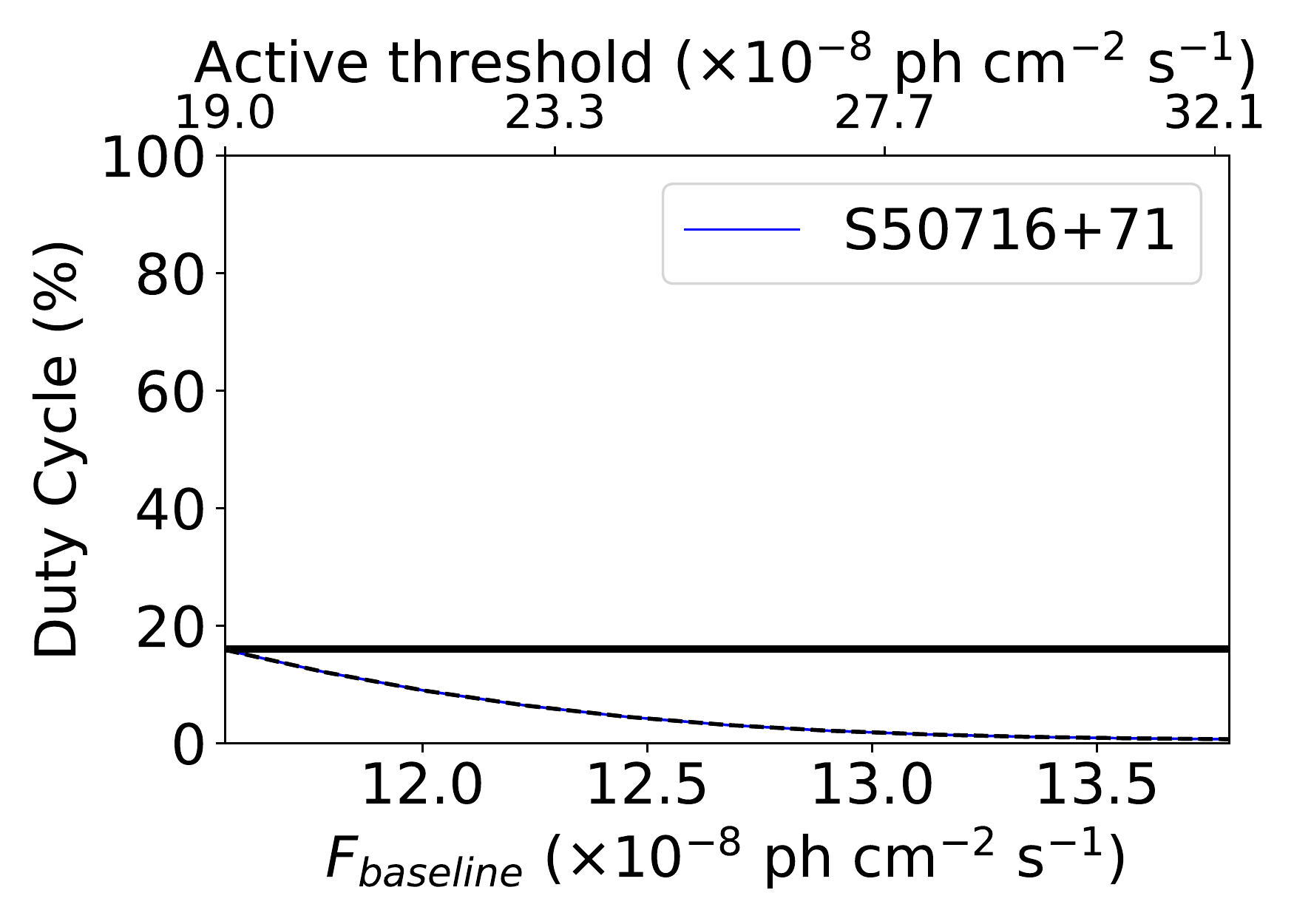}}
\subfloat{
\includegraphics[width=0.29\textwidth]{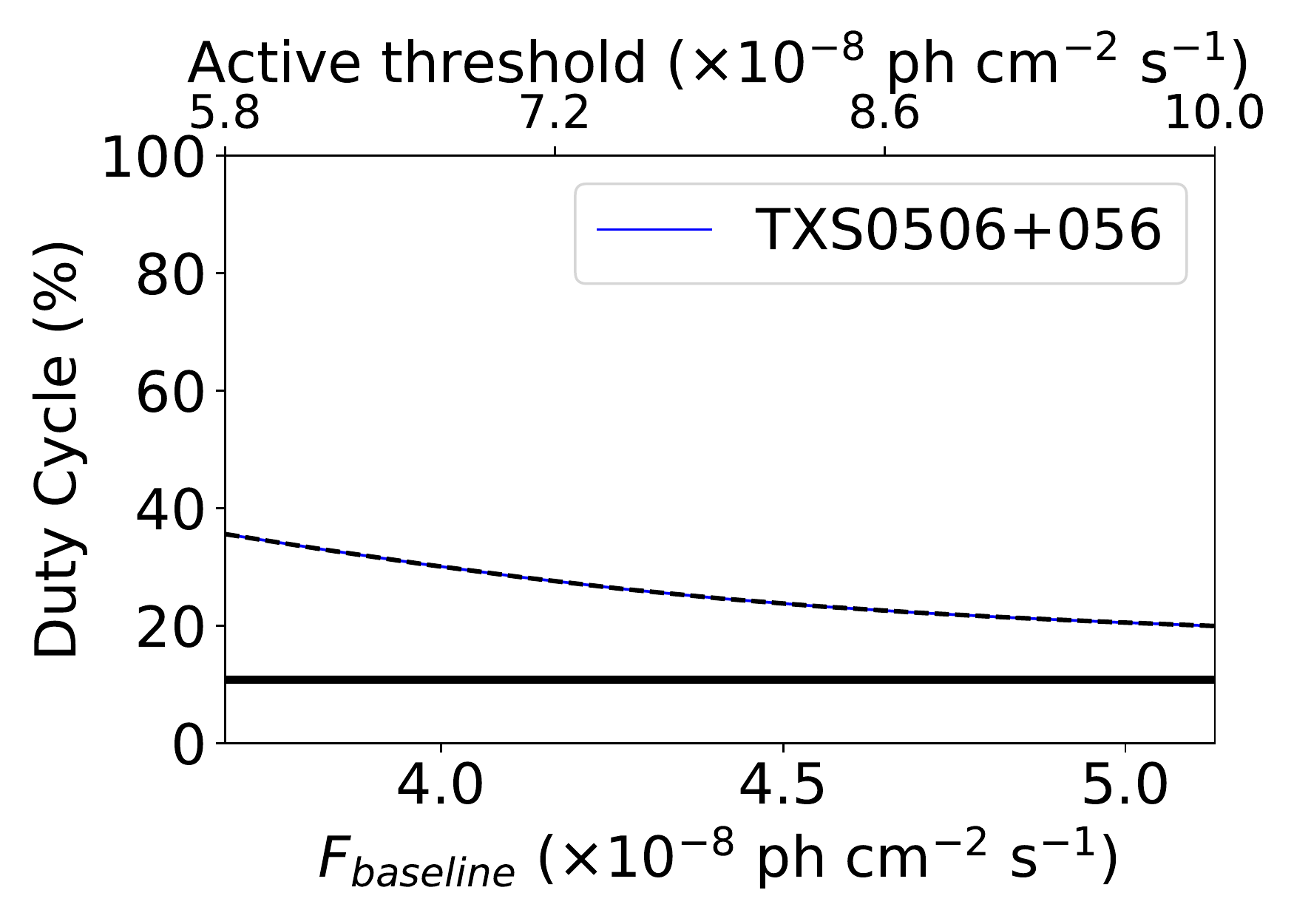}}
\qquad  %%%%%%%
\subfloat{
\includegraphics[width=0.29\textwidth]{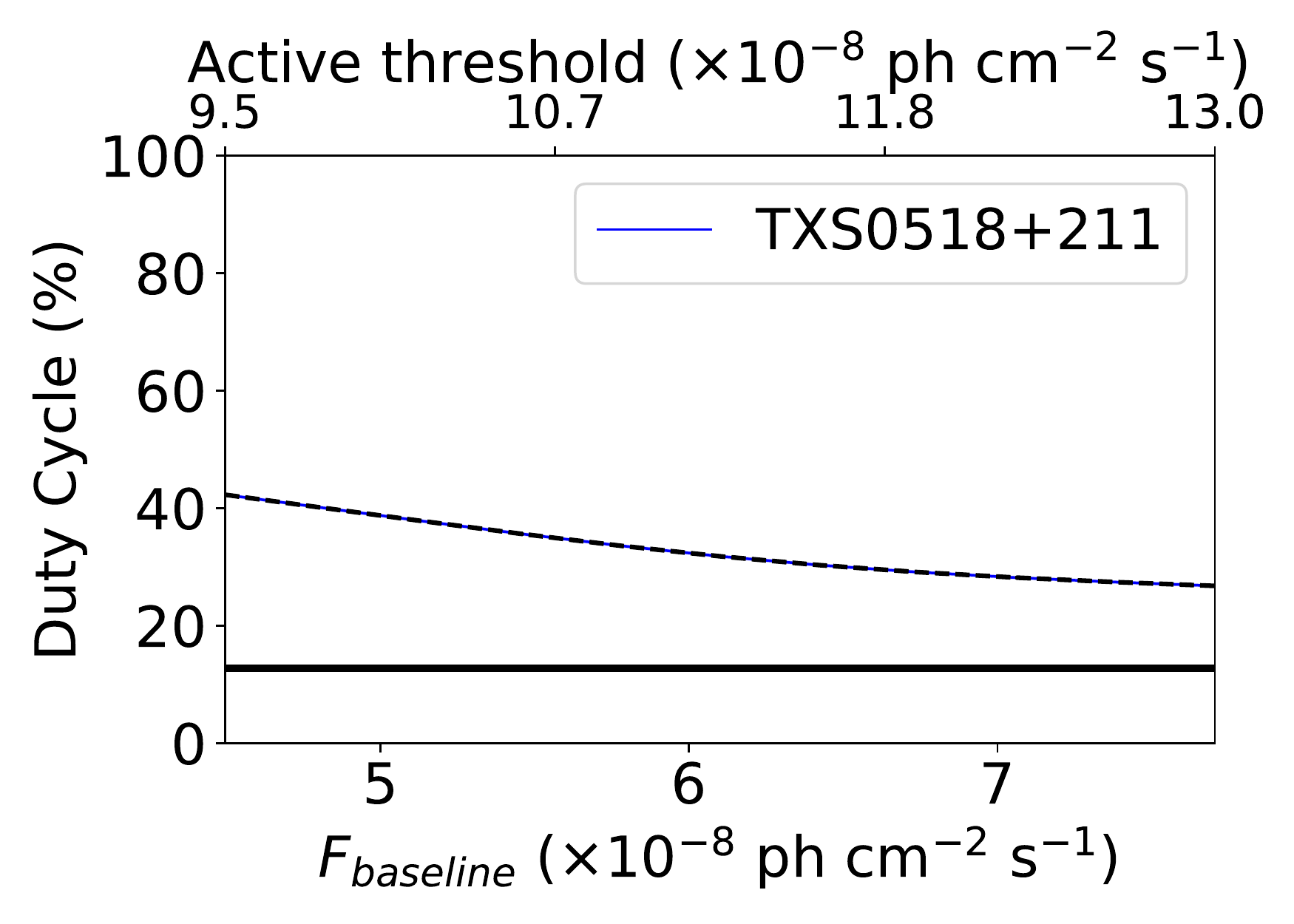}}

\caption{(continued)}
\end{figure*}

\end{appendices}
\end{document}